\pdfoutput=1
\documentclass[a4]{myhab}
\usepackage[pagewise]{lineno}

\usepackage[nottoc]{tocbibind}
\usepackage{graphicx}
\usepackage{fancyhdr}
\usepackage[francais,english]{babel}
\def\ra{\rightarrow }

\def\epem{\mbox{e}^+\mbox{e}^- }

\def\q2{\mbox{Q}^2 }
\begin{document}
\selectlanguage{english}
\input{titlepage.dat}
\input{firstpage.dat}
\cleardoublepage
\pagenumbering{roman}
\chapter*{Preface}
\label{preface}

 This Habilitation is based on selected publications, which represent my major scientific contributions as an experimental physicist to the field of particle accelerators and detectors applied to medical diagnostics and therapy. They are reprinted in Part II of this work to be considered for the Habilitation and they cover original achievements and relevant aspects for the present and future of medical applications of particle physics.
 
 The text reported in Part I is aimed at putting my scientific work into its context and perspective, to comment on recent developments and, in particular, on my contributions to the advances in accelerators and detectors for cancer hadrontherapy and for the production of radioisotopes.
\vskip 0.5 cm
Dr. Saverio Braccini
\vspace*{6pt}

Bern, 25.4.2013

\tableofcontents
\cleardoublepage
\pagenumbering{arabic}     

\chapter*{Introduction}
\label{Introduction}
\addcontentsline{toc}{chapter}{Introduction} 

Particle accelerators and detectors are fundamental instruments in modern medicine to study the human body, to detect and cure its diseases. In the recent years, they reached a high level of sophistication that allowed not only for great discoveries in fundamental physics but also for impressive advances in medical diagnostics and therapy.

Instrumentation issued from fundamental research in physics can be found today in any hospital and nowadays-common medical practices were not even imaginable only a few years ago. While many applications are well established, many others are the focus of scientific research and innovative ideas are continuously appearing at the horizon. In particular, particle accelerators and detectors for cancer hadrontherapy and for the production of isotopes for functional imaging represent cutting-edge fields at the forefront of science and technology. Many disciplines are necessarily involved and common efforts from scientists 
active in research fields such as physics, chemistry, engineering, biology, pharmacy and medicine are necessary not only to pursue knowledge but also to make new diagnostic and treatment modalities routinely available for patients.

In this dynamic and stimulating context, experimental particle physicists can give decisive contributions and bring new ideas through their expertise acquired in fundamental research. This is the main reason why, after 
having been involved in several experimental particle physics projects at CERN,
I have decided ten years ago to focus my scientific efforts on medical applications of physics. It is in the field of particle accelerators and detectors for cancer hadrontherapy and for the production of radioisotopes where I gave my major contributions, and this Habilitation thesis is therefore centred on the achievements in these specific domains.

Accelerated beams of charged hadrons, protons and carbon ions in particular, allow the treatment of solid tumours by external irradiation with unprecedented precision. This modality of radiation therapy is denominated hadrontherapy. To exploit at best the physical and radiobiological properties of charged ions, the development of high performance accelerators and detectors is crucial. In particular, active beam scanning techniques allow {\it painting} the target volume with millimetre precision and innovative systems are under study to optimize the performance and to reduce the size and the complexity of the treatment facilities. Proton and ion linear accelerators ({\it{linacs}}) represent a promising solution to face the challenges of hadrontherapy and, as reported in the first paper reprinted in Chapter~4, their use is proposed for advanced beam scanning. Furthermore, as described in the second paper of Chapter~4, linacs can be coupled to cyclotrons used for other medical purposes, such as the production of radioisotopes for functional imaging or for the production of beams of lower energy carbon ions. This new kind of hybrid accelerator is denominated {\it Cyclinac}. The experience I acquired in the study and in the design of Cyclinacs allowed me in 2007 to be one of the proponents of the new cyclotron laboratory located at the Inselspital, the University Hospital in Bern. This new centre is conceived for top-level radioisotope production for Positron Emission Tomography (PET) and multi-disciplinary research running in parallel by means of a specifically conceived beam transfer line (BTL) terminated in a separate bunker. As reported in the third and fourth publication of Chapter~4, the BTL represents a peculiar feature for a hospital-based cyclotron facility. The Bern cyclotron and its beam line were designed, constructed, and commissioned under my coordination. This laboratory,
whose beam line is a unique facility in Switzerland,  is now fully operational and the first research programs are ongoing.

Highly performing particle detectors are crucial for scientific research as well as for applications. In particular, the accurate monitoring of the beams during irradiations for medical purposes is of prime importance. In hadrontherapy, the precise on-line measurement of the position, the intensity and the shape of the clinical beams is crucial, together with the knowledge of the range of the accelerated ions. To address these problems, I contributed to design, develop and test three novel particle detectors, as reported in the publications reprinted in Chapter~5. 
Segmented ionization chambers constitute an effective solution
for on-line monitoring of the clinical beams during hadrontherapy treatments, especially in the case of {\it spot} scanning. 
These developments led to the conception and to the construction of a specific strip ionization chamber developed for the Gantry~2 at the Paul Scherrer Institute (PSI), the most advanced centre for {\it spot} scanning worldwide. In hadrontherapy, the dose given to the healthy tissues surrounding the tumour has to be minimized and methods are under study to improve the knowledge of the particle range; Proton radiography is one of the most promising ones and we developed at LHEP an innovative technique based on nuclear emulsion films.
Furthermore, ion beams have to be precisely controlled along the beam lines, either for hadrontherapy or for the production of isotopes. For this purpose, I proposed a new device based on doped silica and optical fibres. The first prototype is installed on the research beam line of the Bern cyclotron laboratory and is performing already to specifications.

The present document is structured into two parts, each containing three Chapters. In Part I, Chapter~\ref{ch1} presents an introduction to accelerators and detectors applied to medicine, with particular focus on cancer hadrontherapy and on the production of radioactive isotopes. In Chapter~\ref{ch2}, my publications on medical particle accelerators are introduced and put into their perspective. In particular, high frequency linear accelerators for hadrontherapy are discussed together with the new Bern cyclotron laboratory. Chapter~\ref{ch3} is dedicated to particle detectors with particular emphasis on the three instruments I 
contributed to propose and develop. In Part II, the mentioned publications are reprinted in Chapters~4 and~5, together with three review papers on the present and future of cancer hadrontherapy reprinted in Chapter~6.

\input{part_I.dat}
\chapter{Particle Accelerators and Detectors applied to Medicine}
\label{ch1}

The applications of nuclear and particle physics are fundamental in modern medicine and deeply influenced its development. In particular, the conception and the realization of specific particle accelerators and detectors determined crucial advances in medical imaging and cancer radiation therapy. These instruments are today at the basis of scientific progress in these domains.

\begin{figure}[t!]
 \centering
   \includegraphics[height=7 cm]{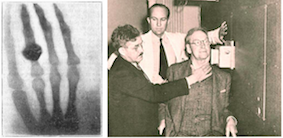}
     \caption{
     The paper announcing the discovery of X-rays made by W. C. R\"ontgen in November 1895 already contains their first medical application, the radiography of the hand of his wife obtained only one month later, in December 1895 (left). Ernest and John Lawrence performed the first salivary gland cancer treatments using one of the first cyclotrons in 1936 (right).  
     }
       \label{fig:Roentgen-Lawrence}
\end{figure}

Since the very beginning at the end of the nineteenth century, radiation sources of increasing energy and power together with specific instruments to detect subatomic particles led to great discoveries in fundamental physics and, in parallel, to revolutionary and often unexpected medical applications. These developments changed the life of many individuals and produced a direct benefit to the society.
A textbook example is the discovery of X-rays by Wilhelm Conrad R\"ontgen in 1895~\cite{Roentgen}. Only one month after the first observation of this new and at that time mysterious kind of penetrating radiation, R\"ontgen had the idea that different tissues of the human body could be characterized by different stopping capabilities. This intuition is at the basis of the first radiography, which is reported 
in Fig.~\ref{fig:Roentgen-Lawrence} (left). Already a few months after the discovery, this novel clinical methodology allowed saving a large number of human lives, especially in battlefields. Radiography surely represents the first fundamental contribution of radiation physics to medicine and the beginning of a fascinating scientific adventure, which is still continuing nowadays.

Right after the discovery of X-rays, many fundamental findings followed, opening the way to the understanding of the structure of matter and of its fundamental interactions. In 1896, natural radioactivity was discovered in uranium minerals by Henri Becquerel~\cite{Becquerel} and, two years later, polonium and radium were discovered by Pierre Curie and Maria Sk\l odowska Curie~\cite{Curie}. Although the effects of ionizing radiations in biological tissues were not known at that time, the idea of using radiations to cure cancer was soon conceived and some superficial tumours were successfully treated by applying a radium source directly in contact with the skin. These pioneering treatments signed the beginning of cancer radiation therapy. A basic problem came out immediately: to be effective, radiations must be selectively directed towards the tumour target, sparing at best the surrounding healthy tissues. This concept is today called {\it local control} and represents the driving force that, from these first attempts, brought to the most recent developments in cancer radiation therapy as, for example, the use of precise pencil proton or ion beams in cancer hadrontherapy.

The first medical application of ion accelerators came only a few years after the conception and construction of the first cyclotron by Ernest Lawrence and Stan Livingston in 1931~\cite{LawrenceLivingston}. John Lawrence -  medical doctor and brother of Ernest - injected into humans an isotope of phosphorus - the $\beta^-$ emitter $^{32}$P - produced with one of the first cyclotrons to cure leukaemia. These early clinical studies signed the birth of nuclear medicine. Furthermore, the Lawrence brothers irradiated patients affected by salivary gland tumours using neutron beams produced by 5~MeV accelerated deuterons on a beryllium target~\cite{LawrenceAebersold}, as shown in Fig.~\ref{fig:Roentgen-Lawrence}~(right). Since fast neutrons produce nuclear fragments, these treatments represent the first use of ions in cancer therapy.

In the same years, the discovery of the properties of slow neutrons in inducing radioactivity in matter by Enrico Fermi and his group in Rome in 1934~\cite{Fermi} opened the way not only to nuclear energy but also to large productions of radioisotopes for medicine. In 1935, George Charles de Hevesy performed the first studies on the metabolism of rats using $^{32}$P~\cite{Hevesy}, demonstrating that the formation of bones is a dynamic process in which about 30\% of the phosphorus in the skeleton is replaced in about twenty days. This pioneering studies paved the way for the use of artificially produced radioactive isotopes in life sciences. 

The discovery of $^{99m}$Tc by Emilio Segr\`e and Glenn Seaborg~\cite{Segre} in 1938 has been fundamental for nuclear medicine. This isotope is characterized by a half-life of 6 hours and is naturally produced by the decay of $^{99}$Mo, which has a much longer half-life of 66 hours. This allows the realization of $^{99}$Mo-$^{99m}$Tc generators~\cite{Tucker}, which make $^{99m}$Tc daily available in hospitals. The conception of the gamma camera detector~\cite{Anger} and the development of a large variety of technetium based compounds led to the large diffusion of Single Photon Emission Tomography (SPECT), which accounts today for more that 80\% of the examinations in nuclear medicine.

With the aim of producing high-energy electron beams for fundamental physics, in 1947 William Hansen, together with the brothers Sigurd and Russel Varian~\cite{Varian}, developed the first linear electron accelerator, which is today the main instrument used in cancer radiation therapy. All modern hospitals are in fact equipped with one or more of these devices and about twenty thousands patients every ten million inhabitants are treated every year by means of these machines in western countries. It is interesting to remark that of the about twenty thousands accelerators nowadays running in the world, about one half are linacs used in radiation therapy~\cite{Maciszewski}.

From the first Geiger-M\"uller tubes developed in 1908~\cite{Geiger}, particle detectors underwent a continuous progress that led to the most modern technologies used today in particle physics experiments, such as ATLAS and CMS at the Large Hadron Collider (LHC) at CERN. These developments gave fundamental contributions to science and, in the medical field, made possible innovative non-invasive imaging modalities that allow exploring the human body, as it were {\it transparent}. I have personally contributed to this adventure, first conceiving, constructing and testing high precision muon chambers for the ATLAS detector~\cite{Braccini_ATLAS} and later designing, realizing and operating the novel medical detectors discussed in Chapter~\ref{ch3}.

Computed Tomography (CT) was first developed in 1972~\cite{HounsÞeld} and it is based on the use of X-ray beams passed through the patient's body at different angles to obtain cross-sectional images - the so-called {\it slices} - of the tissues being examined. By putting together successive cross-sectional images, it is possible to obtain a precise three-dimensional reconstruction of the morphology, which represents a fundamental bit of information for studying and detecting a large variety of pathologies. For this reason CT is today one of the most common diagnostic techniques and, in particular, CT three-dimensional imaging is used as the basis for treatment planning in radiation oncology.

The first full body Magnetic Resonance Imaging (MRI) scan was performed in 1977~\cite{Damadian} and, since then, this technique developed remarkably, also due to the advances in the conception and realization of large superconducting magnets for experiments in fundamental physics. MRI is based on the magnetic resonant effects of protons immersed in a non-homogeneous magnetic field that is used to obtain a spatial modulation of the characteristic response electromagnetic signal. With this technique, high contrast and high resolution - of the order of 0.5 mm - images are obtained. Since the resonance signal produced by protons is influenced not only by their density in tissues but also by the magnetic properties of molecules located in their proximity, MRI can be used to obtain information on
the metabolism by means of functional Magnetic Resonance Imaging (fMRI). MRI represents today a standard in medical diagnostics and a very powerful research tool.

Positron Emission Tomography (PET) is one of the most important functional imaging techniques. It based on the injection in the patient of specific $\beta^+$ radio-tracers. Thanks to the development and to the diffusion of compact cyclotrons, efficient scanners based on novel scintillator detectors and specific radio-tracers, PET is fundamental in several clinical and research applications, mainly in oncology, cardiology and neurology. Following the pioneering studies performed in the fifties and in the sixties, the first clinical PET apparatus was developed in 1975~\cite{Ter-Pogossian}. Differently with respect to SPECT - which makes use of gamma emitters such as $^{99m}$Tc - PET is based on the simultaneous detection of the two characteristic back-to-back 511 keV annihilation photons. This technique allows for high accuracy of the order of 1 mm. To give valuable clinical information, the up-take volumes of the radio-tracer have often to be precisely correlated with their localization 
in the body.
For this reason, based on a PET scanner prototype developed in collaboration with CERN, the first PET-CT scanner was constructed in 1998~\cite{Townsend}. Since then, a very fast diffusion followed and PET-CT is nowadays standard in nuclear medicine. Along this line the combination of PET with MRI is today a very active field of research.

The methods and the techniques I have here shortly summarized demonstrate the continuous interplay between fundamental nuclear and particle physics and medical applications.
These developments are proceeding along several lines. Among them, two represent the direct application of the most recent advances in particle accelerator and detector physics: cancer hadrontherapy and artificially produced isotopes for diagnostics and therapy. It is in these two fields where I have concentrated my scientific efforts in the last years and, to set the basis for the material presented in the following two chapters, they will be introduced here in more detail.

The most advanced application of particle accelerators to medicine is surely represented by cancer hadrontherapy that was first proposed by Robert Wilson in 1946~\cite{Wilson_1}, when powerful enough accelerators and accurate imaging techniques were simply a dream. Former student of Ernst Lawrence, Wilson performed accurate measurements of the range and of the energy deposition of 4 MeV protons in air in 1941~\cite{Wilson_2} and observed that they deposit most of their energy near the end of their path - in what is now called the Bragg peak - as already observed by William Henry Bragg with alpha particles in 1904~\cite{Bragg}. There was nothing unexpected about this measurements but, when Wilson came in contact with John Lawrence and other physicians, he had the idea of irradiating tumours by exploiting the properties of the Bragg dose distribution produced by accelerated ions. Since then, a remarkable scientific and technological development transformed Wilson's vision into reality and hadrontherapy can be considered nowadays an established clinical practice and a challenging field of research in full evolution~\cite{AmaldiBraccini}.

Cancer is one of the most diffused diseases and one person over three experiences this pathology during life. Due to the progress of modern medicine, the overall survival probability increased from less than 50\% in the seventies to about 70\% in present times, reaching values higher than 95\% for certain selected pathologies. These encouraging results are mostly due to the improvement of scientific knowledge as well as to the development and wide availability of effective diagnostic and treatment modalities. About two-thirds of all cancer patients are treated using ionizing radiations, often in combination with other clinical options such as surgery or chemotherapy. The above mentioned improvement in the survival probability is also due to the advances of cancer radiation therapy towards the optimization of local control. 

In cancer radiation therapy,
radiations are usually administered by teletherapy, a modality in which beams produced outside the patient's body are directed toward the tumour volume. Brachytherapy is a less common clinical practice and is based on radiation sources temporary inserted or permanently implanted inside the tumour. In general, radiation therapy has curative purposes but it is important to stress the role of palliative treatments, which have the aim of reducing the tumour volume and to release the patient from pain.

With the aim of improving the local control, cancer radiation therapy is signed by a profound evolution that, especially in the last twenty years, led to the introduction of novel radiation sources, irradiation techniques and treatment planning systems. The use of high-energy photon beams produced by 6-12 MeV linear accelerators (linacs) progressively replaced 1 MeV photons emitted by intense $^{60}$Co sources. 
In a clinical linac based system, accelerated electrons collide with a high Z target where photons are produced by bremsstrahlung. To limit the effect of the soft part of the spectrum and to obtain a flat distribution,
lead flattering filters are located after the target~\cite{IAEA_ROP}.
High-energy photons allow to better reach deep-seated tumours, to increase the dose to the target and, consequently, to improve the tumour control rate. It is important to stress the fact that the dose received during a treatment by the so-called Organs at Risk (OAR) is of paramount importance and represents the limiting factor in cancer radiation therapy. In order to selectively irradiate deep-seated tumours, multiple beams from several directions, usually pointing at the geometrical centre of the target, are used. This is achieved by employing a mechanical structure containing the linac, which rotates around a horizontal axis passing through the isocentre, the so-called isocentric gantry. The isocentre usually coincides with the geometrical centre of the tumour target.
The most recent Intensity Modulated Radiation-Therapy (IMRT)\cite{Webb} makes use of up to 10-12 photon beams, often non-coplanar by means of a robotic couch for patient positioning and with the intensity varied by means of a computer-controlled multi-leaf collimator. Further developments allow for the continuous irradiation of the tumour while the gantry is rotating. This is the case of Tomotherapy~\cite{Mackie} or Rapid Arc~\cite{Lagerwaard} in which irradiation with rotating modulated beams is performed in parallel with high-energy photon imaging. All these techniques are based on sophisticated computer controlled treatment planning systems which, based on CT imaging, allow determining the dose to the target and to the surrounding healthy tissues together with the clinical beams to be delivered to the patient. Furthermore, treatment planning systems and fractionation schemes - which typically foresee treatments of 30-40 sessions of about 2 Gy each - have been improved thanks to the recent developments in radiation biophysics.

\begin{figure}[t!]
 \centering
   \includegraphics[height=13 cm]{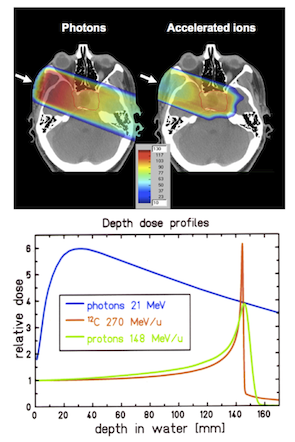}
     \caption{
     Dose distribution delivered by a single photon field and by a singe beam of accelerated ions modulated in a Spread Out Bragg Peak (SOBP) (top). Dose distributions are compared for protons produced by 21 MeV electrons and mono-energetic beams of protons and carbon ions with the same range in water (bottom).
     }
       \label{fig:Photons-Ions}
\end{figure}

As presented in Fig.~\ref{fig:Photons-Ions}, the absorbed dose due to a beam of MeV photons is characterized by a roughly exponential decrease after an initial increase with a maximum reached at a depth of 2-3 cm of soft tissue. The relative deep location of the maximum is due to the relatively long range of the secondary electrons and positrons released in the patient by photon interactions. This is the so-called build-up effect, which gives the clinical advantage of a lower dose to the skin. On the other hand, due to the exponential decrease, the dose is about one third of the maximum at a depth of 25 cm. To treat deep-seated tumours, this fundamental fact leads to an unavoidable dose to the surrounding healthy tissues and represents the major limiting factor of photon beam radiation therapy for many clinical cases.

To overcome this limitation, hadrontherapy is based on a different kind of radiation
and can be considered the present frontier of cancer radiation therapy. The basics, the developments and the perspectives of this discipline are discussed in detail in several comprehensive collection of reviews (see for example~\cite{Kooy}) and review papers (see for example my publications~\cite{AmaldiBraccini}~\cite{Braccini_1}~\cite{Braccini_2}). 
In particular, carbon ion therapy is treated in detail in Refs.~\cite{AmaldiKraft} and ~\cite{Kraft}. Aiming at giving a concise overview, only the basics aspects and facts are discussed here.

Based on the physical and radiobiological properties of accelerated ions, hadrontherapy allows for higher precision treatments with respect to photons, as shown in Fig.~\ref{fig:Photons-Ions}. Charged hadrons are characterized by a definite range in matter, which is function of their initial energy, and by a relative low dose all along their path sharply increasing only before the stopping point in the Bragg peak region. The width and the relative height of the Bragg peak are particle dependent. As shown in Fig.~\ref{fig:Photons-Ions}~(bottom), due to the larger energy straggling, protons are characterized by a wider Bragg peak distribution with a consequent less sharp distal fall off with respect to carbon ions. On the other hand, carbon ions undergo fragmentation due to nuclear interactions occurring all along their path in matter. Since the fragments have almost the same velocity as the primary ion, due to their smaller charge, they are characterized by a longer range, responsible for the tail extending over the fall off.

\begin{figure}[t!]
 \centering
   \includegraphics[height=13 cm]{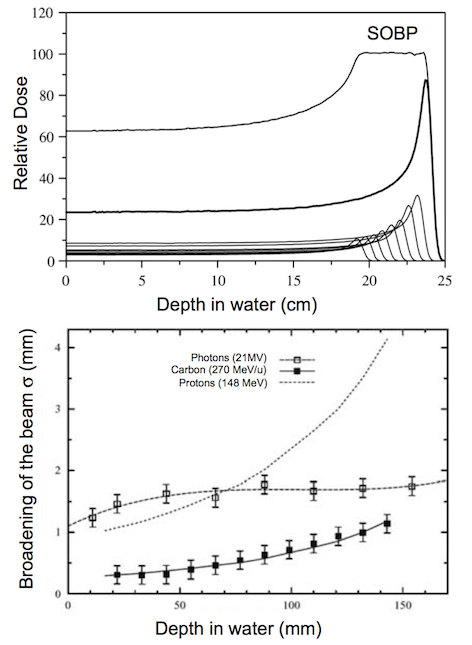}
     \caption{
     The Spread Out Bragg Peak (SOBP) is a flat dose distribution obtained by the superposition of several Bragg peaks corresponding to different energies and intensities (top). The lateral penumbra for photons, protons and carbon ion clinical beams is reported as a function of depth in water (bottom).
     }
       \label{fig:Straggling-Penumbra}
\end{figure}

Either in the case of protons or of carbon ions, the Full Width Half Maximum (FWHM) of the Bragg peak is of the order of a few millimetres, too small for treating tumours having typical dimensions of a few centimetres. For this reason, several Bragg peaks of different intensity and penetration depths are employed in a clinical beam to obtain a homogeneous flat dose distribution corresponding to the target volume in the so-called Spread Out Bragg Peak (SOBP), as shown in Fig~\ref{fig:Straggling-Penumbra}~(top). Multiple scattering determines the lateral dose distribution of the clinical beam. As presented in 
Fig~\ref{fig:Straggling-Penumbra}~(bottom)~\cite{Kraft}~\cite{Jackel}, protons and carbon ions are characterized by a sharper lateral penumbra with respect to photons, especially at large depths. The distal fall off and the lateral penumbra are usually defined as the distance between points at 20\% and 80\% of the maximal dose and represent the two fundamental properties used in treatment planning to protect the Organs At Risk (OAR), which are often located in the proximity of the tumour.

During treatments, charged particle beams are administered by means of two main methods: passive spreading and active scanning. The latter is very innovative and allows for ÒpaintingÓ the tumour volume with unprecedented precision. In this regard, due to their beam characteristics, linear accelerators can be very fruitfully employed, as proposed in the papers reprinted in Chapter4. These two dose delivery modalities will be compared and discussed in more detail in the next Chapter.

The penetration depth in tissues and the dose to be delivered directly determine the main characteristics of the accelerator. In order to treat deep-seated tumours, maximum depths of the order of 25 cm in soft tissues have to be reached. This directly translates into the maximum energy of proton and carbon ion beams, which have to be employed, typically 200 MeV and 4500 MeV (i.e. 375 MeV/u), respectively. As far as beam currents are concerned, the limit is set by the amount of dose to be delivered to the tissues, typically 2 Gy per litre per minute, as in the case of conventional radiation therapy. This translates into beam currents on target of the order of 1 nA and 0.1 nA for protons and carbon ions, respectively.

Living organisms are made of cells and the basic information of life is contained in the DNA, which is the key element of cell reproduction. The size of a cell is in the $\mu$m range while the diameter of the double helix of the DNA is about 2-3 nm. Radiations produce damage to cells via ionization, a process that requires energies of the order of 20-30 eV. These basic figures allow introducing the main parameter used to quantify the effects of radiations on living bodies and to naturally determine its scale. The Linear Energy Transfer (LET) in a medium is defined as $dE/dl$, where $dE$ is the amount of energy locally deposited when a length $dl$ is traversed. A kind of radiation capable to produce direct damage to the DNA must produce one or more ionizations inside the double helix and has therefore to be characterized by a LET of at least 10 eV/nm, i.e. 10 keV/$\mu$m. Below this value, radiations are defined as {\it low LET} and produce damage to the DNA mostly indirectly, through the formation of free radicals such as H$_2$O$^+$ or OH$^-$ molecules ({\it indirect action}). These aggressive ions are first formed from water, which constitutes 80\% of a cell, and, successively, migrate to damage the DNA. This is the case of X-rays, of MeV electrons produced by photons used in radiation therapy and, to a large extent, of protons used in proton therapy. On the other hand, {\it high LET} radiations are capable to produce dense ionization columns, provoking direct damage to the DNA ({\it direct action}). This is the case of high-energy ions to which astronauts are exposed or of accelerated carbon ions used in cancer hadrontherapy.
Due different cell damage mechanisms, different biological effects correspond to different LET. To quantify this effect, the Relative Biological Effectiveness (RBE) is introduced as the ratio between the dose delivered by a standard radiation with respect to the dose from a test radiation for a well defined biological effect, denominated end point. As standard radiation, 250~kVp X-rays are often considered.

The radio sensitivity of the tissues is influenced by the presence or absence of oxygen, which plays an important role in the formation of free radicals. Especially for low LET radiation, the presence of oxygen may lead to a large enhancement of the biological effects. This is expressed by the Oxygen Enhancement Ratio (OER), which, for a given biological effect, is the ratio between the delivered dose without and with oxygen. 
OER values up to 3 are observed for low LET radiation while, being capable of provoking direct damage to the DNA, high LET radiations are characterized by much lower values, usually near to one.

While protons have radiobiological characteristics similar to photons employed in radiation therapy, carbon ions are characterized by a higher RBE and allow the treatment of tumours, which are usually resistant to conventional radiation therapy and proton therapy. Since the LET is not constant along the path of the ion in matter, the RBE changes and reaches a maximum in the Bragg peak region. For this reason, the SOBP for carbon ions has to be carefully modulated in order to obtain a flat dose distribution by means of a non-flat physical dose distribution, once the biological effects - i.e. the RBE - have been taken into account. This is a very delicate issue and an actual theme of research. In clinical practice, several treatment planning systems are available based on different models. Accelerated ions having an intermediate mass between hydrogen and carbon are characterized by an intermediate LET. They could represent promising clinical options for future case-by-case personalized treatments.

From being practiced only in laboratories for nuclear and particle physics, often sharing beam time and resources with fundamental research, proton therapy is becoming today a common clinical option in radiation oncology. About 100000 patients have been treated worldwide, about 40 hospital-based centres are treating patients on a daily basis and about 20 are under construction or in an advanced project phase~\cite{PTCOG}. On the other hand, carbon ion therapy is an active research field, facing a challenging phase of intense 
scientific development and clinical trials~\cite{AmaldiKraft}. About 10000 patients have been treated with carbon ions so far, mostly at the Heavy-Ion Medical Accelerator in Chiba (HIMAC) in Japan. In Europe, the experience of GSI opened the way to the construction of other centres as the Heidelberger Ionenstrahl-Therapiezentrum (HIT) in Germany and the Centro Nazionale per Adroterapia Oncologica (CNAO) in Italy~\cite{LibroBianco}.

\begin{figure}[t!]
 \centering
   \includegraphics[height=9 cm]{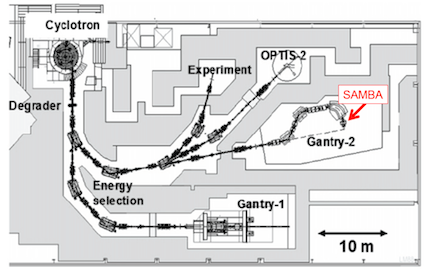}
     \caption{
   Layout of the PROSCAN proton therapy centre at PSI. It features a 250 MeV superconducting cyclotron, two treatment rooms with rotating gantries, one treatment room with a fixed horizontal beam for eye treatment (OPTIS 2) and one room for research activities. The Energy Selection System (ESS) is present only with cyclotrons and consists on a long magnetic channel, equipped with movable absorbers and slits, needed to vary the beam energy. The Gantry~2 is optimized for advanced spot scanning and is equipped with the SAMBA strip ionization chamber, which is discussed in detail in Chapter~\ref{ch3}.
     }
       \label{fig:PROSCAN}
\end{figure}

As presented in Fig.~\ref{fig:PROSCAN}, a modern hadrontherapy centre is based on a complex facility in which the ion accelerator is connected to several treatment rooms by means of beam transport lines. The particle accelerator is the ÔheartÕ of the system and represents the only radiation source. For proton therapy, cyclotrons and synchrotrons are used while for carbon ions only large (20~m diameter) synchrotrons are employed at present. The treatment rooms are equipped with fixed beams - usually horizontal but vertical and 45$^\circ$ are also in use - or with gantries. 
A gantry is a rotating structure supporting the magnets for beam transport and the dose delivery system. In this way, the patient can be irradiated from any direction, exactly like in conventional X-ray radiation therapy. 
For proton therapy, a gantry is about 10~m high and 100000~kg heavy. For carbon ion therapy, only one isocentric gantry has been constructed so far for the HIT centre, mainly devoted to clinical trials aimed at evaluating the need of this kind of apparatus. This 600000~kg and 25~m long structure represents the limit of the actual technology and new solutions are needed to improve the feasibility of carbon ion gantries.

The tumour target is in general characterized by volume changes - usually shrinkages - during radiation therapy and some very common pathologies Ð as lung cancer Ð are subject to movements of the order of 1 cm due to respiration. These effects are at present usually neglected in the treatment planning and the development of tools and techniques to take them into account and perform the necessary corrections represents an important research challenge. Along this line, Image Guided Radiation Therapy (IGRT) aims at guiding the clinical beams on the basis of the on-line localization of moving tumour targets and Intensity Modulated Proton Therapy (IMPT) at fully exploiting the potentialities of the Bragg dose distribution by means of pencil beams of variable intensity. In this framework, the dose distribution and the patient positioning systems are fundamental components. They are the focus of cutting-edge multidisciplinary research in which particle accelerator and detector physics play a crucial role. A sound example of these recent scientific developments is the Gantry~2~\cite{Pedroni} constructed for the PROSCAN project (Fig.~\ref{fig:PROSCAN}) at the PSI. This new device is designed to scan the tumour with parallel beams in a two-dimensional surface and is able to rotate only by 180$^\circ$ to allow for an easier access to the patient and for a more compact structure. The possibility to use X-rays parallel to the proton beam and passing through a special aperture in the 90$^\circ$ magnet is foreseen to optimize the treatment of moving organs. The monitoring of the scanned beams of the Gantry~2 is realized by
a high precision strip ionization chamber, which is discussed in detail in Chapter~\ref{ch3}.

In hadrontherapy the role of particle detectors is essential, not only to monitor and measure the dose delivered to the patient. The accurate control of the beams in the accelerator, at the extraction points and along the beam lines is crucial to obtain an optimal and reliable treatment. For this reason, I dedicate part of my scientific activity to the development of 
beam monitor detectors as the device based on doped and optical silica fibres, discussed in Chapter~\ref{ch3}. For an accurate treatment planning, the precise knowledge of the range of the ions inside the patient body is mandatory. In proton therapy, it is possible to obtain images of the patient using high-energy pass-through protons, giving direct information on the density of the traversed tissues. This technique is denominated proton radiography and represents the focus of several research groups. It will be discussed in Chapter~\ref{ch3} together with the novel method based on nuclear emulsion films, which was developed 
at LHEP in Bern.

Treatment planning is at present based on CT but molecular imaging is expected to play an important role in the near future~\cite{Nestle}. In particular, the use of PET radiotracers will allow the detection and the localization of hypoxic tumour tissues, which are more resistant to radiation. This possibility opens the way to selective treatments using different kinds of radiations as, for example, photons in combination with carbon ions.

\begin{figure}[t!]
 \centering
   \includegraphics[height=10 cm]{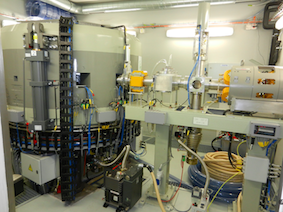}
     \caption{
    The 18 MeV proton cyclotron installed at the Inselspital in Bern in the framework of the SWAN project. One of the out-ports for radioisotope production are visible together with the characteristic beam transfer line for research activities. 
      }
       \label{fig:Bern-Cyclotron}
\end{figure}

PET is both a medical and a research tool. Being able to provide specific and very often unique information on important functions of living bodies, such as blood perfusion, oxygen use, and glucose metabolism, it is employed more and more either on humans or on animals. To exploit at best the large scientific potential of PET imaging,
the availability of a large variety of PET radioisotopes and radio-tracers is crucial. A cyclotron laboratory for radioisotope production represents therefore the fundamental instrument for an advanced clinical and research centre. In this context, the SWAN~\cite{Braccini_SWAN} - which stands for SWiss hAdroNs - project was promoted by the Inselspital and by the University of Bern.
It aims at the establishment of a multi-disciplinary clinical and research centre equipped with accelerators for PET radioisotope production~\cite{Braccini_SWAN_Isot} and cancer hadrontherapy~\cite{Braccini_SWAN_Had}. I am one of the initiators of this project and, in particular, I have designed the new cyclotron laboratory for radioisotope production and proposed its peculiar beam transfer line for multi-disciplinary research activities (Fig.~\ref{fig:Bern-Cyclotron}).

PET is a very vast discipline, which comprehends many aspects of physics, from detectors and data acquisition systems for scanners to accelerators and detectors for radioisotope production. Several recent reviews treat the argument of the production of PET radioisotopes in detail (see for example Refs.~\cite{IAEA_465} and~\cite{IAEA_468}) and only a short introduction to the main arguments related to my scientific activities are given here.

The main radioisotope used for PET imaging is $^{18}$F and the most common compound is the 18-fluoro-deoxy-glucose (FDG). FDG is an analogue of glucose and they are both transported into the cells in the same way. Once the cell membrane is passed, they both undergo phosphorylation but, contrary with respect to phosphorylated glucose, phosphorylated FDG cannot be further metabolized and remains trapped inside the cell~\cite{Mettler_Guiberteau}.
Tumour cells use a large amount of glucose as energy source and are therefore characterized by a large FDG uptake.
FDG is used to investigate a large number of pathologies, cancer and brain diseases in particular, and at present accounts for the largest majority of PET examinations. The half-life of $^{18}$F is 110 minutes, allowing the transportation of FDG and other $^{18}$F based compounds within a relatively large area around the production site. For these reasons, a PET radioisotope production laboratory has to be necessarily optimized for $^{18}$F production.

Other relevant PET isotopes are: $^{11}$C (20 min.), $^{13}$N (10 min.) and $^{15}$O (122 sec.). These elements play a fundamental role in the mechanisms of life and are physiologically relevant. 
They have a considerable research potential and the possibility to produce them is mandatory for a clinical and research facility. 
Being characterized by much shorter half-lives with respect to $^{18}$F, they have necessarily to be used inside the centre where they are produced.

PET isotopes are proton rich and are usually produced by bombarding a specific target with proton beams. The use of deuterons is limited to a few specific cases~\cite{IAEA_468}. For the production of $^{18}$F, a H$_2$$^{18}$O enriched liquid target is used while for the other above mentioned radioisotopes gas targets are also employed. Solid targets are less common, as for the case of $^{64}$Cu production. 
For these reasons, a good production centre must be equipped for all kind of targets. In particular, specific predispositions have to be put in place for solid and gas targets to comply with the highest scientific and radiation protection standards. 

The accelerated beams must have sufficient energy to produce the desired nuclear reactions and enough current is required to give good yields for the future use of the produced radioisotopes. For these purposes, proton cyclotrons represent the best solution due to the suitable energies, the large currents, the good quality and stability of the beams, and the lower complexity and cost with respect to other types of particle accelerators. 

Nuclear reactions for the production of radioisotopes are governed by complex mechanisms and several models have been developed. The ÔcompoundÕ nucleus model, which was first proposed by Bohr in 1936~\cite{Bohr}, can be used to derive the basic physical quantities. According to this model, the accelerated particle and the target nucleus form an intermediate state - the compound nucleus - which successively disintegrates into the decay products. The compound nucleus {\it forgets} how it was formed, except for the implications given by the fundamental conservation laws. In classical physics, a charged particle induced nuclear reaction is possible only if the projectile of charge $z$ has energy larger than the Coulomb barrier
\begin{equation}
B=\frac{zZe^2}{r}
\end{equation}
where $Z$ is the atomic number of the target nucleus, $e$ the charge of the electron and $r$ its nuclear radius. Due to quantum mechanics tunneling effects, lower energies are possible. The Coulomb barrier sets the scale of the energies needed for radioisotope production. For the reaction 
\begin{equation}
^{18}\mbox{O}(p,n)^{18}\mbox{F}
\end{equation}
the Coulomb barrier corresponds to 3.8 MeV and similar values can be obtained for the production of the other above mentioned PET radioisotopes. Once the incident particle has overcome the Coulomb barrier, there is another condition for a nuclear reaction to take place since the available total energy must be larger than the total mass of the decay products. Furthermore, the cross section for the reaction must be large enough to obtain a good yield. For PET isotopes, the production cross sections are characterized by several peaks in the 10-20~MeV region~\cite{IAEA_468}. For these reasons, PET cyclotrons operate in this energy range.

The beam currents used in radioisotope production are directly related to the cross section and to the required amount of activity. They are limited by the amount of power that the target can stand without being damaged. In the case of commercial $^{18}$F liquid targets of the most common horizontal type, the beam current has to be limited to about 75~$\mu$A.
High power $^{18}$F liquid targets have been recently developed~\cite{Strangis}~\cite{Stokelya}, able to stand currents up to about 150~$\mu$A. However, due to cooling limitations, the production yield is found to diminish above 100~$\mu$A. It has to be remarked that high power may lead to the formation of impurities which may have negative implications in the synthesis of the radio-tracers.Ó

To estimate the produced activity, the decay during irradiation has to be taken into account and saturation effects are relevant already for bombardment times of the order one  half-life. The production rate R, i.e. the number of nuclei formed per second at a time $t$ after the beginning of the irradiation, is given by
\begin{equation}
R=\frac{dn}{dt}=\frac{n_t I}{e}(1-e^{-\lambda t})\int^{E_f}_{E_i}\frac{\sigma(E)}{\left(\frac{dE}{dx}\right)}dE
\end{equation}
where $n_t$ is the number of target nuclei per cm$^3$, $I$ the beam current, $\lambda$ the decay constant of the produced radioisotope equal to $\frac{\ln 2}{T_{1/2}}$, $E_i$ and $E_f$ the initial and final energy of the incoming proton. At saturation, the production rate equals the decay rate, and, assuming an average value for the cross section of 100 mb, one obtains the value of 9 GBq/$\mu$A. This is the typical value for $^{18}$F production. 

PET cyclotrons are usually able to produce activities of the order of 100-200~GBq of $^{18}$F in 60-90 minutes of irradiation at 50-80~$\mu$A total beam current. 
After the synthesis, this corresponds to about 50-100~GBq of FDG.
As an example, this quantity is enough for serving 2-4 PET centres examining about 10 patients each and located at one hour distance from the production centre. The typical injected dose for one examination is 400-500 MBq.

PET cyclotrons usually accelerate negative H$^-$ ions. The advantage with respect to positive ion accelerators is represented by a much easier extraction by carbon foil stripping with efficiency close to 100\%. 
Using two stripper foils and with an optimized beam dynamics, it is possible
to extract the beam into two different out ports at the same time.  This dual beam extraction 
allows for the simultaneous bombardment of two targets. In this way, a larger production or the simultaneous production of two different kind of radio-nuclides  can be obtained.
In the H$^-$ ion, the electrons are very weakly bound, being the binding energy only 0.7 eV. For this reason,
stripping may occur during acceleration with consequent loss of beam intensity and production of unwanted activation.  To limit this effect, a good vacuum is crucial in this kind of machines.

According to the survey performed by the IAEA in 2006~\cite{Cyclotron_Directory}, 262 cyclotrons for the production of isotopes are running in the 39 member states of the agency. The real number is certainly higher and it is increasing continuously by approximately 50 units per year, mainly due to the installation of new PET radioisotope production centers~\cite{Ramamoorthy}. In the countries of the European Union, about 100 cyclotrons for the production of radioisotopes are in operation. In particular, 33 are located in Germany, 21 in Italy, 18 in France,  and 4 in Switzerland~\cite{Stamm}~\cite{AmaldiMagrin}.
Besides medical applications, cyclotrons for the production of isotopes represent the primary tool for a wide range of biological, environmental and industrial research activities.
Several industrial  solutions are available providing beams in the range 3-70~MeV~\cite{Schmor}. As it will be discussed in the next Chapter, industrial solutions have often to be adapted and optimized for
specific research needs.

PET radioisotope production facilities usually consist of a single cyclotron and a single bunker. They are equipped with targets directly mounted on the accelerator after the stripping foils. 
Since one or more production cycles per day are performed, the beam area is practically not accessible for research activities due to radiation protection hazards provoked by the high residual radioactivity.
Since PET examinations are performed in the morning or in the afternoon, production runs daily during the night and early in the morning. This duty cycle leaves a considerable amount of beam time 
available for other activities.

PET cyclotrons are characterized  by a large research potential beyond PET applications. In particular, they can be
optimized to produce beams ranging from fractions of nA to several $\mu$A
for novel detector, material science, radiation biophysics, radiation protection, radiochemistry and radiopharamcy research activities~\cite{BracciniScampoli}. To exploit this potential without interfering with production, I proposed in 2007 
the realization of a specifically conceived beam transfer line and the construction of a second bunker for the PET radioisotope production facility in Bern. 
This solution represents a unique feature for a hospital-based centre.
The new Bern cyclotron laboratory (Fig.~\ref{fig:Bern-Cyclotron}) was commissioned in 2012 and is now fully operational, as discussed in the next Chapter.

\chapter{Particle Accelerators for medical Diagnostics and Therapy}
\label{ch2}

This Chapter is focused on my contributions to particle accelerator physics applied to cancer hadrontherapy and to the production of PET radioisotopes.
Linacs are fast cycling accelerators and their beam structure is optimal for beam scanning. They are suitable to face the actual challenges of hadrontherapy.
Cyclotrons are the main instrument used for the production radioisotopes for PET imaging and, at the same time, have a high potential for
multi-disciplinary research. This concept is at the basis of the new Bern cyclotron laboratory.
Although very different, cyclotrons and linacs can be combined to form a new kind of accelerator complex, the cyclinac. Cyclinacs are proposed to
perform radioisotope production and proton therapy with a single apparatus and to realize an innovative linac based single-room proton therapy facility. 
Applications to carbon ion therapy are also proposed.
These issues are discussed here in close connection with the publications reprinted in Chapter~4.

\section{Linacs and Cyclinacs for Hadrontherapy}

Hadrontherapy is nowadays facing a challenging phase aimed at improving the local control of the tumours and at 
attacking cancers that cannot be treated with other modalities~\cite{AmaldiBraccini}~\cite{Kooy}. This is the case of radio-sensitive pathologies located very near the OARs, of hypoxic tumours and of pediatric radiation oncology, where modest doses to the healthy tissues may provoke heavy secondary effects and sequelae, including radiation induced cancers. A special case is the treatment of moving organs which is the focus
of several recent developments.
To reach these ambitious goals, the superior physical and radiobiological properties of accelerated ions have to be exploited at best and hadrontherapy has to be made available to the largest number of patients.
In this context, research in accelerator physics is essential to conceive, study and eventually realize innovative solutions.

\begin{figure}[t!]
 \centering
   \includegraphics[height=12 cm]{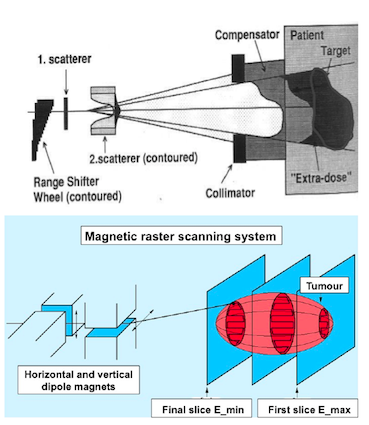}
     \caption{
     Passive spreading is based on scatterers to widen and flatten the beam, on a range shifter wheel (propeller) to obtain the SOBP and on personalized collimators and compensators to conform the dose distribution to the tumour volume (top). Active scanning is based on the use of a pencil beam of variable intensity to paint a slice of the tumour continuously or using successive spots. The irradiation of successive slices results in a three-dimensional uniform dose distribution (bottom).
     }
       \label{fig:Scattering_Scanning}
\end{figure}

As mentioned in the previous Chapter, the SOBP can be obtained using two main techniques:
the most common passive spreading and the nowadays challenging active scanning (Fig.~\ref{fig:Scattering_Scanning}). Passive spreading is relatively simple and requires the construction of personalized collimators and compensators for every delivered radiation field~\cite{Gottschalk}. 
Extra doses are produced at the edges of the tumour, leading to a non-optimal conformality. To reduce these effects, further approaches, still based on passive devices, have been developed in Japan, as the layer stacking technique~\cite{Kanematsu} and the most recent cone type filter method~\cite{Ishizaki}. 
Passive spreading does not make use of a fundamental characteristic which distinguishes ions from photons: the electric charge.
Based on this property, active scanning dose delivery systems have been developed in the recent years. They are based on electromagnetically driven pencil beams without the need of any passive personalized device. 

Two are the main active scanning techniques. The {\it raster scanning} has been developed at the GSI Helmholtzzentrum f\"ur Schwerionenforschung~\cite{Haberer} and is based on a pencil beam of about 6 mm FWHM with variable intensity, which is moved across the tumour to paint a slice at a fixed penetration depth. The beam energy is then reduced in succession to cover the entire volume. Developed at the Paul Scherrer Institut (PSI)~\cite{Pedroni_2}, the {\it spot scanning} technique consists in chopping the beam from a cyclotron using a fast kicker to obtain discrete dose spots of given energy, intensity and direction which are used to obtain a uniform three-dimensional dose distribution. More information on these two techniques can be found in Ref.~\cite{Kooy}.
It is important to remark that active scanning allows for a much higher conformation of the dose and, through IMPT, to cover complex concave tumor volumes with a minimal dose given to the healthy tissues.
Furthermore, much less neutrons and secondary particles are produced with respect to passive spreading, leading to a significant reduction of the corresponding unwanted dose. 
This new technique is surely superior but difficult to implement, as demonstrated by the fact that only about 2\% of the patients have been treated with active scanning so far.  

Among the many challenges of today's hadrontherapy~\cite{AmaldiBraccini}, a very fascinating scientific theme is the treatment of organs which are affected by movements due to respiration~\cite{Bert_Durante}.
These movements are of the order of 10 mm with a frequency of approximately 0.2 Hz. If not compensated, they may provoke remarkable under and over dosages and spoil the high precision offered by hadrontherapy. The compensation can be obtained through three different approaches.
The easiest one consists on synchronizing the delivery of the dose with the patient expiration ({\it respiratory gating}). This method is already employed in conventional radiotherapy but has the draw back of an inefficient use of the beam time. The second and the third approaches are possible only with active scanning and can be combined. They are the focus of several actual research developments.
If the beam delivery is fast enough, the tumor can be painted
many times in three dimensions so that the movements of the organs can cause acceptable dose variations within about 3\% ({\it multi-painting}).
Furthermore, the movements of the organs can be detected and followed by means  
of a suitable system. This information can be used as a feedback to adjust on-line the transverse and longitudinal locations of the following spots ({\it beam tracking}). 

To implement multi-painting and beam tracking, innovative solutions are needed in many fields, including particle accelerator physics. An ideal beam would be characterized by fast scanning capabilities together with fast changes of the energy. In hadrontherapy, cyclotrons and synchrotrons are nowadays in use. 
The beam produced by cyclotrons is characterized by a fixed energy and by a 30-100 MHz pulsed beam which can be considered continuous when compared with the human respiration period. 
To vary the energy of the clinical beam, mechanically moving absorbers are used in the long and complex Energy Selection System (Fig.~\ref{fig:PROSCAN}). With this device, the energy can be varied in about 100 ms.
The beam produced by synchrotrons is characterized by a spill time of about 1 s, during which the beam is extracted for therapy, and by a filling and accelerating time of about 1-1.5 seconds in which the beam is not available. The energy can be varied from spill to spill. The beam periodicity is similar to the one of the respiration cycle, which generally represents a disadvantage for the irradiation of moving organs.

\begin{table}[t!]
\centering
\begin{tabular}{|l|c|c|c|}\hline
Accelerator & Is the beam  & Is the output energy  & Which is the time  \\
   & always available? & variable? &  to vary E$_{clin}$ (ms)? \\
\hline
Cyclotron  & Yes  & No  & 100 \\
Synchrotron  & No  & Yes  & 1000 \\
Linac  & Yes &  Yes  & 2-5 \\
\hline
\end{tabular}
	\caption{Proprieties of the beam of various accelerators relevant for hadrontherapy.
	E$_{clin}$ is the energy of the clinical beam reaching the patient.
	}
	\label{tab:comparison}
\end{table}

An innovative solution is represented by fast cycling high frequency linear accelerators. Compared to presently operated cyclotrons and synchrotrons (Table~\ref{tab:comparison}), they can
provide pulsed beams with an optimal repetition rate, intensity and energy modulation for applications in hadrontherapy. Along this line, fast cycling synchrotrons have also been recently proposed~\cite{iRCMA}.

On the basis of the high potential of linacs for hadrontherapy,
the TERA Foundation - of which I was the Technical Director from 2003 to 2006 - initiated and still pursues an intense research program to which I have strongly contributed~\cite{Braccini_RAST2}~\cite{Braccini_CYCLINACS}. Among the many research activities in which I was involved,
the conception and the design of an innovative linac based proton therapy facilitiy - denominated Institute for advanced Diagnostics and RAdiation therapy (IDRA) - is discussed here in more detail. 

\begin{figure}[t!]
 \centering 
   \includegraphics[height=7 cm]{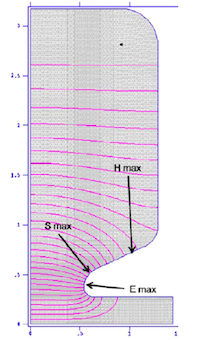}
   \caption{
   Design of a half accelerating CCL cavity of a high frequency linac. 
  The red curves represent the electric field lines of
the accelerating mode. The arrows indicate the regions where the Pointing vector S
and the electric and magnetic fields (E, H) are maximal, in the so called {\it nose}.
         }
       \label{fig:acc_cell}
\end{figure}

Several are the problems to assess for the realization of a linac that will be suitable for an hospital-based proton therapy facility optimized for beam scanning. Besides the very stringent requirements on the beam characteristics, the accelerator must be compact, reliable and consume a reasonable amount of power. To realize a compact linac, a high average electric accelerating field is needed. For proton therapy, energies up to about 200~MeV
have to be reached, implying average accelerating fields of at least of 10~MV/m to limit the length of the accelerator to 20~m. Since the shunt impedance increases with the frequency as $Z\sim f^{1/2}$, the choice of a high frequency allows
for higher electric accelerating fields for the same power consumption. For this reason, the frequency of 3 GHz was chosen, although unusual for proton linacs since it implies several design difficulties. In particular, small resonant cavities have to be realized with the implication of a small iris for the beam. 
For proton therapy, this does not represent a serious problem since, as already observed in the previous Chapter, currents of the order of 1 nA or less are needed.  High accelerating fields imply a very accurate design of the accelerating cavities, especially in the so called {\it nose} region where discharges have to be avoided.
On the other hand, the choice of 3~GHz gives the advantage of the availability of standard klystrons to provide the needed radio-frequency (RF) power.
The design of a typical high frequency cavity is shown in Fig.~\ref{fig:acc_cell}.

\begin{figure}[t!]
 \centering 
   \includegraphics[height=5.5 cm]{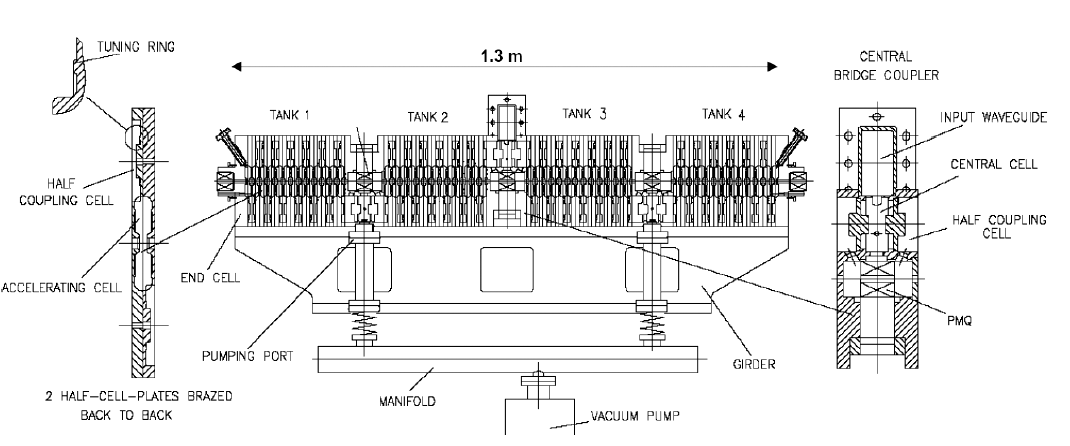}
   \includegraphics[height=6. cm]{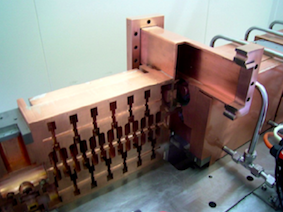}
   \caption{
   Mechanical design of the four {\it tanks} of the LInac BOoster (LIBO) prototype, forming one {\it unit} made of two {\it modules}. Each tank is made of a number of basic elements machined with high accuracy in copper and called {\it half-cell-plates}. Permanent Magnetic Quadrupoles (PMQ) are located between two successive tanks to assure
   focussing by means of a FODO lattice (top).
   The first LIBO prototype after it was cut out to examine the brazing procedure. The CCL structure formed by accelerating (AC) and coupling (CC) cells is visible (bottom). 
         }
       \label{fig:LIBO}
\end{figure}

For enhancing the compactness and for obtaining a larger shunt impedance, the Coupled-Cavity Linac (CCL) -~sometimes also referred as Side Coupled Linac (SCL)~- structure working in the $\pi/2$ mode was chosen. The structure of the accelerator and the first prototype~\cite{Braccini_RAST2}~\cite{Amaldi_LIBO} are shown in Fig.~\ref{fig:LIBO}. This standing wave accelerator has a biperiodic structure composed by on-axis accelerating cells (AC) and off-axis coupling cells (CC). The CCs are located off-axis to reduce the length of the linac and have the fundamental role to allow the RF power to be distributed to all the accelerating cavities.
In the $\pi/2$ mode, the electric field follows the scheme [1, 0, -1, 0, 1, 0, $\dots$] so that a proton must transit from an AC to the next in a time equal to $T/2$, where T is the RF period. For this reason, the distance
$d$ between two successive accelerating cells has to be
\begin{equation}
d=\frac{vT}{2}=\frac{\beta\lambda}{2}
\end{equation}
where $v$ is the average velocity of the protons and $\lambda$ the wavelength of the RF. Being the RF fixed, this implies that the distance between successive ACs must change along the linac and becomes larger at higher energies.
For the first prototype, $d$ is 2.8~cm at 200~MeV ($\beta=0.57$) and 1.7~cm at 60~MeV ($\beta=0.34$). For constructive reasons, it is not possible to vary $d$ from cell to cell and the linac is organized in
{\it tanks} in which $d$ is kept constant. Four tanks constitute an accelerating unit fed by one single klystron coupled via a bridge coupler (Fig.~\ref{fig:LIBO}). The copper structure of the accelerator is
made of half-cells brazed together. Due to the mechanical construction difficulties and to the fact that the shunt impedance decreases with $\beta$, the first proof of principle prototype was realized 
starting from 62~MeV. This prototype successfully accelerated protons and, during power tests, reached an accelerating gradient of 28.5~MV/m, much larger than the design value of 15.8~MV/m~\cite{Amaldi_LIBO}.

The choice of 62~MeV was also due to the fact that this is the typical energy of cyclotrons used for the treatment of ocular uveal melanomas and that
the linac could be used as a booster to treat deep seated tumours.
The combination of a cyclotron and a linac has been denominated {\it cyclinac}.
Following the success of the first prototype, a second more challenging one was designed and built starting from 30 MeV~\cite{Braccini_RAST2}. 

CCL structures become highly inefficient at low $\beta$ due the low shunt impedance and they have to be coupled to another kind of accelerating structure as, for example, a Drift Tube Linac (DTL) preceded
by a Radio Frequency Quadrupole (RFQ). 
Moreover, at very low $\beta$, $d$ becomes very short and the mechanical structure impossible to realize and tune to the right frequency. 
The tuning of the cavities is performed by successive measurements aimed at finding the optimum configuration of the cell by successive machining and changes in the length of specific copper rods ({\it tuners}).
With this procedure, a precision of 1/1000 at 3~GHz can be obtained.
After a careful study based on simulations and calculations, the starting energy of 30 MeV -~corresponding 
to $\beta=0.25$ and  $d=1.25$~cm~- was considered the optimum.

There is another important reason to choose 30 MeV as the starting energy of the linac.
Considering the fact that the proposed linac is designed for medical applications, 
an unusual and novel solution was proposed by the TERA Foundation:
the use of a 30 MeV high current cyclotron for radioisotope production as the injector.
This idea led to a further development of cyclinacs. This new combination allows the production of radioisotopes for diagnostics 
(PET and SPECT) and 
for radioimmunotherapy by means of the cyclotron in parallel with advanced scanning proton beam teletherapy by means of the linac used as a booster.

As discussed in the previous Chapter, cyclotrons for the production of PET isotopes are characterized by energies ranging from 10 to 20 MeV, too low to be used as the injector of a CCL linac.
30 MeV cyclotrons are more complex and less common machines. Nevertheless, they are commercially available and are usually devoted to the production of SPECT isotopes, $^{123}$I and $^{201}$Tl in particular.
They are able to extract the beam at different energies and
they can produce very large currents, up to 500~$\mu$A or more. For these reasons, they have also a large research potential, an interesting feature for an academic clinical environment.
These machines can also be used for the production of PET isotopes.

\begin{figure}[t!]
 \centering 
   \includegraphics[height=9 cm]{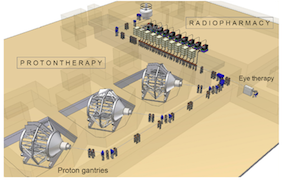}
     \caption{
    The layout of the Institute for advanced Diagnostics and RAdiation therapy (IDRA) features a 30 MeV cyclotron, a linac of the LIBO type and three treatment rooms equipped with rotating gantries and a fixed beam line for the treatment of eye tumours.
     }
       \label{fig:IDRA}
\end{figure}

IDRA is shown in Fig.~\ref{fig:IDRA}. Its name was chosen since 
in medieval latin the name {\it Idra} ({\it Hydra} in English)
designated a mythologic creature with many very vital heads~\cite{Aberdeen}, just as the high-current cyclotron in the proposed facility.

The design of IDRA demanded to assess several specific issues which were never studied before:
the cyclotron-linac coupling, 
the linac structure starting from 30~MeV
and the realization of a beam optimized for the treatment of moving organs by means of spot scanning. In particular, the energy variation and the multi-painting capabilities had to be carefully studied.
Moreover, the possibility to treat ocular pathologies with beams of 62-65~MeV had also to be possible.

Since the linac is a pulsed accelerator with a duty cycle of about $10^{-4}$, its acceptance
is definitely not appropriate for the almost continuous beam produced by the cyclotron used as an injector.
To avoid the large unwanted irradiation during the down-time of the linac and all the consequent radiation protection issues,
the ion source of the cyclotron has to be pulsed to match the RF structure of the linac. 
Furthermore, during proton therapy runs,
the intensity of the ion source has to be changed from spot to spot to obtain at least a course variation of the intensity of the beam injected in the linac. 
30 MeV cyclotrons have an external ion source on which specific pulsed electronics can be installed. The {\it einzel} electrostatic lenses located in the Low Energy Beam Transfer line (LEBT)
can also be used to obtain further variations of the intensity of the beam injected into the cyclotron.  
To improve the matching and increase the precision of the intensity variation, a fast kicker and a fast quadrupole doublet located at the injection
of the linac can be employed. Simulations and calculations showed that an overall precision in the beam intensity within 5\% could be obtained with a variation in the number of protons
per pulse $N$ of
$N_m/200<N<N_m$, where $N_m$ corresponds to the maximum current of the clinical beam.

The design and the optimization of the linac were focussed on the main figures of merit~\cite{Braccini_RAST2}: the effective shunt impedance, the quality factor $Q_0$, the field uniformity and the power efficiency. 
Transverse and longitudinal beam optics were also optimized by means of specific solutions.
The main parameters of the linac of IDRA are reported in Table~\ref{tab:LIBO}.

\begin{table}[t!]
\centering
\begin{tabular}{|l|c|}\hline
Accelerated particles &	protons \\
Type of linac &	CCL \\
RF Frequency [MHz]	& 2998.5 \\
Input energy  [MeV]	& 30 \\
Output energy  [MeV] &	230 \\
Total length of the linac  [m] &	18.5 \\
Cells per tank / tanks per module	& 16/2 \\
Number of accelerating modules  &	20 \\
Thickness of a half cells in a tank [mm] &	6.3-14.6 \\
Diameter of the beam hole  [mm] &	7.0 \\
Normalized transversal acceptance [mm mrad] &	1.8 $\pi$ \\
Number of Permanent Magnetic Quadrupoles	& 41 \\
Length of each PMQ [mm]	& 30 \\
PMQ gradients [T/m] 	& 130-153 \\
Synchronous phase [deg]	 &  -15 \\
Peak power per module (with 20\% losses)  [MW] &	3.0 \\
Effective shunt impedance ZT$^2$ (inj.-extr.)  [M$\Omega$/m] &	30-90 \\
Axial electric field (inj.-extr.)  [MV/m]	  & 15-17 \\
Number of klystrons (peak power=7.5 MW)  &	10 \\
Total peak RF power for all the klystrons [MW]	  & 60 \\
Klystron RF efficiency &	0.42 \\
Repetition rate [Hz]	& $\leq$200 \\
Duration of a proton pulse [$\mu$s]	 & 1.5 \\
Max. number of protons in 1.5 $\mu$s (2 Gy l$^{-1}$ min$^{-1}$)	& $4\times10^{7}$ \\
Effective duration of each RF pulse [$\mu$s] & 	3.2 \\
RF duty cycle	&  $3.2\times10^{-4}$ \\
Plug power at 100 Hz + 100 kW auxiliaries [kW]  &	150 \\
\hline
\end{tabular}
	\caption{The main parameters of the linac specifically designed for IDRA.}
	\label{tab:LIBO}
\end{table}

The linac is composed of 20 accelerating modules.
Calculations show that the effective shunt impedance is roughly proportional to the length of each module and varies from 30~M$\Omega$/m for the first to 90~M$\Omega$/m for the last one.
This fact allows to inject almost the same power in all the modules with minimal variations of the accelerating electric field $E_z$, which varies from 15 to 17 MV/m. In this way, a compact linac 
of 18.5 m length is obtained. The relative small value of the synchronous phase of -15$^\circ$ strongly reduces defocussing effects due to RF but limits the phase acceptance.
This is not a problem since low currents (nA) are needed for therapy and large currents (more than 100 $\mu$A) can be provided by the cyclotron.

The transverse beam optics~\cite{Zennaro} is realized by 41 Permanent Magnet Quadrupole (PMQ) magnets and requires some care since the output energy of the linac has to be varied and a beam
of 65 MeV has to be extracted for eye therapy, as shown in Fig.~\ref{fig:IDRA}. The use of PMQs is an elegant and simple solution with the draw back of the impossibility to adjust the optics of the FODO
according to the beam energy. To maximize the acceptance in a so wide range of conditions, a large betatron phase advance per focussing period is appropriate. 
On the other hand, a low value allows for a higher focalization. 
Since the beam emittance decreases with $\beta\gamma$ along the accelerator,  the betatron phase advance $\sigma_t$ is varied from $74^\circ$ at the injection to $44^\circ$
at the end of the linac. The gradient of the PMQs is varied from 130 to 153 T/m, accordingly. 
This solution minimizes the losses and allows the transport of a beam of 65 MeV through the full linac.

Multiparticle simulations~\cite{Zennaro} showed that the linac acts like a filter and only a negligible fraction of the beam (always less then 7\%) is not fully accelerated.
Moreover, the fraction of the injected beam that is not fully accelerated has an energy much lower then the fully accelerated beam and can be therefore easily separated by means of the dipole magnets
generally used to guide the beam to the treatment rooms. The energy spread of the accelerated beam is very small, about 0.45~MeV.

Particular care has been devoted to the design of the ACs, especially in the {\it nose} region~\cite{Braccini_RAST2}~\cite{Tesi_Puggioni}. In a standing wave cavity, the ratio between the maximum field on the surface
of the cell and the accelerating field in the gap has to be kept under control to avoid discharges. According to calculations and measurements on prototypes, this ratio can be varied in the range 5-8 allowing for 
accelerating fields of 30 MV/m at 3~GHz. This gives a safe margin of operation to the linac.
 
Another important issue that pertains to CCL structures operating in the $\pi/2$ mode is the {\it stop band}. For the linac of IDRA, it was studied 
by means of a model based on a circuit representation of coupled electric RC oscillators~\cite{Braccini_RAST2}~\cite{Tesi_Puggioni}. ACs and CCs resonate at the frequencies $\omega_a$ and $\omega_c$, respectively, when considered alone. When $N$ cells are coupled,
they form a system of $N$ coupled oscillators. This system is characterized by  $N$ different oscillation modes $\omega_1, \omega_2, \dots, \omega_N$, one for each phase difference between two neighbor cells $\phi_1, \phi_2, \dots, \phi_N$. The $\omega$ versus $\phi$ curve is denominated the dispersion curve of the structure. 
If $N$ approaches to infinite, the dispersion curve shows a discontinuity for $\phi=\pi/2$
and the frequencies
\begin{equation}
\frac{\omega_a}{\sqrt{1-k_a}}\le\omega\le\frac{\omega_c}{\sqrt{1-k_c}}
\end{equation}
do not allow to excite the structure. Here, $k_a$ and $k_c$ are the AC-AC and CC-CC coupling constants, respectively. This phenomenon is called the stop band.
To achieve a stable $\pi/2$ mode configuration, the stop band has to be closed by choosing the coupling constants in such a way that the upper and lower frequency of the stop band coincide.
When the stop band is closed, the slope of the dispersion curve in the proximity of the $\pi/2$ mode is found to be proportional to the AC-CC coupling constant.
It is therefore important to keep this parameter large in order to separate at maximum the $\pi/2$ mode from the neighboring modes.
In the case of the linac of IDRA, the stop band was reduced down to 1 MHz which is better than the overall required precision of 1/1000 on 3 GHz.
With the chosen configuration, the $\pi/2$ mode can be excited safely and with high precision.

\begin{figure}[t!]
 \centering 
   \includegraphics[height=5 cm]{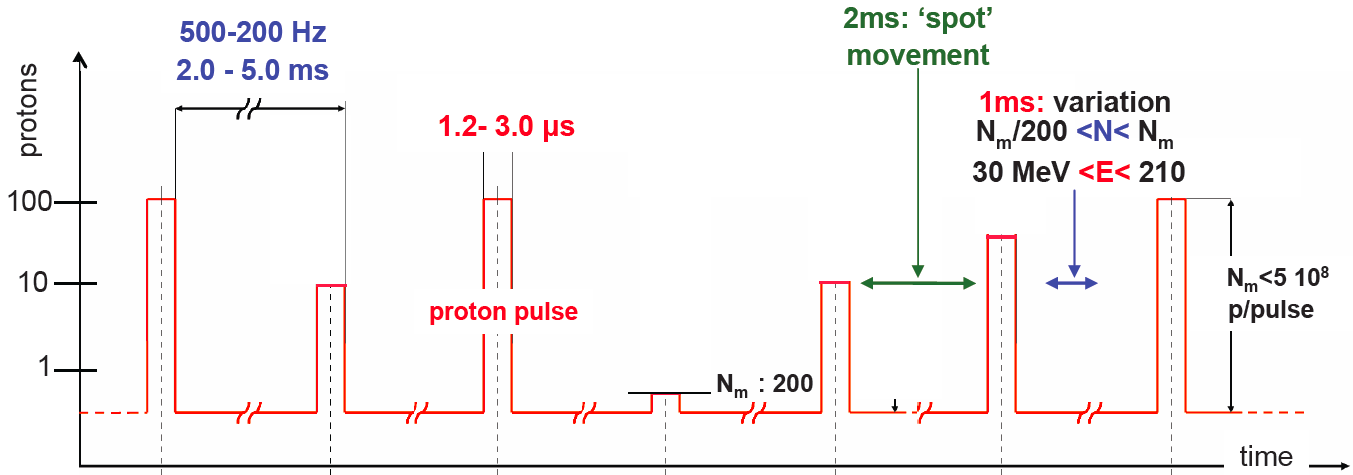}
     \caption{
     Beam structure of the linac of IDRA. It
     allows for multi painting and beam tracking by means of a high repetition rate, variable intensity and fast energy variation from spot to spot.
     }
       \label{fig:linac_beam_structure}
\end{figure}

\begin{figure}[t!]
 \centering 
   \includegraphics[height=9 cm]{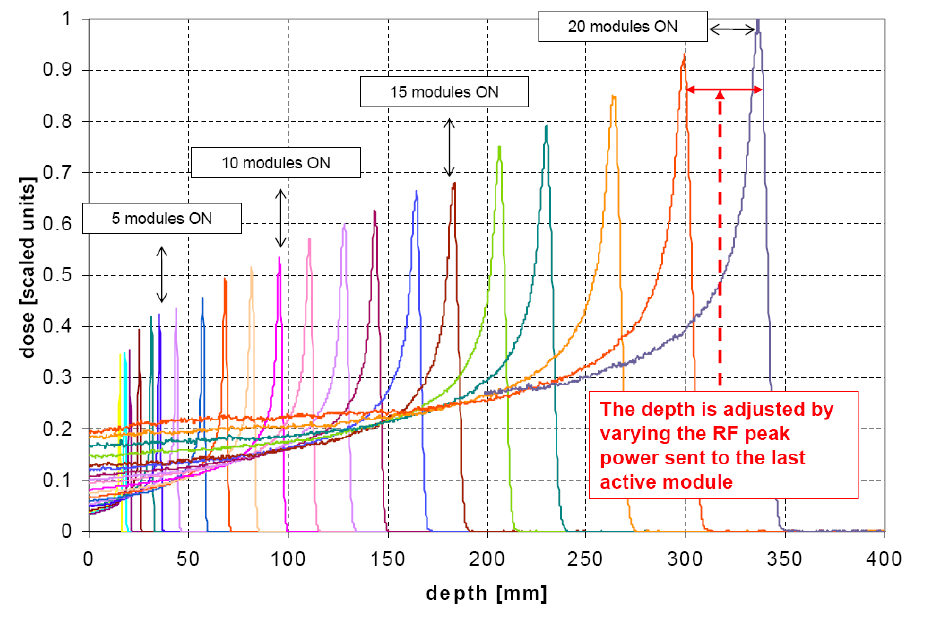}
     \caption{
     Proton depth dose distribution when the number of the active accelerating modules is varied one by one. To avoid super-positions a different normalization is used for each curve. 
     }
       \label{fig:linac_energy}
\end{figure}

The linac of IDRA is able to produce a beam with the structure reported in Fig.~\ref{fig:linac_beam_structure}. The beam is composed by 1.5~$\mu$s long discrete spots of variable intensity from $N_m$ to $N_m/200$ with the repetition rate of 200 Hz, where $N_m$ is the maximum number of protons per pulse. Higher repetition rates and longer spots are possible but imply a larger power consumption.
$N_m$ is calculated to be able to deliver 2~Gy~l$^{-1}$s$^{-1}$, the typical dose delivery rate in particle therapy. The factor 200 has been chosen to allow for a precision in the delivery of the dose in the SOBP well below 2.5\%, typical value required in radiation therapy.  
During the 5 ms between two successive spots, the energy of the beam is varied by switching off a certain number of modules starting from the last one
and the transverse position changed by means of scanning magnets (Fig~\ref{fig:Scattering_Scanning}).
The energy variation and the corresponding positions of the Bragg peaks are shown in Fig.~\ref{fig:linac_energy}.
Due to these unique features, linacs represent an innovative and advanced tool for beam scanning, multi-painting and beam tracking for the treatment of moving organs.

In the proposed linac, the energy can be in principle varied spot-by-spot between the cyclotron output (30~MeV) and the maximum value (230~MeV). In practice, this is not the case either for clinical reasons
or for the finite momentum acceptance of the beam transfer line located downstream of the linac. For the treatment of moving organs, variations in the proton range of $\pm$10~mm are needed and 
the typical momentum acceptance of the transfer line within $\pm$1.5\% is enough for the case of deep seated tumors. To study this effect more quantitatively, a fit on experimental data leads to the formula
\begin{equation}
R=0.00176\left( E\mbox{[MeV]}  \right)^{1.82}\mbox{cm}
\end{equation}
where $R$ is the proton range and $E$ the kinetic energy. This implies that
\begin{equation}
\frac{\Delta R}{R}=1.82\:\frac{\Delta E}{E}
\label{eq:R_E}
\end{equation}
and, considering that the protons are almost non-relativistc,
\begin{equation}
\frac{\Delta R}{R}=3.64\: \frac{\Delta p}{p}
\end{equation}
where p is the momentum. Variations of $\pm$10~mm are well within the momentum acceptance
for 175~MeV protons, which correspond to a penetration depth of 20 cm.
For tumours located at a depth of 10 cm, 120 MeV are needed and  variations only up to $\pm$5.5~mm are possible. To enhance the range variation, a 10 cm absorber located very close to the patient
could be used together with larger beam energies.

To obtain a continuous energy variation, two different methods were considered and 
studied~\cite{Zennaro}~\cite{Braccini_RAST2}, both acting only on the last active module: the variation of the peak power and the change of the phase with respect to
the rest of the linac. Both the solutions can be implemented by acting on the
low-level phase circuits of the RF driver and provoke an increase in the energy spread. This effect has to be carefully assessed since an increase of the energy spread directly translates into an increase of the uncertainty of the range, according to Equation~\ref{eq:R_E}. 
As stated above, the intrinsic energy spread of the linac is about 0.45~MeV with a negligible dependence on the total energy if modules are completely on or off, as shown in Fig.~\ref{fig:linac_energy}.
When a continuous energy variation is performed,
simulations show that  the energy spread increases up to 0.6~MeV in case of the variation of the peak power in the last active module. The increase is up to 0.9~MeV by changing the phase of the last active module. For this reason, the first method was preferred. It has to be noted that an energy spread of 0.6~MeV produces a range variation 
of 1.4 mm and 1.1 mm at 200~MeV and 150~MeV, respectively. This effect is acceptable since it has to be compared to the natural
straggling of 4.3~mm and 2.3~mm  at 200~MeV and 150~MeV, respectively.

\begin{figure}[t!]
 \centering
   \includegraphics[height=6.5 cm]{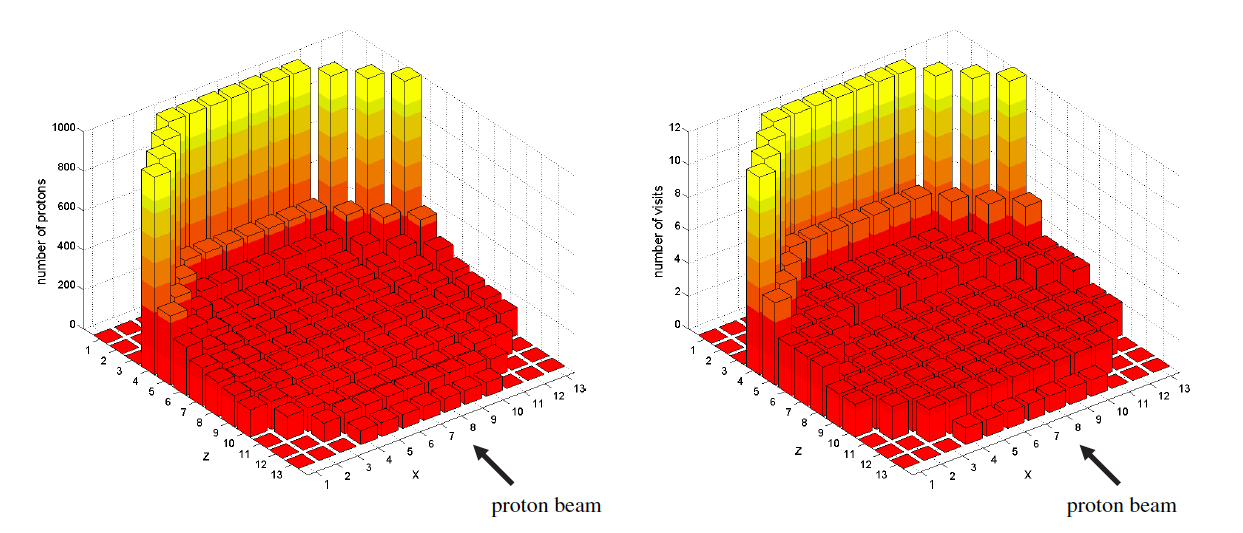}
     \caption{
     Number of protons (in arbitrary units) delivered in each voxel of the central transversal slice needed to obtain a ±1.25\% uniform dose distribution inside a 6.2~cm radius spherical volume (1 liter) centered at 20~cm depth in water (left). Number of visits needed to obtain a flat equivalent dose distribution with the condition that any missing visit does not change the total local dose by more than 3\% (right). 
     The coordinates $z$ and $x$ are given as a number of voxels, $z$ is the longitudinal and $x$ the transversal coordinate.
     }
       \label{fig:painting_protons}
\end{figure}

To asses the distribution of the dose, to evaluate the effects of the errors and the time needed for a treatment with spot scanning and multi painting, a study was performed by considering the irradiation of 
a spherical tumor of one liter, centered 200 mm deep in water~\cite{Braccini_RAST2}~\cite{Braccini_CYCLINACS}. 
For this study, one direction for the irradiation was considered. 
At 200~mm water depth, the lateral spread of the Bragg peak due to multiple scattering has a FWHM of 11.5~mm which gives an overall FWHM of 13.5~mm
when combined with a proton pencil beam having a FWHM of 7~mm. 
This corresponds to a 6.4~mm 80\%-20\% lateral fall-off. Following the same specifications used for spot scanning at PSI~\cite{Pedroni_2}, the distance between the spots has been chosen to be 75\%
of the FWHM. 
The results are reported in Fig.~\ref{fig:painting_protons}.
The proximal voxels need many less visits to obtain the SOBP with respect to the distal ones that are visited 12 times.
In this way, 
each spot is painted 3.5 times in average so that about  5200 spots in total are needed.
For the calculations, the condition that a missing spot produces an error on the local dose of less than 3\% was imposed.
At 200~Hz, a tumor of one liter can be repainted 12 times in less then 30 seconds,
when the dead times due the lowering of the magnetic fields in the delivery magnets
are taken into account~\cite{Amaldi_Braccini_arXiv}.
The maximum number of protons in one voxel is needed only for the distal spots and corresponds to N$_m$=$4\times10^7$, which is well feasible considering the high current provided by the cyclotron.
When an organ is painted 12 times
with spots containing a number of protons which fluctuate from one visit to the next by $\pm$5\%, a $\pm$1.5\% effect on the dose accuracy is induced. 
This result has to be compared with the
 $\pm$2.5\% uniformity of the dose required in radiation therapy.

\begin{figure}[t!]
 \centering
   \includegraphics[height=6.5 cm]{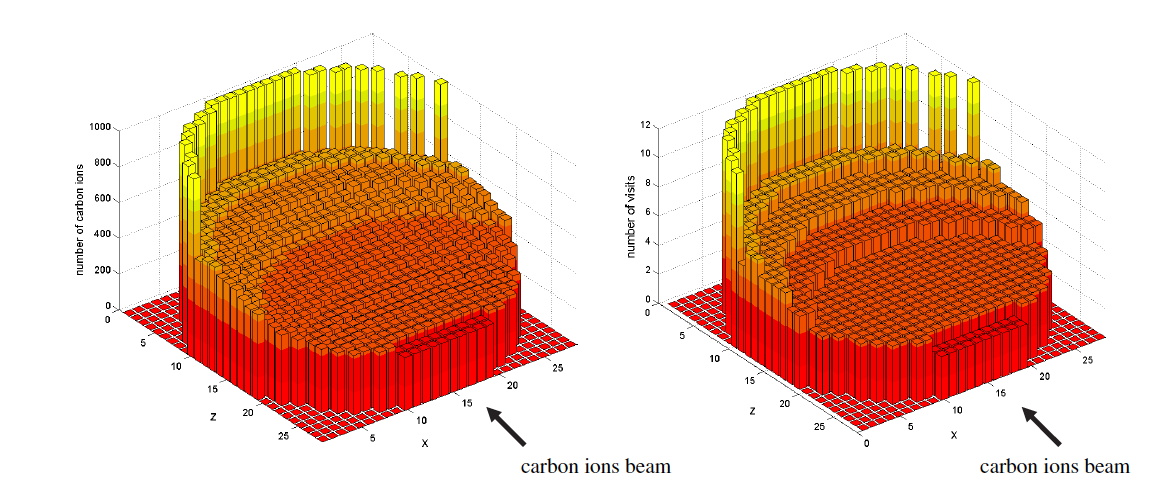}
     \caption{
     Number of carbon ions (in arbitrary units) delivered in each voxel of the central transversal slice needed to obtain a ±1.25\% uniform biological dose distribution inside a 6.2~cm radius spherical volume (1 liter) centered at 20~cm depth in water (left). Number of visits needed to obtain a flat dose distribution with the condition that any missing visit does not change the dose by more than 3\% (right).  The coordinates $z$ and $x$ are given as a number of voxels, $z$ is the longitudinal and $x$ the transversal coordinate. With respect to protons, due to the smaller FWHM of the beam, the number of spots for each dimension is double.
     }
       \label{fig:painting_C_ions}
\end{figure}

Cyclinacs are proposed also for carbon ion therapy~\cite{Braccini_RAST2}~\cite{Braccini_CYCLINACS}. At present only synchrotrons are employed for the running centers and for the current clinical projects.
Due to their large cost and complexity, innovative alternative solutions are under study. One possibility is represented by superconducting cyclotrons, which are nevertheless big and complex.
Compact superconducting cyclotrons with a beam energy limited to 300~MeV/u corresponding to 17~cm depth in water, are also proposed. This kind of accelerator allows for the treatment of 
a relatively large but limited spectrum of pathologies. For example, it has been estimated that 85\% of all head and neck cancers could be treated. 
In this context, a linac able to boost the energy to 400~MeV/u  would be beneficial to eliminate any limitation due to the penetration depth.
For this purpose, a specific linac was studied by TERA. This kind of accelerator is similar with respect to the linac of IDRA but presents specific problems to be addressed.
In particular, a new kind of ion source is needed to provide the necessary beam current which can compensate the losses due to the cyclotron-linac coupling.
The distribution of the dose needs also a particular care since the Bragg peak produced by carbon ions penetrating at 200~mm (330 MeV/u) is about 3.1~mm, 4 times sharper with respect to protons having the same penetration depth.
Furthermore, the RBE of carbon ions changes with the penetration depth and a non uniform distribution of the physical dose has to be implemented. 
Considering the energy spread due to the linac,
a pencil beam with 5.9 mm FWHM, corresponding to a fall-off of 2.8 mm, was considered. The value of the fall-off is longitudinally too narrow but the unique features of
this kind of linac allow for producing the necessary spread by slightly varying the energy in different visits of the same voxel. The results of this study are reported in Fig.~\ref{fig:painting_C_ions}.
With respect to protons, a much larger number of spots is needed to cover the same spherical volume, about 80000. This implies the need of a larger repetition rate. At 400~Hz, about 220 seconds are
needed to perform a dose delivery with a 12 visit repainting leading to effects of less then $\pm$1.5\% on the dose accuracy~\cite{Amaldi_Braccini_arXiv}.
A high repetition rate and a high gradient are fundamental features for a carbon ion linac for therapy. Along this line, a high frequency linac at 5.7~GHz was studied~\cite{Braccini_CYCLINACS}.
On this basis, the first RF cavities were constructed and tested~\cite{TERA_HF_NIM}.

\begin{figure}[t!]
 \centering
   \includegraphics[height=9 cm]{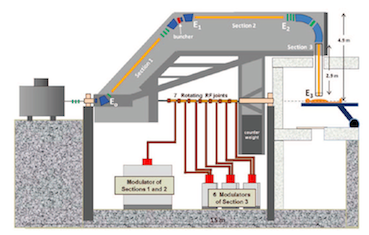}
     \caption{
     TULIP (Turning LInac for Prototherapy) is a linac-based single-room proton therapy facility optimized for fast cycling
spot scanning. The version presented here is based on two 5.7 GHz high gradient linacs connected by a buncher.
The dose is delivered by scanning by means of two transverse magnetic fields
to an area equal to 20$\times$20~cm$^2$. 
     }
       \label{fig:TULIP}
\end{figure}

High gradient linacs are beneficial also for proton therapy. As pointed out in Ref.~\cite{AmaldiBraccini}, the number of treatment rooms needed for photon, proton and carbon ion therapy scales as
$8^2$, 8 and 1, respectively. For proton therapy, the need is estimated to be about 7 treatment rooms for 10 millions inhabitants and several medium sized hospitals are starting to consider proton therapy nowadays.
For these reasons, single-room proton therapy facilities are the focus of intense research and development programs.
Along this line and on the basis of the experience gained with the design of IDRA, a linac based single-room proton therapy facility was proposed and patented by TERA~\cite{TULIP_Patent}~\cite{Braccini_RAST2}.
This system is shown in Fig.~\ref{fig:TULIP}.
This apparatus consists of a cyclotron used as the injector of a system composed by two high-frequency
and high-gradient linacs mounted on a rotating structure. The active dose delivery system is based on the apparatus designed for IDRA and, due to the rotation around the patient, it is
similar to a helical Tomotherapy. Since protons are used instead of photons, this system has the 
advantage of a much better sparing of the healthy tissues surrounding the tumour target. This innovative instrument presents many 
challenges, such as the mechanical structure, the beam transport through different accelerating structures
and the distribution of the radio-frequency (RF) power to the rotating linacs.

Another application of cyclinacs was recently proposed by TERA and PSI~\cite{Schippers}. The superconducting cyclotron 
installed at PSI is designed for the treatment of deep seated tumors and produces 250~MeV protons (Fig.~\ref{fig:PROSCAN}).
A linac booster could enhance the energy of the protons to  350~MeV and will open the way
to new treatment modalities.
The {\it shoot-through} technique is based on the sharpness of the lateral penumbra of high energy protons which do not stop in the patient's body. This modality
is an actual theme of clinical research, especially for the treatment of inter-cranial pathologies.
Furthermore, higher energy beams passing through the patient's body  can be fruitfully employed for proton radiography, an important imaging technique that will be discussed in the next Chapter.

\section{The new Bern Cyclotron Laboratory and its Research Beam Line}

The Bern cyclotron laboratory~\cite{Braccini_Bern_Cyclotron} is situated in the campus of the Inselspital and represents the first phase of the SWAN project~\cite{Braccini_SWAN}.
This new facility
has been conceived to fulfill two main goals. The first one is the production of PET radio tracers for the Inselspital and for other healthcare institutions
by means of
the most modern industrial Good Manufacturing Practice (GMP) technologies.
As observed in the first Chapter, PET cyclotrons are characterized by a high potential for research. 
The second goal is to perform 
cutting-edge multidisciplinary scientific activities in parallel with radioisotope production.

\begin{figure}[t!]
 \centering
   \includegraphics[height=9 cm]{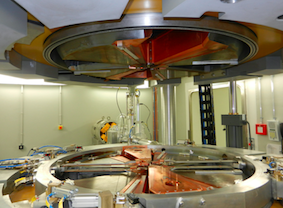}
     \caption{
   The Bern cyclotron opened during commissioning. The 30$^\circ$¡ {\it dees} are visible together with the magnetic poles producing the alternating magnetic field for vertical focussing ({\it hills} and {\it valleys}). 
   The two {\it flaps} for the D$^-$ operation mode and some of
   the eight exit ports are also visible together with the targets for $^{18}$F production. The exit port number four is connected to the BTL whose first quadrupole doublet is located in the cyclotron bunker.
     }
       \label{fig:cyclotron_opened}
\end{figure}

I have been involved in this project since the very beginning, being in charge of the choice of the accelerator, of radiation protection as Radiation Protection Officer (RPO) and of the conception of the research facility~\cite{Braccini_SWAN_Isot}. 
Besides scientific programs based of the the produced radioisotopes, 
valuable developments on accelerator physics, novel particle detectors, radiation protection, radiation biophysics and materials science can be performed~\cite{BracciniScampoli} 
only if a regular access to the beam area is possible. Furthermore, beams of variable shape and intensity must be available.  
Considering the severe access limitations due to radiation protection in radioisotope production machines and the fact that suitable research beams cannot be produced only with the cyclotron, 
I proposed in 2007 the construction of an external Beam Transport Line (BTL) terminated in a second bunker. Although more complex and unusual for a hospital-based facility, this solution allows performing
production and
research independently and efficiently by means of a fully dedicated apparatus. 

The beam line vault is connected via underground pipes to a physics
laboratory situated on the same floor.
Specific pipes have been installed also for the passage
of highly radioactive solid targets without the need of
opening the bunker doors.
A dedicated radiochemistry and radiopharmacy
laboratory is situated on the floor above the cyclotron~\cite{Zhernosekov_Braccini}~\cite{Tuerler_Braccini} together with the GMP production plant.
Radioisotopes produced in any of the targets of the cyclotron can be transferred to any of the hot cells located in the upper floor 
by means of shielded capillaries.

\begin{table}[t!]
\centering
\begin{tabular}{|ll|}\hline
Constructor	& Ion Beam Applications (IBA), Belgium \\
Type	                 & Cyclone 18/18 HC \\
Accelerated particles	& H$^-$ (D$^-$ on option) \\
Energy	& 18 MeV (9 MeV for D$^-$) \\
Maximum current	& 150 $\mu$A (40 $\mu$A for D$^-$) \\
Number of sectors & 4 \\
Angle of the dees & 30$^\circ$ \\
Magnetic field & 1.9 T on the hills and 0.35 T on the valleys \\ 
Radio frequency & 42 MHz \\
Weight  & 24000 kg\\
Dimensions & 2 m diameter, 2.2 m height \\
Ion sources	& 2 internal PIG H$^-$  \\
Extraction ports	& 8 (one of which connected with the BTL) \\
Extraction	& Carbon foil stripping (for single or dual beam) \\
Strippers	& Two per extraction port on a rotating carrousel \\
Isotope production targets	& 4 $^{18}$F$^-$, $^{15}$O ($^{11}$C and solid target are foreseen) \\\hline
Beam Transport Line (BTL)	& 6.5 m long; \\
 	& two quadrupole doublets (one in each bunker);  \\
 	& XY steering magnet; upstream collimator;   \\
	& 2 beam viewers; neutron shutter \\ \hline
\end{tabular}
	\caption{
	Main characteristics of the Bern cyclotron and its beam transport line.
	}
	\label{tab:Bern_Cyclotron}
\end{table}

After a phase of study and design, the construction started in 2010. In June 2011, the civil engineering infrastructure was ready and
the cyclotron was successfully {\it rigged} inside its bunker, after having passed a series of functional tests performed at the factory.  
Following the construction and test of all the service subsystems, the installation of the cyclotron and of the beam line proceeded on schedule and the first proton beams were 
accelerated and extracted in February 2012.
An intense optimization phase followed and, 
at the end of 2012, the whole facility was fully operational. 
  
The {\it heart} of the facility consists of the IBA Cyclone 18 MeV cyclotron shown in Fig.~\ref{fig:cyclotron_opened}.
The main technical data are reported in Table~\ref{tab:Bern_Cyclotron}.
The cyclotron is equipped with two H$^-$ ion sources, a redundancy aimed at maximizing the efficiency for daily medical radioisotope production.
It provides large beam currents up
to 150~$\mu$A in single or dual beam mode. Extraction is obtained by stripping the negative ions by means of 5~$\mu$m thick pyrolytic carbon foils.
In dual beam mode, the beam is extracted by two strippers located approximately at an angular distance of 180$^\circ$. The relative position of the last orbit can be tuned
in such a way that different currents can be extracted in the two out ports. Although difficult to obtain, this feature was required for the Bern cyclotron in order
to operate two different targets producing two different kind of radioisotopes. It is important to remark that high current negative ion cyclotrons are possible thanks to
the development of specific materials for the strippers, able to stand temperatures of about 1500~$^\circ$C or more produced by the energy released
by the protons and, mainly, by the stripped electrons. 

An optimized vertical focussing and longitudinal phase stability are
crucial for high current cyclotrons. This is realized by alternating four 60$^\circ$ sectors at high magnetic field ({\it hills})
with four 30$^\circ$ sectors at low magnetic field ({\it valleys}). This implies that the orbits are not circular and that a complex and iterative mapping procedure is necessary
to obtain a precise magnetic filed. It has to be noted that a fine tuning of the cyclotron can be performed only on site by measuring the accelerated beam. In particular,
the magnetic field of the Bern cyclotron was carefully regulated 
in the central region to obtain the optimum transmission of the low energy ions produced by the ion source. An iterative procedure was applied
aimed at modifying the gap of the magnet in the central region ({\it shimming}).

The average magnetic field of 1.4~T gives the cyclotron frequency 
\begin{equation}
\nu_c = \frac{qB}{2\pi m}
\end{equation}
of 21~MHz. Since this cyclotron is equipped with two {\it dees}, only RF frequencies which are even harmonics of the cyclotron frequency lead to acceleration. 
As shown in Table~\ref{tab:Bern_Cyclotron}, the RF frequency of 42~MHz corresponds to the second harmonic for H$^-$ ions.
This kind of cyclotron is able to accelerate also deuteron beams, as tested at the factory.
In the case of D$^-$ ions, the cyclotron frequency is about one half and, being the RF fixed, the machine works in the fourth harmonic. To correct for effects due to the mass defect of
deuterons, perturbations to the magnetic field are produced by specific movable components inserted in the upper part two of the valleys ({\it flaps}), as shown in Fig.~\ref{fig:cyclotron_opened}. 

Acceleration occurs when the ions cross the borders of the dees and the energy gain per turn $E_k$ is given by the formula
\begin{equation}
\Delta E_k = V_{dee} N q \sin\left(\frac{h \alpha}{2} \right)
\end{equation}
where  $V_{dee}$ is the RF peak voltage, $N$ the number of accelerating gaps corresponding to twice the number of the dees, $q$ the charge of the accelerated ion,
$h$ the harmonic mode and $\alpha$ the azimuthal dee angular length. With a RF peak voltage of 30~kV, this leads to values of $E_k$ of 60~keV and 100~keV for H$^-$ and D$^-$, respectively.
In this way, H$^-$ ions reach the extraction energy of 18 MeV after 300 turns while 9~MeV D$^-$ are extracted after 90 turns.

The performances obtained with the Bern cyclotron for radioisotope production are excellent. In particular, 500~GBq of $^{18}$F  can be produced with dual beam
irradiations lasting less than 70 minutes. Considering the total efficiency of the chemical synthesis (usually about 60\%), this corresponds to about 250 GBq of FDG 
produced in a single run. This result has to be compared with the typical figures given in the
first Chapter. It has to be noted that with this apparatus larger productions would be possible and that  500~GBq represents the legal radiation protection limit for a laboratory of this kind 
(type B) in Switzerland.

\begin{figure}[t!]
 \centering
   \includegraphics[height=9 cm]{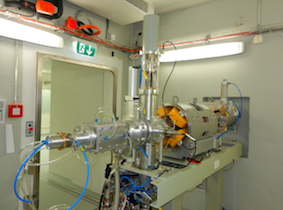}
     \caption{
   The BTL is terminated in the beam line vault where a neutron shutter is located before the second quadrupole doublet. A beam viewer and a {\it four finger} collimator
   allow to position the beam before its use for scientific purposes.
     }
       \label{fig:BTL}
\end{figure}

\begin{figure}[t!]
 \centering
   \includegraphics[height=9 cm]{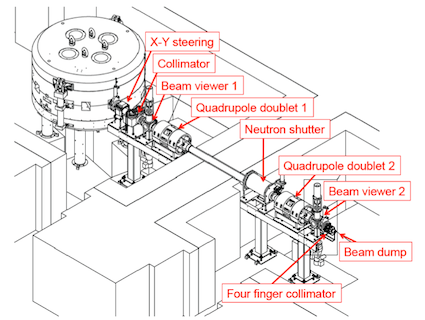}
     \caption{
   Schematic view of the BTL of the Bern cyclotron laboratory, in which all the main elements are highlighted.     }
       \label{fig:BTL_scheme}
\end{figure}

Beam lines are quite rare for 18~MeV cyclotrons. They are usually only about 2~m long, located in the same bunker as the cyclotron and dedicated to the bombardment of solid targets.
This kind of apparatus was not suitable for our purposes. At the beginning of the SWAN project, IBA was designing a 6~m long beam line, aimed 
at the production of radioisotopes by means of high current beams, of the order of 50~$\mu$A or more. For the aims of the Bern project, I proposed the 
construction of a BTL able to transport the maximum current of 150~$\mu$A with more than 95\% transmission. 
High transmission is crucial to limit unwanted activation and to protect scientific equipment sensitive to radiation.
Furthermore, this performance is important for future developments since remarkable progress is made on targets able to stand very large currents.
At the same time,
currents down to fractions of nA are needed for research purposes. 
Furthermore, beams of different sizes ranging from a few millimeters  to a few centimeters diameter on
target have to be obtained.
To fulfill these ambitious goals, specific solutions were implemented and the initial industrial design modified.
These original developments are described here in more detail~\cite{Braccini_CAARI}.

The BTL is shown in Figs.~\ref{fig:Bern-Cyclotron},~\ref{fig:cyclotron_opened} and~\ref{fig:BTL} and
a schematic view is presented  in Fig.~\ref{fig:BTL_scheme}. The FODO lattice of the BTL is realizeded by two horizontal-vertical (H-V) quadrupole doublets, the former located
in the cyclotron bunker and the latter in the BTL one. An unavoidable design constraint is represented by the distance between the two quadrupole doublets
due to the 180~cm thick wall separating the two bunkers. This thickness is due to radiation protection since large fast neutron fluxes are generated during isotope production, for example
via (p,n) reactions. For the same reason, a movable cylindrical neutron shutter is located inside the beam pipe when the BTL  is not in use. In this way,
the penetration of neutrons to the second bunker through the beam pipe is minimized.
Taking into account the space needed for the vacuum pumps, beam diagnostics, collimators and other devices, the beam line is 6.5~m long. 

\begin{figure}[t!]
 \centering
   \includegraphics[height=9 cm]{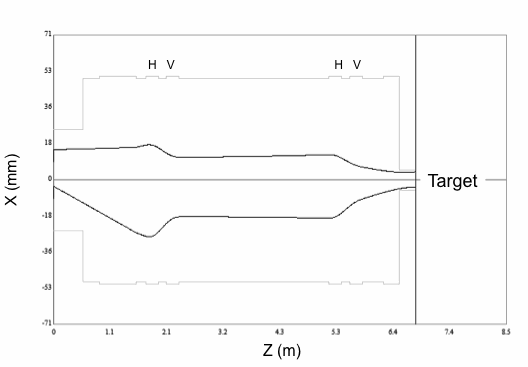}
     \caption{
     Simulation of the beam envelope in the XZ
plane in the original industrial design of the BTL.
The beam is centered on the first viewer and has a tilt
angle of 7 mrad with respect to the optical axis. The XY
steering magnet - located in front of the first quadrupole
doublet - is able to perform a substantial correction.
However, distortion effects still remain, leading to a non
optimal transmission. The horizontal (H) and vertical (V)
focusing quadrupoles are indicated.
       }
       \label{fig:beam_dynamics_1}
\end{figure}

\begin{figure}[t!]
 \centering
   \includegraphics[height=9 cm]{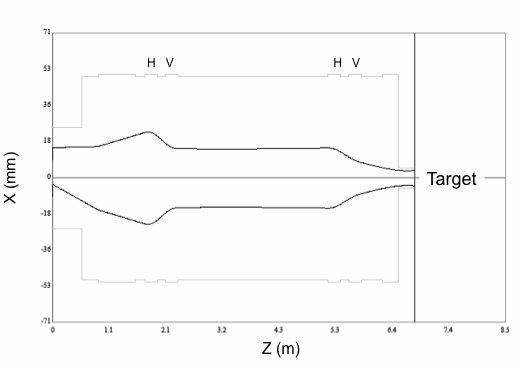}
     \caption{
     Simulation of the beam envelope with the modified design of the BTL. The beam is
extracted from the cyclotron in the same conditions as in
Fig.~\ref{fig:beam_dynamics_1}. With the XY steering magnet located as near as
possible with respect to the cyclotron, it is possible to obtain
a centered and parallel beam with negligible distortion
effects and an optimized transmission.
       }
       \label{fig:beam_dynamics_2}
\end{figure}

Since the trajectory of the ions is curved inside the cyclotron and straight in the beam line,
the injection of the beam into the BTL is critical, especially if a transmission larger of 95\% is required.
To study this issue, simulation studies were performed with Beamline Simulator~\cite{Braccini_CAARI}~\cite{BeamLineSimulator}.
It has to be noted that 
the transfer of the extracted beam is much more difficult in the accelerating (horizontal) plane as compared to the vertical plane.
The simulations showed that
the beam has to be centered with respect to the BTL better than 1~mm and, at the same time, its parallelism with respect to the axis has to be
better than 1~mrad to ensure high transmission.
In the case of larger deviations, the beam is subject to an overly large dipolar field when passing through the
quadrupoles and the beam envelope becomes highly distorted. 

To obtain a centered beam, the accuracy of the mechanical alignment of all the elements of the BTL is crucial. This was realized by means of a laser installed inside the
cyclotron to reproduce the ideal beam path and to give a precise reference for the positioning of the two quadrupole doublets.  
To allow for corrections of the beam direction after the extraction, a XY steering magnet able to bend the 18~MeV beam of maximum $\pm$7~mrad vertically or horizontally is used.
In the initial design, this magnet was located in front of the first quadrupole doublet, about 1.5~m away with respect to the extraction point.
Measurements of the beam profile performed with gafchromic film detectors showed that the beam exited the cyclotron with an angle of 7~mrad when centered on the first beam viewer.
As demonstrated by measurements and by the simulation reported in 
Fig.~\ref{fig:beam_dynamics_1}, these conditions led to the non optimal transmission of about 80\%, well below the required 95\%.
These measurements were performed using standard carrousels for the strippers and with an optimized azimuth angle of the strippers aimed at directing 
the beam right in the centre of the steering magnet. The XY steering magnet had to be set at its maximum current to correct for the exit angle.
 
 To improve the transmission, the position of the XY steering magnet can be optimized by locating it as close as possible with respect to the extraction point, as shown in Fig.~\ref{fig:beam_dynamics_2}.
 This modification was implemented and the transmission improved significantly to more than 90\%.
 Also in this case, the XY steering magnet had to be set at its maximum current, a situation which did not allow for further improvements with a standard set-up.
 
Another parameter can be used to tune the centering of the beam in the horizontal
plane: the radial position of the stripper. Together with the azimuth angle, it determines the position and the angle with which the beam leaves the cyclotron.
Once the radial position of the stripper is fixed, the azimuth angle can be adjusted to find the optimal exit angle.

\begin{figure}[t!]
 \centering
   \includegraphics[height=4.5 cm]{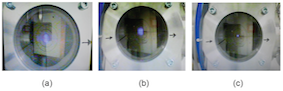}
     \caption{The beam spot extracted from the cyclotron is measured by the first beam viewer located before entering the FODO lattice (a).
     By adjusting the optics of the beam line, the beam spot on target can be adjusted according to the specific needs. Beam spots of about 1~cm (b) and 3~mm (c) diameter are reported here, as measured with the beam viewer located
     after the FODO lattice.
      }
       \label{fig:Beam_Spots}
\end{figure}

Variations of the radial position of the stripper induce changes on the
output energy of the beam. As it was already observed, orbits are characterized by successive increases of 60~keV with the RF voltage set at 30~kV.
Since the radius $\rho$ depends on the kinetic energy $E$ as
\begin{equation}
\rho=\frac{\sqrt{2Em}}{qB}
\end{equation}
a radial increase of the position of the stripper of 1~mm corresponds to an increase of the extracted energy of about 60~keV.
Furthermore, since extraction is performed on the {\it hills} of the magnet, changes in the radial position of the stripper determine changes in the
path length inside the region where the magnetic field is maximal.
These two effects lead to a high sensitivity of the angle with which the beam leaves the cyclotron with respect to the radial position of the stripper.
These effects are difficult to simulate, mainly due to the stray field of the cyclotron magnet. 

For the IBA 18 MeV cyclotron, the strippers are mounted on a rotating carrousel with a fixed radial position. Thus
only the azimuth angle is adjustable, once the dimensions of the carrousels are chosen.
To determine the optimal radial position of the stripper for an optimum injection in the BTL, several carrousels with different radii were constructed and 
beam profile measurements were performed using gafchromic film detectors.
The optimum was found at a radius 5~mm larger than the standard one. With this modification together with the new position of the steering magnet, excellent results were obtained.
A transmission larger than 97\% was measured at the maximum current of 150~$\mu$A.
Below 10~$\mu$A, the transmission was found to be about 100\%.
A transmission of 99\% was obtained when operating in dual beam with a total summed current of 150~$\mu$A.
In this mode, only a part of the beam is extracted towards the BTL and the beam emittance is consequently smaller.
Beams of different shapes can be obtained at the end of the BTL ranging from a few mm to a few cm, as shown in Fig.~\ref{fig:Beam_Spots}.
It has to be noted that all these measurements were obtained with a negligible current on the XY steering magnet, demonstrating the optimum centering of the beam.

 \begin{figure}[t!]
 \centering
   \includegraphics[height=7 cm]{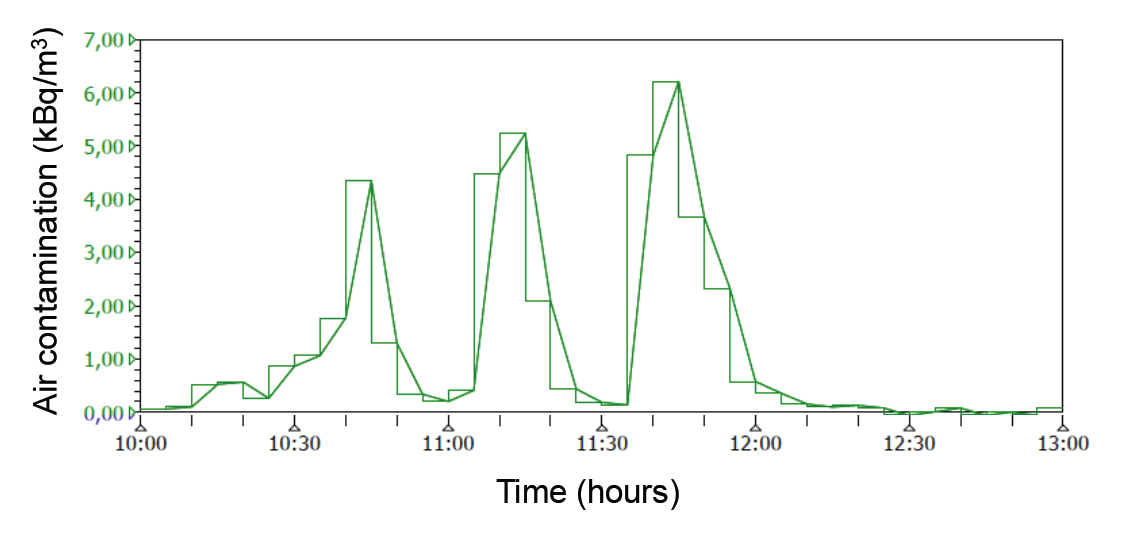}
     \caption{
   Measurement of air contamination on the main exhaust of the facility due to three successive irradiations with 18~MeV protons extracted into air at an intensity of about 50~nA for 60 seconds. The peaks correspond to
    10, 13 and 15~MBq, respectively, mostly due to $^{11}$C produced by the reaction $^{14}$N (p, $\alpha$) $^{11}$C with a contribution of about 10\% of $^{13}$N from the reaction 
    $^{16}$O (p, $\alpha$) $^{13}$N.
       }
       \label{fig:Air_Contamination}
\end{figure}

In a complex installation where very high
quantities of radioactive substances are produced, radiation protection is essential.
The Radiation Protection Monitoring System (RPMS) of the Bern cyclotron 
is composed of about 50 sensors and is both a safety and a research tool.
Ambient dose gamma radiation levels, air
contamination, and neutron fluxes are continuously monitored on-line and can be accurately studied.
 Special attention has been devoted to the monitoring of the emission of radioactive
gases into the atmosphere by means of sophisticated sensors able to detect air contamination levels down to 10~Bq/m$^3$, as shown in Fig.~\ref{fig:Air_Contamination}~\cite{Braccini_CYCLEUR}.
Along this line, novel compact sensors for the measurement air contamination are under development in Bern~\cite{EnvAir} using both gas and solid state sensors.

The commissioning and the optimization of the cyclotron and of the BTL were successful, either for production or for multi-disciplinary research activities.
This process
clearly put in evidence the important role of beam monitoring devices. Along this line, the first research activities focus
on the development of innovative beam monitor detectors, as presented in more
detail in the next Chapter. Furthermore, this enterprise is fostering research in accelerator physics in Bern~\cite{PMQ_RFQ}~\cite{ECPM_PSI}~\cite{Ion_Beams_12}.

\chapter{Particle Detectors for medical Applications of Ion Beams}
\label{ch3}

This Chapter is focused on my contributions to particle detectors used for the measurement and the control of ion beams in medical applications. The main feature of 
hadrontherapy is its precision. The control of the beams during treatments and the knowledge of the range of the ions are
fundamental for an optimum treatment. Especially for active beam scanning, 
segmented ionization chambers represent a valuable solution for on-line monitoring. Proton radiography is a powerful technique to assess the range of the accelerated ions
and innovative solutions are under study for its application in a clinical environment.
Beam monitoring along transport lines or in front of the target is fundamental for all applications, cancer hadrontherapy and radioisotope production in particular.
These issues are discussed here
in close connection with the publications reprinted in Chapter~5.

\section{Segmented Ionization Chambers for Beam Monitoring in Hadrontherapy}

The on-line measurement of the intensity, the position and the shape of the clinical beam just before entering the patient's body is of paramount importance in hadrontherapy treatments.
This information is used to monitor the amount and the spatial distribution of the delivered dose. 
In the case of active scanning and moving organs, these measurements can be used to apply corrections during the irradiation by means of an appropriate feedback system. 

In parallel with the development of accelerators for medical applications, I dedicate part of my scientific activity to
novel particle detectors. During the time I was with the TERA Foundation, I contributed to the design, the construction and the test of segmented ionization 
chambers for the on-line monitoring of therapeutic ion beams. This activity continued after I moved to the University of Bern. Two detectors are discussed in detail in this 
paragraph: a pixel and a strip ionization chamber, named MATRIX~\cite{Braccini_MATRIX} and SAMBA~\cite{Braccini_SAMBA}~\cite{Ferretti}, respectively.
The Strip Accurate Monitor for Beam Applications (SAMBA) detector was developed for the Gantry 2 at PSI where it is currently in use, as shown in Fig.~\ref{fig:PROSCAN}.

To measure the integral of the dose, parallel plate ionization chambers are usually used.
For passive scattering, the flatness of the clinical field is generally controlled by means of ionization chambers subdivided in four or five sectors.
In case of active scanning, the position of the pencil beam is normally monitored using wire chambers located in front of the patient and covering the whole clinical field, typically
about 20~$\times$~20~cm$^2$. 

An innovative alternative is represented by highly segmented pixel or strip ionization chambers, able to provide spatial information and to measure the fluence at the same time. 
If the electrodes are made very thin, scattering effects can be minimized with respect to standard systems. 
Due to the absence of multiplication,
the signals produced by the low intensity therapeutical beams have to be collected and treated by a dedicated sensitive electronics and read-out system.
This problem is enhanced by the fact that each strip or pixel sensor necessarily detects only a fraction of the ionization signal produced inside the fiducial volume of the detector.
 
To better describe this problematic, let's consider the proton therapy beam of the PSI. The dose is delivered by spots 
of variable temporal length depending on the amount of dose to be delivered. The intensity can be varied from 0.2 to 2~nA. A typical situation is represented by
an average current of 0.2~nA with spots 
of 5~ms, corresponding to a total of about 6$\times$10$^6$ protons per spot. In a ionization chamber with a gap $d$ between the two electrodes,
the gain factor $G$ represents the average number of electron-ion pairs produced by one single proton traversing the chamber.
The gain factor is given by
\begin{equation}
G=\frac{dE}{dx}\frac{d}{W}
\end{equation}
where $W$ is work function, the average energy to produce an electron-ion pair. For dry air $W$ is 35~eV.
Considering a planar ionization chamber operating in air  with $d$ of 1~cm and a
beam energy of 160~MeV, the total charge of 150~pC is produced by each spot.
Due to the segmentation of the detector, charges down to 10~pC have to be measurable with a good precision.

 \begin{figure}[t!]
 \centering
   \includegraphics[height=9 cm]{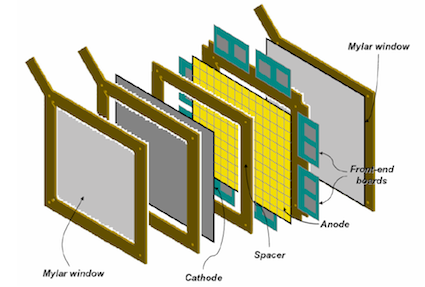}
     \caption{
     Schematic view of the MATRIX pixel ionization chamber.
         }
       \label{fig:MATRIX_Scheme}
\end{figure}

For the read-out of segmented ionization chambers, the TERA Foundation and the Italian National Institute for Nuclear Physics (INFN) in Turin developed a specific 
very large scale of integration (VLSI) chip, named TERA~06.
This device has 64 independent channels based on a recycling integrator architecture. Each channel is equipped with a 16-bit digital counter and
is connected to one strip or pixel of the detector. The collected charge is continuously integrated by a capacitor of 0.6~pF.
When the voltage of the capacitor is larger than a programmable threshold value,
the capacitor is discharged by the injection of pre-defined quanta of charge. In this way,
the collected charge is measured by counting the quanta of charge needed to bring back the voltage of capacitor under threshold.
The use of recycling integrators allow reducing dead time effects to a negligible level.
The sensitivity of the measurement can be adjusted by varying the quantum of charge in the range between 100 and 800~fC. 
The maximum read-out frequency of 5 MHz limits the maximum current of each channel to 4~$\mu$A. For this chip,
the linearity of the measurement of the charge was measured to be better than 1\% in the full range of operation and the noise to be 
the less than one count per channel per second.
The tested minimum read-out time for  one channel was of 1~$\mu$s.

\begin{figure}[t!]
 \centering
   \includegraphics[height=7.5 cm]{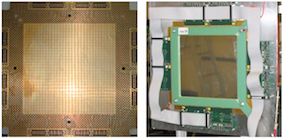}
     \caption{
The segmented anode with the 1024 pixels and the guard ring (left). The MATRIX chamber fully
assembled. The fibreglass frame and the eight TERA~06 electronic boards for the read-out are visible (right).         }
       \label{fig:MATRIX_Detector}
\end{figure}

The data acquisition systems of MATRIX and SAMBA~\cite{Braccini_Pitta_readout} are based on specifically designed TERA~06 boards, each hosting two TERA~06 chips and capable 
to handle 128 channels in total.
For the measurements presented here, the maximum sensitivity of 100~fC was chosen for the quantum of charge, which corresponds to 100 counts for a total collected charge of 10~pC.
 
 \begin{figure}[t!]
 \centering
   \includegraphics[height=8 cm]{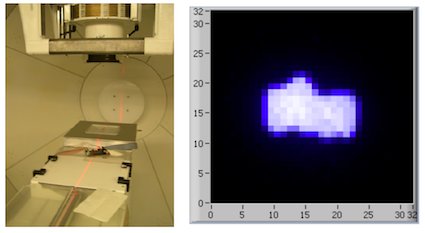}
     \caption{
     MATRIX installed in a gantry room of the Loma Linda University Medical Center (left). Measurement of the 
     two-dimensional dose distribution for a field shaped for the treatment of a prostate cancer. The two dimensions are
     measured in number of pixels (right).
         }
       \label{fig:MATRIX_LLUMC}
\end{figure}

The structure of MATRIX is shown in Fig.~\ref{fig:MATRIX_Scheme} and is based on five successive fibreglass frames.
The external ones are equipped with glued mylar foils which allow to
maintain the planarity of the internal cathodic and anodic electrodes in case a
gas is used at non atmospheric pressure. The sensitive volume of the detector
corresponds to the air gap between the anode and the cathode foils. This gap can
be easily changed by substituting the spacer. A gap of 5~mm was chosen, which for a bias voltage of 400~V
gives an
electric field of 800 V/cm in the sensitive volume.

The gap and the applied electric field determine the collection time of the electrons and of the ions. The drift velocity $v$ is given by
\begin{equation}
v=\mu\frac{\epsilon}{p}
\end{equation}
where $\mu$ is the charge mobility, $\epsilon$ the electric field and $p$ the pressure of the gas filling the sensitive volume. The mobility in air is of the order of 10$^{-4}$~m$^2$atm/Vs 
for ions and about 1000 times smaller for electrons.
In this way, collection times of the order of 1~ms and
1~$\mu$s are obtained for ions and electrons, respectively.
Since these detectors are based on the measurement of the negative charge collected by
the segmented anodes, the timing characteristics are well suitable for spot scanning where the duration of a spot is of the order of a few milliseconds.

The anode and cathode foils are glued to the
respective frames by means of a specific procedure able to guarantee a good planarity. MATRIX has a
sensitive area of 21~$\times$~21~cm$^2$ subdivided in 32~$\times$~32 pixels of 6.5~$\times$~6.5~mm$^2$. The 1024 pixels are read-out by 8 TERA~06 boards.
The anodic electrode was manufactured at CERN using techniques derived from the construction of printed circuit boards (PCB). 
In order to minimize the amount of material traversed by the beam, the electrode is made of a thin 50~$\mu$m
kapton foil, with a deposition of 17~$\mu$m copper on each side. Aluminium depositions are at present under study to allow for a even lower amount of material.
In the case of hadrontherapy,
simulations show that negligible effects are produced to the beam as far as the dose distribution is concerned using both copper or aluminium.
To minimize noise effects, the sensitive area of the anode is surrounded by a guard
ring kept at the same potential as the pixels. The other side of the anode
foil contains the pixel-pad connections used to transport the signals to the read-out electronics. 
The anode and the fully assembled MATRIX detector are shown in Fig.~\ref{fig:MATRIX_Detector}.

Tests of the detector were first performed by X-ray irradiation in order to verify the
correct operation of all the pixels and of the corresponding read-out chains. Following the successful 
beam tests at the 40~MeV cyclotron of the Joint
Research Centre of the European Commission in Ispra (Italy), MATRIX was tested with passive scattering clinical beams at the
Loma Linda University Medical Center in
California, as shown in Fig.~\ref{fig:MATRIX_LLUMC}. 

In Loma Linda, the measurement campaign~\cite{Braccini_MATRIX} was focused on the monitoring of several two-dimensional maps of clinical fields (Fig.~\ref{fig:MATRIX_LLUMC}~(right)) and on the pixel-by-pixel inter-calibtation procedure.
The Bragg peak produced by a 149~MeV proton beam
was measured by interposing successive layers of water equivalent absorbers between
the nozzle and the chamber. The range in water was estimated to be 154~mm in good agreement with the simulation and with previous measurements performed in Loma Linda. 
The effects of the exit window of the accelerator and of the material of the propeller present in the nozzle were taken into account.
The total counts of the chamber were measured also with additional 10~mm water equivalent material with respect to the Bragg peak position. 
In this way, the effect of the dose produced by neutron induced recoil protons can be put in evidence. 
A clear signal was observed and the total counts were estimated to be about 0.1\% of the maximum corresponding to the Bragg peak, in agreement with previously
performed measurements with the same beam and with data reported in Ref.~\cite{Schneider_Agosteo}.
The read-out speed of 2~ms for the full chamber was reached, an essential feature for its use with active scanning systems.

\begin{figure}[t!]
 \centering
   \includegraphics[height=9 cm]{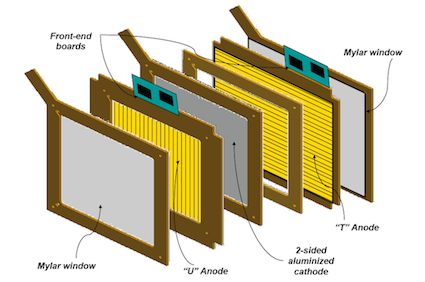}
     \caption{
     Schematic view of the SAMBA detector composed of two orthogonal strip ionization chambers.          }
       \label{fig:SAMBA_Scheme}
  \end{figure}
      
 \begin{figure}[t!]
  \centering
   \includegraphics[height=9 cm]{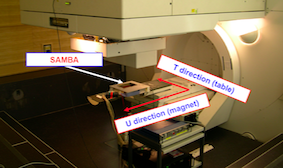} 
     \caption{
     The SAMBA detector under test at the Gantry 1 at PSI. The magnetic sweep of the beam is detected by the 128 strips of the U chamber while the 88 channels of the T chamber are
     sensitive to the movements of the table.
         }
       \label{fig:SAMBA_PSI}
 \end{figure}

 The Gantry 2 at PSI was conceived for
 double parallel magnetic beam scanning over an area of 20~$\times$~12~cm$^2$ with
spots of 8~mm FWHM  and a beam intensity ranging from 0.2 to 2~nA. The temporal length of one spot and the temporal distance between successive spots are both
of the order of a few milliseconds.
For this dose distribution technique,
 the precise and fast determination of the position of each spot represents the most important information that an on-line monitoring detector has to provide. 
 To accomplish this task, a strip ionization chamber is surely more appropriate with respect to a pixel detector since the lower number of channels to be read-out allows for a higher detection speed.
 Furthermore, strips can be made smaller with respect to pixels giving a higher resolution on the determination of the centroid of the spot. Using two orthogonal strip chambers,
 the positions and the projections of all the delivered spots can be measured and stored. If needed, 
 the two dimensional map of the clinical beam can be reconstructed at a later stage, since negligible shape variations are expected between different spots.
 
 The SAMBA detector was designed for the on-line monitoring of the Gantry 2 
 and was conceived, built and tested in the framework of a collaboration between the TERA Foundation and the PSI.
 It consists of two strip ionization chambers, as
 shown in Fig.~\ref{fig:SAMBA_Scheme}.
 
 SAMBA is composed by two chambers (denominated T and U, according to the denomination used for the Gantry~2) with the anodes segmented with 2~mm strips.
 The two chambers are orthogonal one with respect to the other and cover a sensitive area of 25.6~$\times$~17.6~cm$^2$.
 The chamber U and T have 128 and 88 strips, respectively. 
 This is due to the choice of having the same strip width, and therefore the same spatial resolution, for the two chambers.
 Each  chamber is read-out by one TERA~06 board~\cite{Braccini_Pitta_readout}. For the board corresponding to the T chamber, 
 40 channels are not connected to any strip and can be used to monitor the noise in the
 TERA~06 chip. The two chambers operate in air and, to increase the gain with respect to MATRIX, a gap of 10~mm was chosen. 
 The fibreglass frames and the anodes are manufactured with the same technology used for MATRIX. 
To minimize the amount of material traversed by the beam,
a single two-sided  aluminized cathode is located in the middle of the two sensitive volumes.
The two anodes are connected to the ground potential.
A bias voltage of -800~V is applied to the cathode to obtain an electric field of 800~V/cm used to collect the charge produced by ionization.

\begin{figure}[t!]
 \centering
   \includegraphics[height=10 cm]{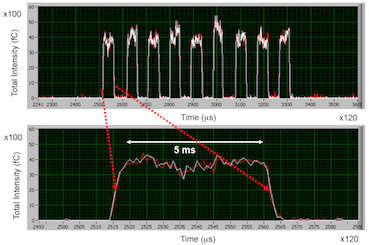}
     \caption{
     Successive spots (top) and zoom on a single spot (bottom) detected by the SAMBA detector. The measurements of the total charge performed by the T and U chambers (white and red, respectively) overlap within the statistical fluctuations. The time interval between full read-out procedures of the two chambers is set to 120~$\mu$s.
             }
       \label{fig:SAMBA_Spots}
\end{figure}

For the beam tests of the SAMBA detector, the Gantry 1 at PSI was used, as shown in Fig.~\ref{fig:SAMBA_PSI}. This apparatus is able to sweep the beam magnetically only in one direction and
two dimensional spot scanning is realized by combing the magnetic sweep with the mechanical movements of the patient's table. The beam energy was set to 160~MeV.

Several measurements were performed to test the timing and dose response of SAMBA. One of those is reported in Fig.~\ref{fig:SAMBA_Spots}, where
a series of 9 spots of a duration of 5~ms and with a temporal separation of 5~ms was directed towards the centre of the detector. The average current in the flat top of the spots was set to 0.2~nA.
 The signal due to the spots is well visible over a negligible background. The total intensity presented in Fig.~\ref{fig:SAMBA_Spots} is measured as the sum of all the counts in the U and in the T chambers,
 respectively. The results of the two chambers are consistent within the statistical fluctuations.
The average value of the flat top is found to be 37 counts (3700~fC) for temporal distance between two successive read-out procedures of the two chambers of 120~$\mu$s.
Once the gain is taken into account, the average current was measured to be 0.20$\pm$0.03~nA, in good agreement with the set-up value of 0.2~nA.
The obtained total read-out time of 120~$\mu$s is suitable for spot scanning. This value can be reduced to about 25~$\mu$s using a specific fast data acquisition (DAQ) system.

\begin{figure}[t!]
 \centering
   \includegraphics[height=5.5 cm]{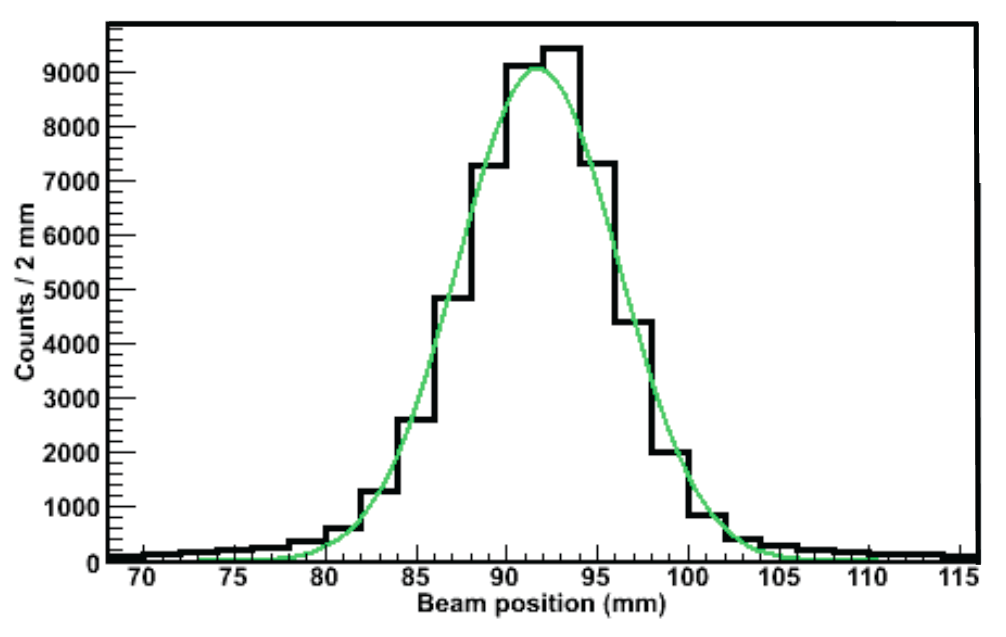} \\
   \includegraphics[height=7 cm]{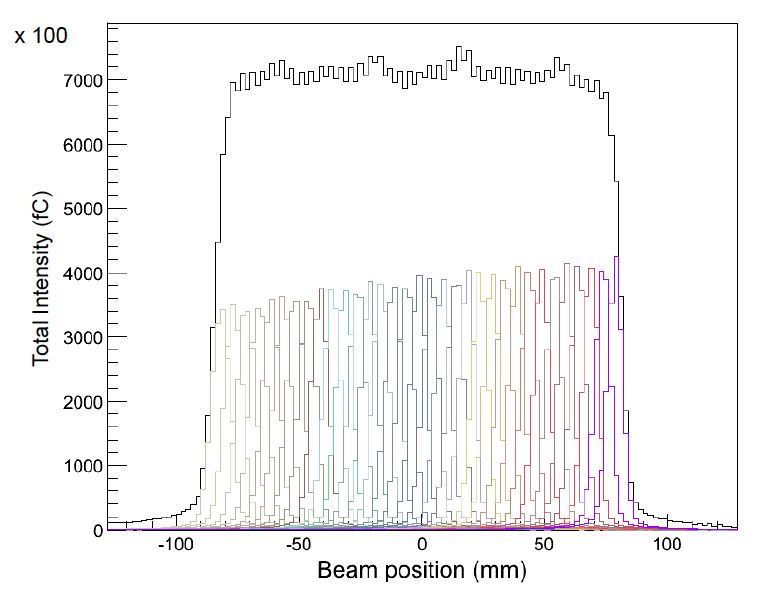}
     \caption{
     One single spot produces a signal on several strips and is well modeled by a gaussian curve (top).
     A flat linear dose distribution is obtained  by superimposing 33 spots spatially separated by 5~mm in the U direction. The movement of the beam is obtained using the fast sweeping magnet (bottom).
         }
       \label{fig:SAMBA_Doseline}
 \end{figure}

Single spots are measured by examining the signal in the strips, as reported in Fig.~\ref{fig:SAMBA_Doseline}~(top). The shape of the projection of the spots can be fitted with a gaussian
function either for the T or for the U chamber. 
As normal for a particle beam,
the distribution of the collected charge presents  a slight excess on the tails with respect to the fitted gaussian. 
The position of the
centroid of the spots can be determined with a precision better than 0.1~mm. 
For the FWHM values ranging from 8 to 10~mm were measured with a precision better than 0.1~mm. Either for the position or for the FWHM, the error is dominated by the beam characteristics 
and not by the intrinsic precision of the detector.

Using the sweeping magnet, series of spots producing lines in the U direction were measured, as reported in Fig.~\ref{fig:SAMBA_Doseline}~(bottom). Here 33 spots separated by 5~mm were 
used to produce a 16~cm long flat dose distribution. The difference between the measurements of the total collected charge performed by the two chambers was measured to be less then 1\%.
The reproducibility of the measurement was estimated to be better then 0.4\% by repeating the full procedure five times.
The modest and repetitive {\it up and down} behavior between adjacent channels is due to a slight different gain of the amplifiers located on different sides of the TERA~06 chip, 
as shown in Fig.~\ref{fig:SAMBA_Doseline}~(bottom). This effect is of the order of 1\%. Once a calibration procedure was applied, the flatness of the dose distribution was measured to be better than 0.6\%,
well within the therapeutic requirement of 2.5\%.

\begin{figure}[t!] 
  \centering
   \includegraphics[height=6.5 cm]{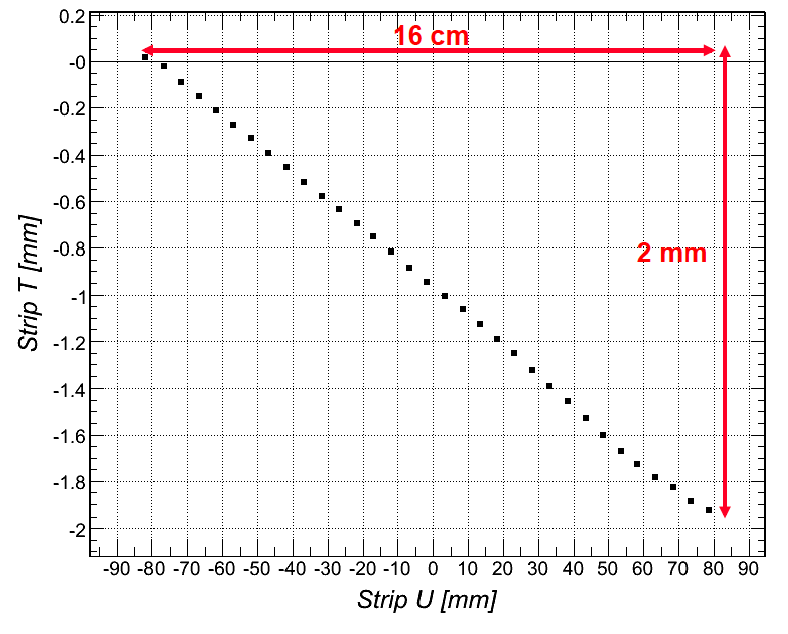} 
   \includegraphics[height=6.5 cm]{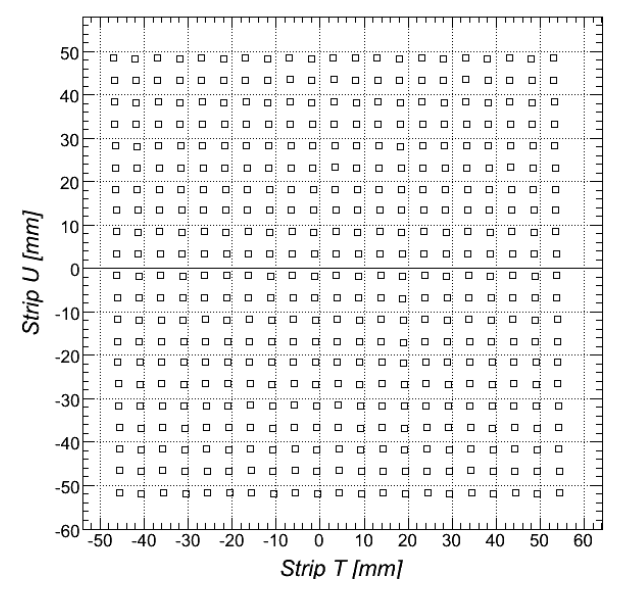}
     \caption{
     The T and U positions of the spots corresponding to the flat linear dose distribution of Fig.~\ref{fig:SAMBA_Doseline}~(bottom) are measured by fitting the projections (left).
     The T and U positions of the spots corresponding to a 10~$\times$~10~cm$^2$ flat square dose distribution. The 441 spots are separated by 5~mm either in the T or in the U direction (right).
         }
       \label{fig:SAMBA_Line_Square}
\end{figure}

The T and U positions of the centroids of all the spots composing the flat linear dose distribution of Fig.~\ref{fig:SAMBA_Doseline}~(bottom) were measured by performing gaussian fits.
The results are shown in Fig.~\ref{fig:SAMBA_Line_Square}~(left). The measurement shows a good linearity. 
A modest deviation of about 2~mm with respect to the table in the T direction was determined. 
This effect was known and due to a modest misalignment of the 90$^\circ$ 
magnet of the gantry with respect to the table. Furthermore, 250~$\mu$m deviations from the linearity were observed at the edges due to non-linearities of the 90$^\circ$ magnet.
Although with a worse precision, these effects were known since the commissioning of the Gantry 1.
It has to be remarked that these small effects have a negligible clinical impact.

By using both the magnetic and the mechanical movement of the table, square flat distributions were measured, as reported in Fig.~\ref{fig:SAMBA_Line_Square}~(right). Here 21~$\times$~21 spots 
at a nominal distance of 5~mm in the T and in the U directions
form a square of 10~$\times$~10~cm$^2$. To evaluate the precision on the positioning of the 441 spots with respect to a given grid, 
 the minimum
of the residual function
\begin{equation}
R=\sum_{i} (T_i^{fit}-T_i^{grid})^2+\sum_{j} (U_j^{fit}-U_j^{grid})^2
\end{equation}
was searched for, where $T_i^{fit}$ and $U_j^{fit}$ were measured by fitting the data and $T_i^{grid}$ and $U_j^{grid}$ were calculated on the basis of the model chosen for the grid.
To describe the grid, the spot-to-spot distances $\Delta T$ and $\Delta U$ and the tilt angle due to the  90$^\circ$ magnet were left as free parameters. By means of this procedure, the best grid was
found to have 4.94~mm and 4.95~mm for $\Delta T$ and $\Delta U$, respectively, very close to the nominal values of 5~mm. 
The two-dimensional mean distance between the measured positions and the corresponding points on the grid
was determined to be 90~$\mu$m. 
It has to be noted that these values are remarkably precise if compared with the precision on the alignment of the patient which is of the order of one millimeter.

\begin{figure}[t!]
 \centering
 \hspace*{-0.6 cm}
   \includegraphics[height=7 cm]{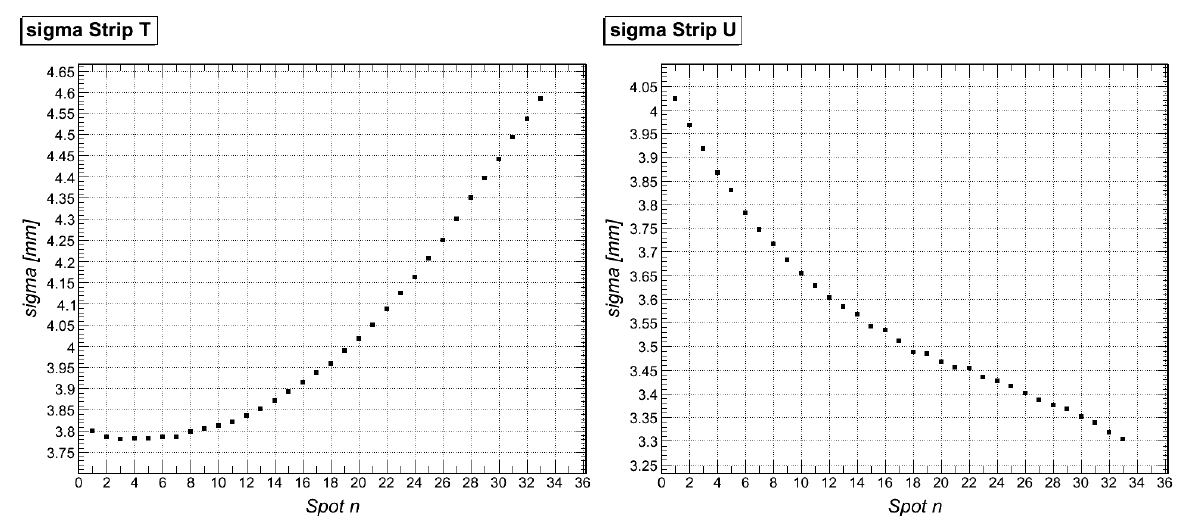}
     \caption{
     Measurement of the standard deviation of all the spots in the T (left) and U (right) directions for the flat linear dose distribution of Fig.~\ref{fig:SAMBA_Doseline}~(bottom).
         }
       \label{fig:SAMBA_Sigma}
\end{figure}

The SAMBA detector allows to precisely monitor the shape of the spots on-line. As shown in Fig.~\ref{fig:SAMBA_Sigma}, the standard deviation of the spots was measured for the 
flat linear dose distribution of Fig.~\ref{fig:SAMBA_Doseline}~(bottom). These measurements show a clear increase of the width in the T direction of about 20\% from the first to the last spot
while a decrease of the same amount was observed in the U direction. These results clearly put in evidence the rotation of the ellipse representing the two-dimensional projection of the beam spot
in the U-T plane 
due to dynamic effects induced in the beam when passing through the magnetic field of the 90$^\circ$ bending magnet.

\begin{figure}[p]
 \centering
   \includegraphics[height =8.5 cm]{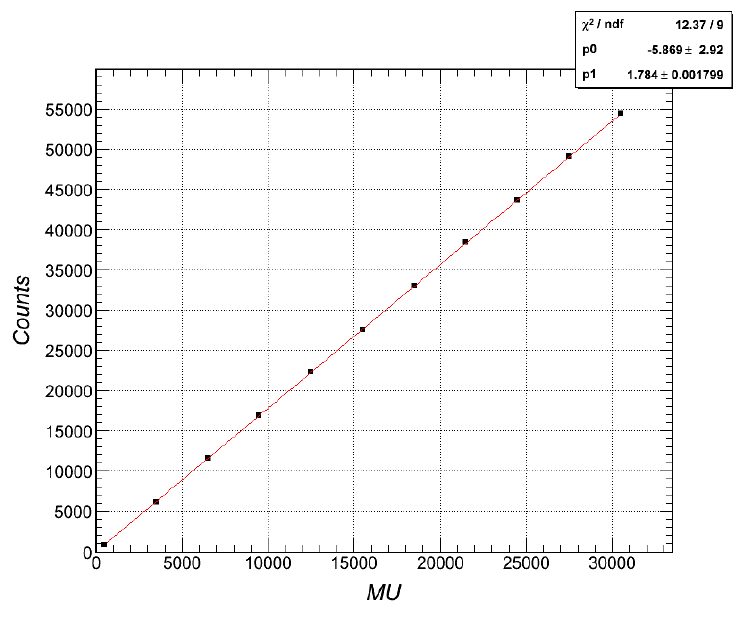}\\
      \includegraphics[height =8 cm]{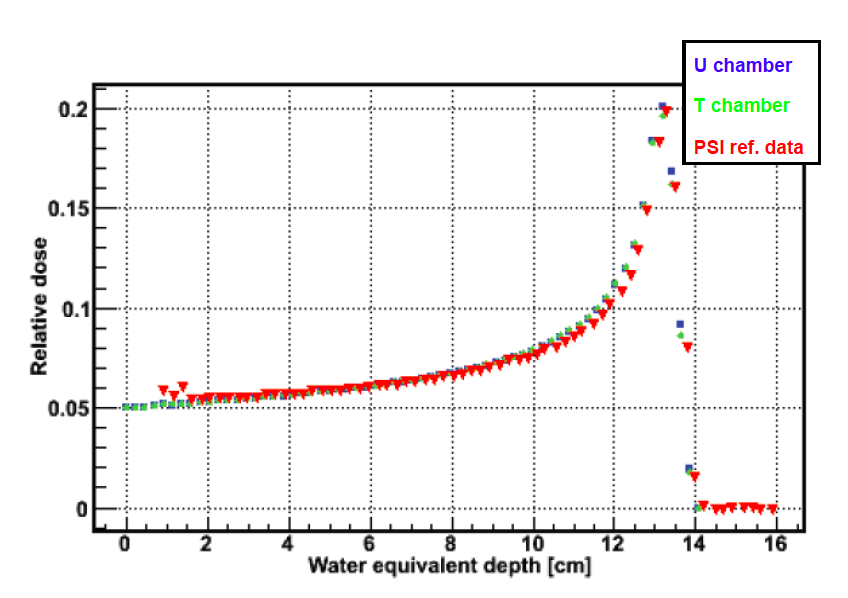}
     \caption{
     The total counts measured by SAMBA show a good linearity with respect to the Monitoring Units (MU) used to measure the delivered dose at PSI (top).
     The Bragg depth dose distributions measured by SAMBA is compared with the reference data used for the Quality Control (QC) at PSI (bottom).
         }
       \label{fig:SAMBA_Linearity_Bragg}
\end{figure}

To evaluate the linearity of SAMBA in the measurement of the delivered dose, a comparison was performed with respect to the standard system used to assess the dose
in the Gantry~1.
With this apparatus, the dose is measured in terms of Monitoring Units (MU)
by means of a planar ionization chamber located upstream. A MU corresponds to about 7000 protons.
A series of spots of increasing temporal length, and therefore of increasing dose, was measured. The total counts of the two chambers were compared with the dose given in terms of MUs.
The results for one of the chambers are reported in Fig.~\ref{fig:SAMBA_Linearity_Bragg}~(top). A very good linearity was found, demonstrating the capability of SAMBA to measure the delivered dose.

Following this result, the Bragg peak distribution of a 138~MeV pencil beam was measured by interposing the water equivalent PMMA layers of the range shifter of the Gantry~1.
Each layer corresponds to 2~mm of water. The data for the two chambers are reported in Fig.~\ref{fig:SAMBA_Linearity_Bragg}~(bottom). 
For the range in water, the value of 133~mm was measured in good agreement with the simulation and with measurements performed with other methods.
The Bragg distribution is compared in Fig.~\ref{fig:SAMBA_Linearity_Bragg}~(bottom) with the reference data daily used at PSI for the Quality Control (QC) procedures.
A good agreement is found and one can conclude that SAMBA represents a powerful instrument for QC procedures in proton therapy.

Using SAMBA,
a much faster QC procedure can be implemented. It has to be noted that the standard procedure implies the use of
a ionization chamber dosimeter operating inside a water phantom that has to be installed and then removed in the treatment room before the beginning of the treatments.
Measurements are taken at different depths by means of a quite time consuming procedure.
On the other hand, SAMBA is able to measure the reference Bragg peak distribution 
 in less than one minute and does not require the installation of any supplementary equipment.

Effects due to charge recombination were also studied by changing the current upstream of the Gantry~1. Spots with total flat top currents of 0.2, 0.4 and 0.6~nA were measured. 
For the same amount of MUs, i.e. for the same amount of protons traversing the detector, the total counts measured by SAMBA were found to decrease 
of 0.3\% and 0.6\% for 0.4 and 0.6~nA, respectively, when compared to the standard intensity of 0.2~nA.
From these measurements one can conclude that the effects due to charge recombination are negligible, once the flat top current is set and only
the modest usual fluctuations of the beam intensity occur during proton therapy treatments.

The development and the tests of the MATRIX and SAMBA detectors demonstrate that 
pixel and strip segmented ionization chambers, alone or in combination,
can be fruitfully used for the control of the clinical beams in hadrontherapy facilities, especially in the case of active beam scanning dose distribution systems.

\newpage

\section{Proton Radiography with nuclear Emulsion Films}

The clinical advantage of hadrontherapy relies on the possibility of modulating the range of the impinging ions to
conform the dose distribution to the target volume with a higher precision with respect to X-rays.
This implies that the uncertainty in
the assessment of the ion range inside the patient's body is crucial for an optimum treatment,
either for passive scattering or for active scanning dose distribution systems.

X-ray Computed Tomography (CT) is presently used to provide the necessary
information to the Treatment Planning System (TPS) for the calculation of the dose distributions.
Since the Hounsfield Units (HU) are obtained from X-ray attenuation, a calibration
curve has to be used to determine proton or carbon ion ranges, which
depend on the density of the traversed material.
This procedure can introduce range uncertainties up to 5\% and limits
the advantages produced by
the ballistic properties of charged ions in matter.

To overcome this limitation,
high-energy protons passing through the patient's body can be used 
to obtain medical images. By means of a mono-energetic proton beam, the measurement of the residual range gives
direct information on the average density of the traversed material. This method is more sensitive  to density differences 
with respect to X-ray imaging, especially for soft tissues.
This technique is denominated proton radiography and can be used for a precise assessment of the range inside the patient's body
with uncertainties below 1\%.
Furthermore, the dose
given to the patient can be significantly reduced with respect to ordinary X-ray radiography, since every single
proton passing through the patient's body brings valuable information.

Proton radiography was proposed in 1968~\cite{Koehler} as an imaging technique. Only a few years later
images of very high contrast were obtained and density differences of 0.5\% were reconstructed with good accuracy.
Images taken from several directions can be combined, leading to a full three-dimensional
reconstruction of the density by means of Proton Computed Tomography (PCT), as first proposed in 1976~\cite{Cormack}.
Despite the good contrast,
the spatial resolution is severely limited by multiple Coulomb scattering and energy straggling.
Due to the limited spatial resolution and since large accelerators are needed to produce high-energy beams, proton radiography and PCT
were almost abandoned after the first pioneering studies in favour of other imaging techniques, as CT or MRI.

The recent development of hadrontherapy determined a renewed interest on proton radiography
and PCT. Several techniques are nowadays the focus of intense research programs
aiming at developing reliable and compact systems to be used in a clinical environment.
As already remarked~\cite{Schippers}, some proton therapy centers are considering to enhance the energy of their beams above the typical 230~MeV by means of a linac booster to allow 
performing proton radiography.

In this framework, I am one of the proponents of a novel technique for proton radiography based on nuclear emulsion films~\cite{Braccini_Proton_Radiography}.
This development relies on 
the long standing expertise of LHEP in Bern on this kind of detectors, mostly for neutrino physics.
This method is different with respect to the standard approaches based on electronic detectors, silicon trackers followed by calorimeters in particular.
Detailed information on recent developments in proton radiography can be found in Ref.~\cite{Braccini_Proton_Radiography} and only the method proposed by our group is described here.

\begin{figure}[t!]
 \centering
   \includegraphics[height=4 cm]{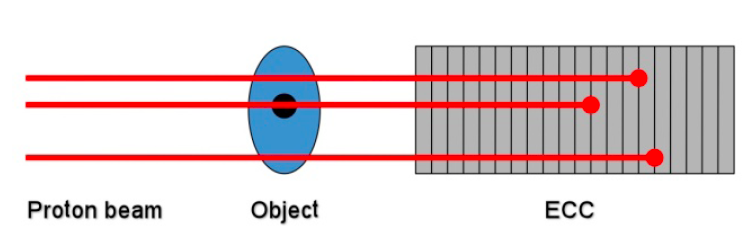}
     \caption{
Principle of proton radiography with nuclear emulsion film detectors. The range of the individual
protons is measured by the number of crossed emulsions in the detector. Higher traversed material densities
correspond to shorter ranges.              
}
       \label{fig:PRAD_scheme}
\end{figure}

The basic principle of proton radiography with nuclear emulsion films
relies on the irradiation of an Emulsion Cloud Chamber (ECC) detector
made of two-sided nuclear emulsions interleaved with tissue equivalent plastic plates.  The whole detector is
located downstream of the object to be imaged, as shown in Fig~\ref{fig:PRAD_scheme}.
This is an end-of-range technique aimed at determining
the position of the Bragg peak inside the ECC. 
The emulsion detector has to be thick enough to contain the end-of-range points of all the
protons. The difference in density encountered by protons translates into a difference on the detected
range. 

The proposed technique is possible due to the recent development of industrially
produced emulsion films and of fast automated scanning analysis systems for modern neutrino physics experiments,
the OPERA neutrino oscillation detector in particular.

\begin{figure}[t!]
 \centering
   \includegraphics[height=6 cm]{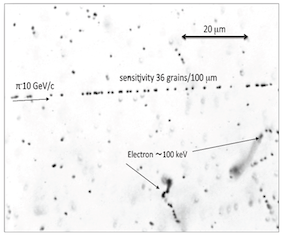}
     \includegraphics[height=6 cm]{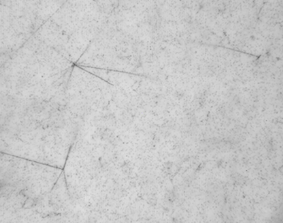} 
     \caption{
A track produced by a 10 GeV/c pion in an emulsion shows the typical characteristics of a track originated by a
minimum ionizing particle (MIP) (left).
Antiprotons annihilate in nuclear emulsion films and produce the characteristic stars
formed by several tacks mostly due to nuclear fragments. Nuclear fragments and protons are highly ionizing particles and can be easily distinguished from MIPs (right).
   }
       \label{fig:PRAD_Emulsion}
\end{figure}

Nuclear emulsions have a very long track record as particle detectors and still represent an
important tool for fundamental discoveries~\cite{Ereditato_Emulsions}. They consist of a gel with silver bromide (AgBr)
crystals where a latent track is formed after the passage of an ionising particle. Tracks can be
visualised and analysed by optical microscopes after a specific chemical development process performed
under controlled conditions.
The chemical development allows the formation of a sequence of silver
grains along the particle track that, with dimensions of 1~$\mu$m or less, are well visible by means of
optical microscopes, as shown in Fig.~\ref{fig:PRAD_Emulsion}. 
Due to their micrometric spatial resolution, nuclear emulsions still represent the most precise
particle tracking device. Even within a single 50~$\mu$ m emulsion layer, they allow full three-dimensional reconstruction
of particle tracks and the measurement of their energy loss. This is performed by evaluating
the silver grain density along the particle track. In this way, highly ionizing protons -~as the ones used in proton radiography~- can be
separated from minimum ionizing particles (MIP). As an example, high ionizing tracks produced by the annihilation of antiprotons in
the same emulsion films used for this work and analyzed at LHEP are shown in Fig.~\ref{fig:PRAD_Emulsion}~(right). 
Once exposed, developed and automatically scanned, the emulsions allow obtaining
digital data on which analysis can be performed by using standard computing techniques.

\begin{figure}[p!]
 \centering
   \includegraphics[height=6 cm]{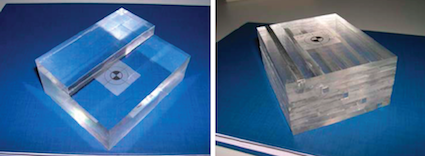}
     \caption{
     The {\it step} (left) and the {\it rod} (right) phantoms used to obtain the first proton radiography images using a nuclear emulsion film detector.
                   }
       \label{fig:PRAD_Phantoms}
   \vspace*{10mm}
   \includegraphics[height=10 cm]{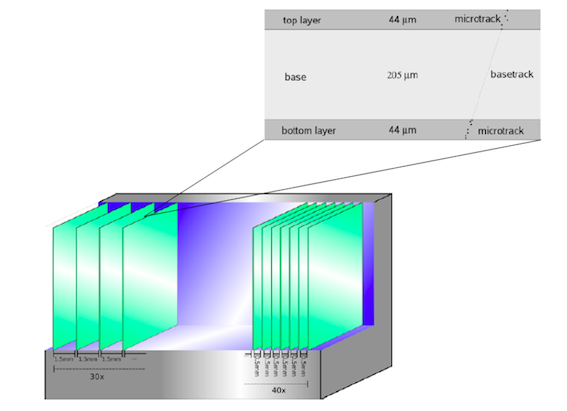}
     \caption{
     Detector structure: Emulsion film plates are interleaved with blocks of absorber (not shown) of
different thickness. The structure of one emulsion film is highlighted in the inset together with the two
micro-tracks produced in the two sensitive emulsion layers.              }
       \label{fig:PRAD_Brick}
\end{figure}

To test the proposed method in a geometry allowing to keep the main parameters under control,
two phantoms were constructed. As shown in Fig.~\ref{fig:PRAD_Phantoms},
the first phantom was made of polymethylmethacrylate
(PMMA) and presented a simple 1~cm step ({\it step} phantom). In this way, protons can traverse only
3 or 4~cm of PMMA. The second phantom was characterized by a total thickness of 4.5~cm and
contained five 5~$\times$~ 5~mm$^2$ aluminium rods positioned at different depths in a PMMA structure ({\it rod}
phantom).

The emulsion films used for this study consisted of 2 layers of 44~$\mu$m thick sensitive emulsion
coated on both sides of a 205~$\mu$m thick triacetate base with a surface of 10.2~$\times$~12.5~ cm$^2$. As
tissue equivalent absorber material, polystyrene plates with the same transverse dimensions as the
emulsions and a thickness of 0.54~mm were used. The density of polystyrene was measured to be 1.05~$\pm$~0.02 g/cm$^3$.

Simulations for several possible configurations of the ECC were performed using GEANT3.
A proton beam with energy of 138~MeV and with a flat
distribution over the 10.2~$\times$~12.5~cm$^2$ sensitive surface was considered. 
Particular attention was paid to optimize the resolution on the determination of the 
proton range and to minimize the number of emulsions to be employed.
The results of the simulations led to the design of the ECC detector shown in Fig.~\ref{fig:PRAD_Brick}.

This apparatus is divided into two parts. The first one is made
of 30 emulsion films, each interleaved with 3 polystyrene absorber plates, allowing tracking of
passing through protons using a minimal number of films. The second part is composed of 40
films, each interleaved with one polystyrene plate, allowing a precise measurement of the proton
range in correspondence of the Bragg peak. The overall dimensions of the detector are 10.2~$\times$~12.5~cm$^2$ 
in the transverse direction, and 90.3~mm along the beam. 

\begin{figure}[t!]
 \centering
   \includegraphics[height=8.5 cm]{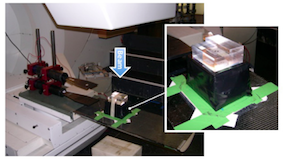}
     \caption{
     First beam tests of proton radiography with nuclear emulsion film detectors at the Gantry 1 at PSI. The detector
with the {\it step} phantom on the top is highlighted in the inset. The scintillator telescope used to verify the
number of incoming protons is visible on the left.
                   }
       \label{fig:PRAD_Gantry1}
\end{figure}

Beam tests were performed with the Gantry 1 at PSI, as shown in Fig.~\ref{fig:PRAD_Gantry1}.
It has to be remarked that a typical clinical beam is not suitable for proton radiography with this method since 
the too high intensity will not allow the reconstruction of the proton tracks. It was calculated that the
number of protons had to be limited to 10$^6$ on a 10~$\times$~10~cm$^2$ surface, corresponding to 100 protons/mm$^2$.
In this way, the emulsions will not be saturated with a good safety margin.
This corresponds to an average dose of the order of 10$^{-5}$~Gy. For this purpose, a special
low intensity beam setting was used and an accurate measurement of the number of protons
per unit area performed with a coincidence scintillating telescope before the irradiation of the ECC.
The two phantom-detector systems were exposed to 138~MeV protons uniformly spread
on the surface by means of spot scanning. The gantry was adjusted to deliver the beam perpendicular to the emulsions.

The emulsion analysis was performed by means of the automatic scanning systems installed at LHEP.
Here a specifically conceived laboratory equipped with six European Scanning System (ESS) microscopes is active for routine scanning of the emulsions of the OPERA experiment.
Scanning speeds
of 10~cm$^2$ per hour can be reached
with prospects for a factor 10 improvement in the next years~\cite{Ereditato_Emulsions}. The scanning
speed represents a crucial point in view of possible routine applications of this technology in a clinical environment.

Track reconstruction is performed through successive steps. Micro-tracks
are first reconstructed as 3D segments located in each sensitive layer, as shown in Fig.~\ref{fig:PRAD_Brick}. At a second stage, base-tracks are searched for as
the lines connecting the two points of the micro-tracks closest to the plastic base. This method allows to minimize possible distortions.
At the energies used for this measurement, protons are characterized by an ionization density much higher than MIPs
and proton tacks can be efficiently selected.  The background due to other particles is negligible. 
The proton density has been measured to be about 70 protons/mm$^2$, in good agreement with
the result obtained with the scintillating telescope before the irradiation.

\begin{figure}[t!]
 \centering
   \includegraphics[height=6 cm]{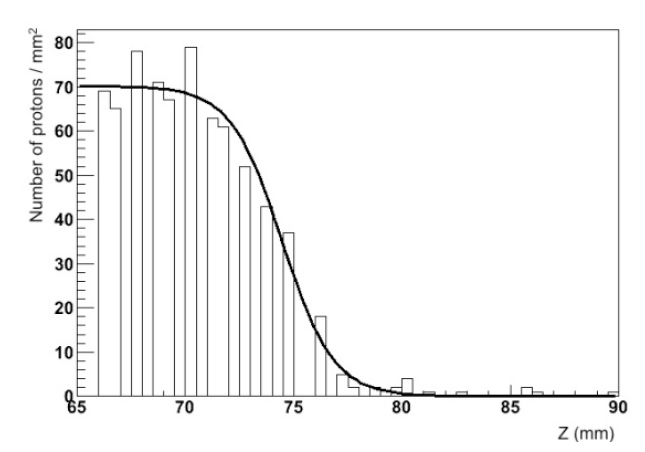}
     \caption{
     The number of identified protons as a function of the $z$ coordinate inside a column of 1~mm$^2$. The
fit allows the localization of the Bragg peak.
              }
       \label{fig:PRAD_Range}
\end{figure}

To obtain proton radiography images of the two phantoms, data were analysed by subdividing the detectors in
columns along the beam direction. The transverse area of each column was 1~mm$^2$, which represents
the size of the pixels of the proton radiography images. Each pixel must contain enough
proton tracks to allow the determination of the range with enough precision. About 70 proton tracks were contained in each pixel, as shown in Fig.~\ref{fig:PRAD_Range}.
The variation of the proton density as a function of the depth has been evaluated for each
column in order to determine the position of the Bragg peak. To calculate the residual range, data were fitted for each column with the
Fermi-Dirac function
\begin{equation}
f(z)=A-\frac{A}{e^{\frac{R-z}{S}}-1}
\end{equation}
where $z$ is the coordinate along the beam, $A$, $S$ free parameters and $R$ represents
the residual range of the protons. One of these fits is reported in Fig.~\ref{fig:PRAD_Range}.

\begin{figure}[t!]
 \centering
   \includegraphics[height=6.5 cm]{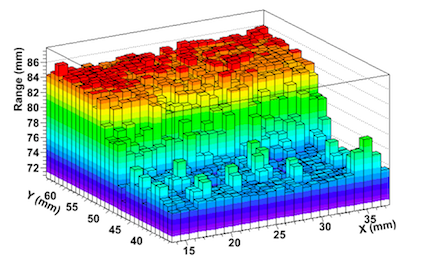}\\
   \includegraphics[height=6.5 cm]{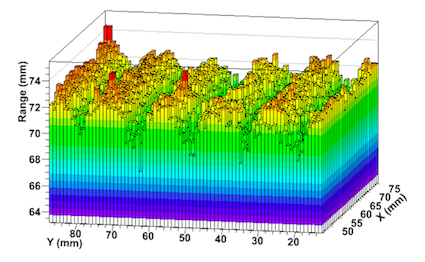}
     \caption{
    The measured proton range as a
function of the transverse coordinates shows a
clear proton
radiography image of the {\it step} (top) and the  {\it rod} (bottom) phantoms. The size of each pixel is 1~mm$^2$.              }
       \label{fig:PRAD_Step}
\end{figure}

Proton radiography images of the phantoms
were obtained by plotting $R$ as a function of the transverse coordinates, as presented in Fig.~\ref{fig:PRAD_Step}.
The shape of the phantoms was well reproduced, demonstrating the validity of the proposed method.
The residual ranges
corresponding to the different traversed densities inside the two phantoms were measured and good agreement was found with the simulation
performed with SRIM. The results are reported in Table~\ref{tab:range_comparison}.
Errors were evaluated by the fitting procedure. 

\begin{table}[t!]
\centering
\begin{tabular}{|l|c|c|c|c|}\hline
Phantom & \multicolumn{2}{|c|}{Traversed material (mm)} &  \multicolumn{2}{|c|}{Range in the detector (mm)} \\ \hline
 & PMMA  & Aluminium  & Monte Carlo SRIM  & Data \\ \hline
Step  & 40 &  -   & 76  & 75.6~$\pm$~0.6 \\
Step  & 30 &  -   & 86  & 84.8~$\pm$~0.8 \\
Rod  & 45  &  -   & 72  & 71.9~$\pm$~0.6 \\
Rod  & 40  &  5  & 68  & 69.6~$\pm$~1.2 \\
\hline
\end{tabular}
	\caption{Measured and simulated values for the proton residual ranges.
	}
	\label{tab:range_comparison}
\end{table}

As already pointed out, the resolution is critical in proton radiography. For the proposed method, it
depends on two main factors:
the size of the pixels in the transverse plane and the effects due to multiple scattering. Since
each pixel must contain enough protons to allow a precise determination of the residual range (of
the order of 50 particles per pixel) and since the number of protons per unit area is limited by the
saturation of the nuclear emulsion films, the size of the pixel was naturally chosen to be 1~mm$^2$.

It has to be remarked that the scanning and reconstruction methods used for this work were those optimized for neutrino physics, where
only minimum ionizing particles with very limited lateral scattering are searched for. By optimizing
the full process to low energy protons used for proton radiography, the use of
at least a factor 5 more protons is estimated to be possible. This will allow for smaller pixels, down to about
0.25~$\times$~0.25~mm$^2$. This value has to be confronted with the intrinsic limitation due to the multiple scattering.
Computer simulations performed with SRIM show that the radial effects in the transverse
plane due to lateral scattering are of 0.5, 1.7 and 3.6~mm for 50, 100 and 150~MeV protons, respectively. The
corresponding residual ranges in the detector are 21.9, 76.1 and 155.0~mm, respectively, typical
of proton radiography. With pixels of 0.25~$\times$~0.25~mm$^2$ the effects of multiple scattering will be dominant 
and smaller pixels will not allow any improvement of the resolution.

The proposed technique has a good potential for proton radiography and clinical beam analysis
since modern emulsion detectors are rather cheap and easy to handle. These characteristics are
surely valuable in a clinical environment where daily routine treatments have absolute priority. 
A clear limitation of emulsion based detectors
is the absence of a real-time response. Although the speed of modern scanning systems is steadily
increasing with good prospects for the future,
it is difficult to imagine that the speed of this technique could ever compete with expensive and sophisticated on-line proton 
imaging electronic devices.

A useful application of nuclear emulsion film detectors
is the characterization of proton beams during the commissioning phase
of proton therapy clinical facilities, when the purchase and operation of expensive and sophisticated
detectors could not be justified for only this purpose. Along this line, the assessment of the peripheral dose in the beam halo region of a proton pencil beam has been
performed by our group using the Gantry~1 at PSI~\cite{Beam_Halo}.

The experience I acquired in nuclear emulsion film detectors by means of this work, allowed me to contribute to other potential applications of this technique, 
either in medical applications~\cite{Emulsions_ICTR-PHE_2012}~\cite{Emulsions_Biodola} or in fundamental physics~\cite{Emulsions_AEGIS}.
In particular, the possibility of measuring both fast and slow neutrons by detecting proton recoils in tissue equivalent absorbers
interleaved with nuclear emulsion films is now under study by our group. The first beam tests were performed with fast neutrons at the CERF facility at CERN~\cite{Neutron_Emulsions}.
Furthermore, I was one of the promoters of an international project~\cite{INET} involving our laboratory and Russian partners centered on the development of innovative nuclear emulsion film
technologies.

\newpage

\section{A Beam Monitor Detector based on doped Silica Fibres}

Beam monitoring is essential in particle accelerator physics. The development, the commissioning, the upgrade and 
a reliable daily use of particle accelerators rely on the knowledge of the beam properties.
Diagnostics devices are usually installed in critical locations, as the injection or the extraction, and in front of the target.
This issue is particularly critical in medical applications,
where ion beams are used to irradiate tumours with millimetre precision or to produce large
quantities of radioisotopes for diagnostics and therapy. 
It has to be remarked that most of the commonly used detectors - such as Faraday cups - are
destructive and their use is not possible without interrupting the irradiation.

The wide spread of medical accelerators is giving impulse to research in the field of
beam monitoring instrumentation to fulfill the requirements of precision,
compactness, robustness and reliability.
Along this line, I dedicate part of my scientific activity to the development of innovative beam monitoring devices to be installed along the
beam lines either for hadrontherapy or for radioisotope production accelerators. In this field,
I contributed to the development and beam test of a non-interceptive beam monitor apparatus~\cite{SLIM} based on the detection of secondary electrons emitted by a thin foil traversed by the beam.
The foil had a diameter of 70~mm and
consisted of
0.1-0.3~$\mu$m of Al$_2$O$_3$ coated on each side with 0.01-0.02~$\mu$m of Al. 
The interaction of the accelerated ions with a so thin foil induces
a negligible perturbation to the beam emittance and produces
$\delta$-rays and electrons below 50~eV,
conventionally called secondary electrons (SE). The SEs are ejected from the foil proportionally to the local beam intensity, with yields calculated to be
about 5\% and 1\% for 20~MeV and 250~MeV protons, respectively. The yield is 
defined as the ratio between the SEs
emitted in backward direction and the number of impinging protons.
The SEs were focalized and accelerated to the energy of 20~keV by means of a specific electrostatic lens.
To measure
the two dimensional profile and the intensity of the beam, a specific sensor was located in the focal plane of the lens to detect the 20~keV electrons. 
A Micro Channel Plate (MCP) coupled with a phosphor (P47) and read-out by a CCD 
was employed. Analog data were digitized by means of a frame grabber. 
A back-thinned pixel silicon sensor was also tested. In the framework of a collaboration with the TERA Foundation, a
further development of this detector is presently ongoing in Bern using the beam line of the new cyclotron.

Although precise and reliable, detectors based on the detection of secondary electrons are relatively large, expensive and require high voltage, sophisticated sensors and
read-out systems. 
Furthermore, in the case of radioisotope production, radiation damage due to neutrons represents a delicate issue for the electronics.
On the basis of this experience, I aimed at developing a much simpler device able to provide information on the beam profile and 
requiring a minimal space on a beam line. These ideas led to the development of the beam monitoring device based on 
doped silica and optical fibres~\cite{Braccini_CAARI}~\cite{Braccini_UniBEam}~\cite{UniBEam_ICTR-PHE_2012}, which is described here in more detail.

\begin{figure}[t!]
 \centering
   \includegraphics[height=10 cm]{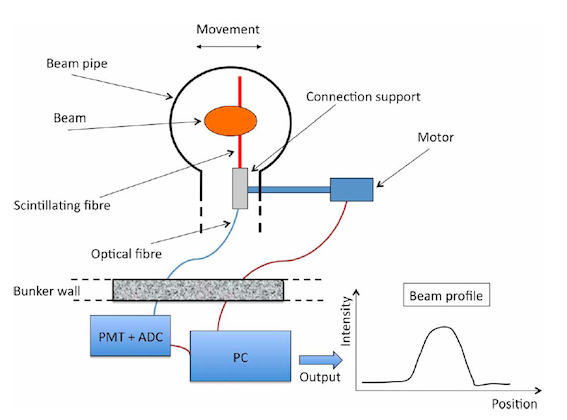}
     \caption{
     Principle and general scheme of a beam monitor detector based on a single moving scintillating
doped silica fibre coupled to an optical fibre.
               }
       \label{fig:UniBEam_Scheme}
\end{figure}

The working principle of this device is depicted in Fig.~\ref{fig:UniBEam_Scheme}. The light produced 
by a single scintillating sensing fibre moved transversally through the beam is transmitted by means of a commercial optical fibre
to a photodetector located several meters away. The position of the fibre is measured and its correlation with 
with the signal given by the photodetector is used to obtain the beam profile. Doped silica fibres are suitable
due to their radiation hardness and resistance to the heat produced by the interaction with the beam.
Since they are very fragile,
difficult to handle and have a considerable attenuation length (of the order of several dB/m) their length has to be limited to
about 10 cm in length. To transport the optical signal produced by the
scintillation induced by the impinging beam particles, the doped fibre is optically connected to a commercial
optical fibre. 
In this way, the signal can be transported to the photodetector over several
meters distance with negligible attenuation (of the order of tens dB/km). 
The photodetector and the related electronics can be located far from the radiation source, thus eliminating any problem related to neutron irradiation.
The connection between the scintillating and the optical fibre is realized inside the vacuum pipe and
a remotely controlled linear translation stage coupled to a vacuum-tight feedthrough or a bellow are used to move the
sensing fibre across the beam.
This kind of detector has the advantage
of being very thin, thus requiring a minimal space along a beam line. Furthermore, a sensor based on fast scintillation is easily adaptable
either to pulsed or continuous beams. For the realization of this detector,
the selection of the scintillating fibre and the connection between the sensing and the optical fibre represent  critical issues.

\begin{table}[t!]
\centering
\begin{tabular}{|l|c|c|c|c|c|c|c|}\hline
Dopant  & Composition  & Core  & Cladding  & Numerical  & Length & Light \\
 &  & diameter  & diameter  & Aperture & & yield \\
 &  & ($\mu$m)  &  ($\mu$m) & (NA)  &  (mm) & (\%)\\ \hline
 & SiO$_2$ (55 \%), &   &  &  & &\\
   &  MgO (24 \%), &  &   &   &   & \\
 Ce$^{3+}$ &   Al$_2$O$_3$ (11 \%), &   Unknown & 520$\pm$20  & 0.25$\pm$0.04    &   120 & 0.6\\
  &   LiO$_2$ (6 \%), &   &   &   &   & \\
  &   Ce$_2$O$_3$ (4 \%)&   &   &   &   &\\ \hline
Sb$^{3+}$  &  (0.5\%) Sb$^{3+}$:SiO$_2$ &  461$\pm$16 & 570$\pm$20  & 0.05$\pm$0.01   & 120  &  0.4 \\ \hline
\end{tabular}
	\caption{Main characteristics of the scintillating doped silica fibres used for the proposed beam monitoring detector. The light yield is expressed in terms of the fraction of the deposited energy.
		}
	\label{tab:fibres}
\end{table}

For beam monitoring, plastic fibres have been proposed for hadrontherapy and
are appropriate only for low currents due to their very low
resistance to heat. 
For high currents of the order of a few $\mu$A, Ce$^{3+}$ doped quartz surfaces were tested up to temperatures of
1600~K. For this reason,
Ce$^{3+}$ doped silica fibres represent a good solution. Furthermore, their 
scintillation light is peaked around 395~nm, a feature assuring a good coupling with
photomultiplier tubes (PMT).

\begin{figure}[t!]
 \centering
   \includegraphics[height=10 cm]{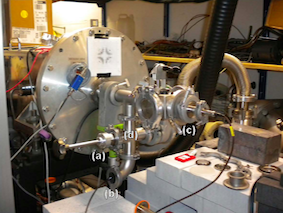}
     \caption{
     The first prototype detector installed on the RFQ accelerator at LHEP. The linear motion (a) and
the optical fibre feedthrough (b) are visible together with the Faraday cup (c) used for measuring the total
intensity of each spill. The V-groove connection supporting the sensing fibre is contained inside the cross
joint (d).
               }
       \label{fig:UniBEam_RFQ}
\end{figure}

A sample of a few Ce$^{3+}$ doped silica fibres from
an experiment performed at LEAR at CERN about twenty years ago was used for the first studies. Since this kind of fibres
is no more commercially available, an alternative solution had to be found. For this purpose, a collaboration with the Institute of Applied Physics (IAP) of
the University of Bern was established in order to explore the possibility to employ doped silica fibres produced with a method based on a silica preform filled with granulated oxides.
Several doping elements were tested and 
Sb$^{3+}$ doped fibres were found to have properties similar to Ce$^{3+}$.

While the scintillation properties of Ce$^{3+}$ doped silica are known, the
response to ionizing radiation of Sb$^{3+}$ doped fibres was never studied before.
To assess the light yield, tests have been performed with a collimated 370~MBq $^{90}$Sr beta
source located about 1~cm away from the sensing fibre. Being the average energy of the emitted
electrons 1.13~MeV, corresponding to a range in glass of 2.2~mm, and the diameter of the fibre about 600~$\mu$m, most of the electrons cross the 
fibre completely. The maximum deposited energy is estimated to be 180~keV. The main characteristics and the measured light yield  
for Ce$^{3+}$ and Sb$^{3+}$ doped fibres are
reported in Table~\ref{tab:fibres}. It is important to remark that this kind of fibres allow detecting low intensity beams down to single particle counting.

A first prototype detector was constructed and tested with the pulsed H$^-$ beam produced by the 2~MeV RFQ accelerator installed at LHEP,
as shown in Fig.~\ref{fig:UniBEam_RFQ}. 
Aiming at realizing a compact, cost
effective and easy to use system, commercial components have been used whenever possible.
The main body of the detector is an ISO KF-40 cross joint on which a vacuum tight linear
motion feedthrough is mounted. 
For this first prototype, the movement was realized manually since radiation protection issues do not represent a problem for 2~MeV H$^-$ beams.
A 5~cm long aluminium support is screwed to the
linear motion feedthrough to hold the sensing fibre and to connect it to the commercial optical fibre.
Inside this support, a V-groove is realized to precisely put in contact the two fibres. 
Contrary to fusion splicing, this solution allows for an easy replacement of the
sensing fibre. Light losses in the range of 20\%~-~50\% were
measured to be constant during the operation of the device.
For read-out and digitization, a photomultiplier and a peak sensing ADC
were used. During the beam tests, a Faraday cup located downstream the
sensing fibre was used to measure the beam current. 

\begin{figure}[t!]
 \centering
   \includegraphics[height=7 cm]{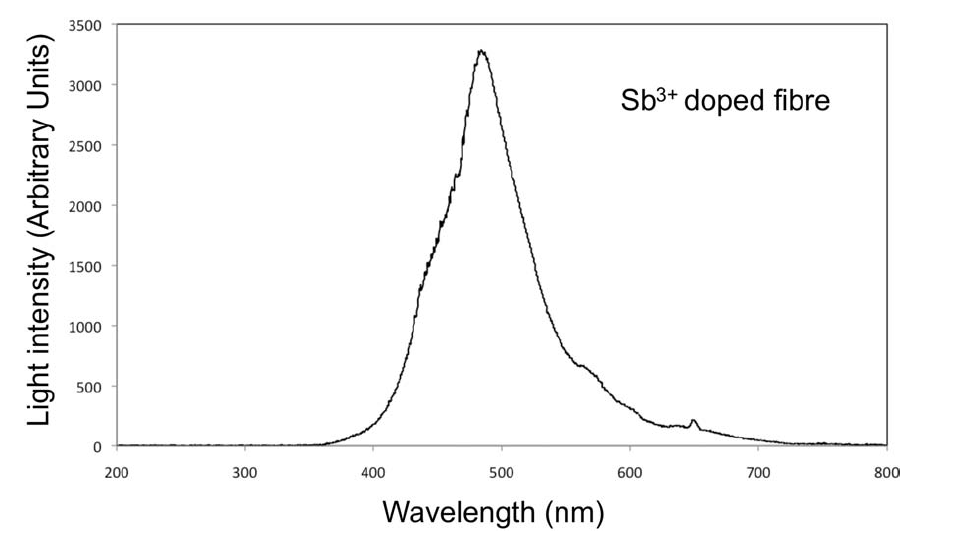}
     \caption{
     Measured spectrum of the wavelengths emitted by a Sb$^{3+}$ doped fibre irradiated with 2~MeV protons.
               }
       \label{fig:UniBEam_Sb_Spectrum}
\end{figure}

The 2 MeV protons loose all
their energy inside the fibre since their range is about 45~$\mu$m in glass, much less with respect to the diameter of the fibre. 
When irradiated by the beam, the
fibre glows, emitting a very clear blue light signal. 
The spectrum of the emitted light is reported in Fig.~\ref{fig:UniBEam_Sb_Spectrum} and shows a 
maximum around 490~nm, allowing for a good quantum
efficiency when coupled to a PMT. 

\begin{figure}[p]
 \centering
 \includegraphics[height=6 cm]{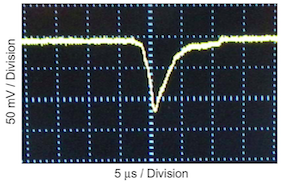}\\
 \caption{
 Typical signal produced by a Sb$^{3+}$
doped fibre during a
proton spill. The nominal duration
of the spill is set to about 3~$\mu$s. The long tail is due to a known feature of the RFQ accelerator.
 }
 \vspace*{0.5cm}
 \label{fig:UniBEam_Signal_Oscilloscope}
   \includegraphics[height=8 cm]{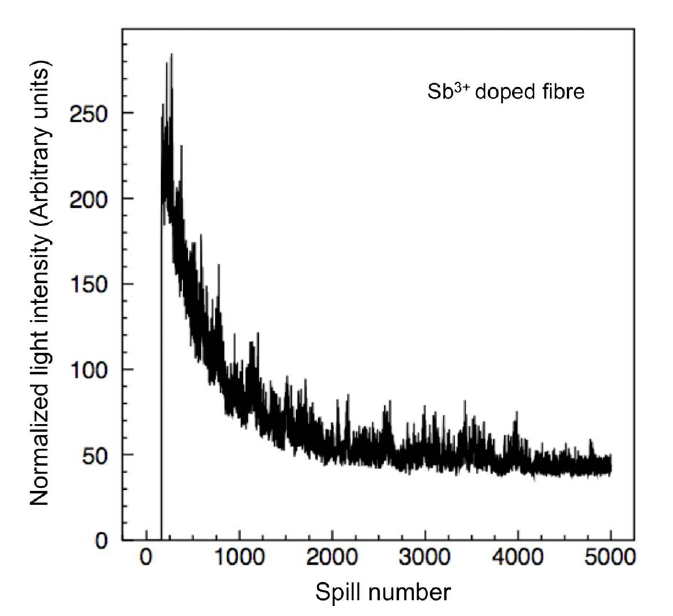}
     \caption{
               Light output for 5000 successive spills of equal intensity. The effect of heating leads to a decrease
of the light output of about a factor five. 
               }
       \label{fig:UniBEam_Signal_Spill}
\end{figure}

For the tests, the beam intensity of the RFQ accelerator was set to about 10$^{11}$ protons per spill, which, at a repetition rate
of 50~Hz, leads to an average total current of about 0.8~$\mu$A. In these conditions about 10$^{10}$ protons
hit the fibre when centred on the beam. The proton spill produces a very clear signal in the PMT, as shown in Fig.~\ref{fig:UniBEam_Signal_Oscilloscope}.
Taking into account the quantum efficiency of the PMT, the signal
was estimated to be generated 
by about 315000 photons reaching the photocathode through the 20~m optical fibre.

The emission of scintillation light depends on the temperature of the scintillator and this phenomenon may have a significant influence
on the output signal. To study this effect, the signals of 5000 consecutive spills were recorded with a fixed position of the fibre.
At the repetition rate of 50~Hz, this corresponds to 100 seconds. The total charge hitting the fibre
is estimated to be 2.7~$\mu$C with a total corresponding delivered dose on the volume of the fibre of
560~kGy. The increase of temperature due to the energy transferred by the beam is calculated to be 8~K/s for a cold fibre.
The heat is mostly dissipated through black body radiation, being conduction mechanisms highly suppressed.
On this basis, the equilibrium temperature can be calculated. For the experimental conditions of the beam tests, the equilibrium temperature of the fibre
is estimated to be around 600~K.
The measurement performed with 5000 consecutive spills puts in evidence a decrease of the light output of about a factor five when long lasting
irradiations are performed, as shown in Fig.~\ref{fig:UniBEam_Signal_Spill}.
This effect may potentially induce distortions of the beam profile if measurements are taken with different temperatures of the fibre.
However, this effect can be significantly reduced
using a motorized device which allows sweeping the beam in less than one
second.  From these measurements, the light loss can be estimated to be  4\% in one second starting from
a cold fibre.

\begin{figure}[t!]
 \centering
   \includegraphics[height=7 cm]{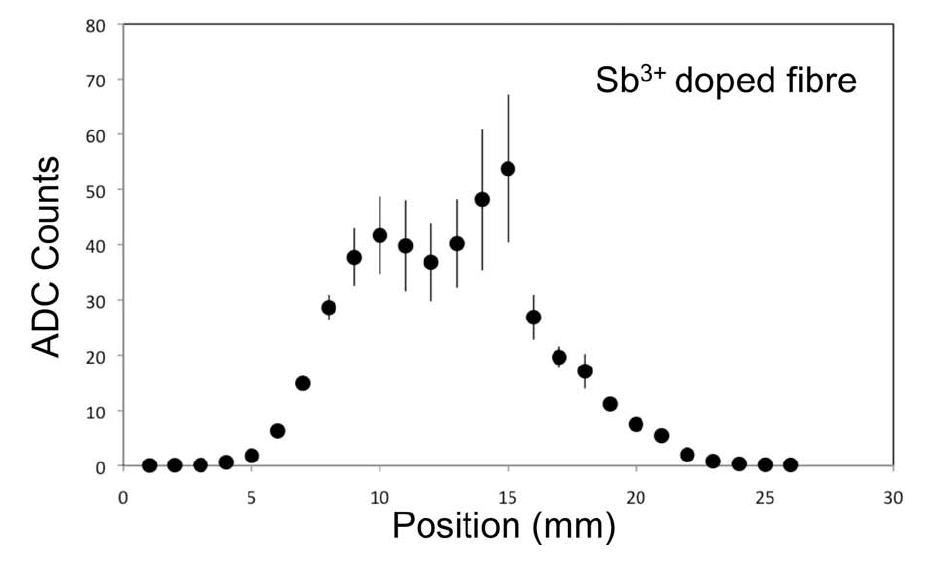}
     \caption{
Beam profile measured with the Sb$^{3+}$ doped silica fibre moved in steps
of 1~mm through the beam. The double structure of the beam is a known effect of the accelerator. The error
bars correspond to the rms of the pulse height distributions and are mainly due to beam intensity fluctuations.
The centre of the beam pipe corresponds to the
position of the fibre at 14 mm.}
       \label{fig:UniBEam_Profile_RFQ}
\end{figure}

Following irradiations lasting more than 100 seconds, it was observed that the light output returns at the initial value if enough time
is allowed for cooling down the fibre to room temperature. Irradiations with up to 50000 spills -
corresponding to 1000 seconds - were performed and no effects due to permanent radiation
damage have been observed. Further investigations on radiation hardness are ongoing using the BTL of the Bern cyclotron.

\begin{figure}[t!]
 \centering
   \includegraphics[height=10 cm]{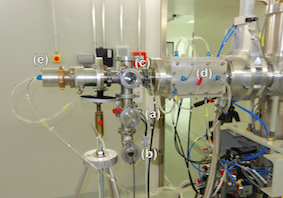}
     \caption{
The second prototype installed in the BTL of the Bern cyclotron. The remotely controlled motor connected to the linear motion feedthrough (a),
the optical fibre feedthrough (b) and the transparent flange (c) used to observe the fibre with a CCD are visible.
The {\it four finger} collimator (d) is used to centre the beam and the beam dump (e)
to measure the total intensity. 
}
       \label{fig:UniBEam_Cyclotron}
\end{figure}

\begin{figure}[th!]
 \centering
   \includegraphics[height=7 cm]{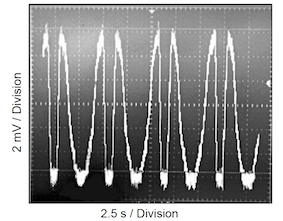}
     \caption{
Signals produced by the 18~MeV proton beam at the intensity of 1~$\mu$A. The Sb$^{3+}$ doped fibre is moved continuously through the beam with a period of 6~seconds
within a spatial range of $\pm$15~mm with respect to the centre of the beam pipe.
The signal produced by the photodiode shows two specular profiles, corresponding to the left-right and right-left passes. The signal is stable and reproducible within the
experimental uncertainties.
}
       \label{fig:UniBEam_Signal_Cyclotron}
\end{figure}

The horizontal beam profile of the 2~MeV pulsed H$^-$ beam from the RFQ accelerator was
measured using Ce$^{3+}$ and Sb$^{3+}$ doped silica fibres by moving them across the beam in steps of 1~mm.
The average peak signal was evaluated for each position by acquiring
1000 consecutive spills after the stabilization of the light
output with respect to the temperature.
The beam profile obtained with  a Sb$^{3+}$ doped fibre is reported in Fig.~\ref{fig:UniBEam_Profile_RFQ}
and shows the characteristic beam structure of the RFQ accelerator. A similar profile was measured with the  
Ce$^{3+}$ doped fibre.
The reproducibility of the measurements was found to be within the experimental
uncertainties, which are dominated by the beam fluctuations. As far as the spatial resolution is concerned, the position of the beam can be measured with
good precision with this first prototype detector.
The
diameters of the sensing fibres used for this device were around 500~$\mu$m (Table~\ref{tab:fibres})
to assure a sufficient rigidity without the use of a specific support. The production of Sb$^{3+}$ doped fibres with a smaller
diameter does not represent a problem and their use could allow for a better precision and produce a smaller
perturbation to the beam.

Following the successful construction and test of the first prototype, the development of a horizontal-vertical two-dimensional beam profiler
for the BTL of the Bern cyclotron is ongoing. This device is equipped with a remotely controlled motorized movement and is optimized to scan
the practically continuous beam extracted from the cyclotron. A second mono-dimensional prototype has been constructed and installed in the BTL,
as shown in Fig.~\ref{fig:UniBEam_Cyclotron}. The fibre can be monitored by a CCD camera located in the proximity of a glass flange.
Due to the amount of light and the continuous temporal structure, a photodiode was used as photodetector. The first preliminary beam profiles
measured with the 18~MeV proton beam at 1~$\mu$A are reported in Fig.~\ref{fig:UniBEam_Signal_Cyclotron}. The beam profile appears very clear over a negligible background.
The specular structure is due to the left-right and right-left passes of the fibre through the beam. The period was set to 6~seconds and the range of the horizontal movement to $\pm$15~mm with respect to
the centre of the beam pipe. In the profile reported in Fig.~\ref{fig:UniBEam_Signal_Cyclotron}, the beam was deliberately cut-out on one side by acting on the steering magnet of the BTL in
order to intercept part of it with the {\it four finger} collimator. The sharp decrease of the signal shows a resolution below 1~mm.
With this second prototype, beam profiles have been measured in the intensity range from 500~nA to 5~$\mu$A. 

A research program is under way based on this new kind of beam monitor detector. It is aimed
at determining the maximum operational intensity and radiation hardens properties of the fibres.
A complete control system including the motor and an ADC based data acquisition system is under development.
The miniaturization of the full apparatus represents a long term goal.
In the framework of this development, new scintillating doped fibres are also under test. In particular, Ta$^{3+}$ doped silica fibres have been selected as potential
sensing fibres. Their light yield has been measured to be 0.5\%, slightly higher with respect to Sb$^{3+}$ doped fibres.

\chapter*{Conclusions and Outlook}
\label{conclusions}
\addcontentsline{toc}{chapter}{Conclusions and Outlook} 

Particle accelerators and detectors are fundamental instruments in modern medicine. Their development for cancer hadrontherapy and 
for medical imaging by means of radioisotopes 
is the focus of actual cutting-edge research programs. I am active in this scientific domain since ten years, 
after having contributed to basic research in particle physics experiments at CERN. Some of my most relevant publications
in the field of particle accelerators and detectors applied to medicine have been selected for this Habilitation thesis. They contain original contributions and ideas together with reviews on the status and perspectives
of this discipline. The following list summarizes the scientific contributions, which are discussed and put in their perspective in Chapters~\ref{ch2} and~\ref{ch3}.

\begin{itemize}
\item Proton and ion linear accelerators ({\it linacs}) constitute a promising solution to face the challenges of hadrontherapy. They
can be coupled to cyclotrons used for other medical applications to form a new kind of hybrid accelerator denominated {\it cyclinac}. I contributed to the study and development
of {\it cyclinacs}, in particular to exploit their unique features for advanced {\it spot} scanning dose distribution.
\item The new Bern cyclotron laboratory represents a multipurpose clinical and research facility for the production of radioisotopes for PET imaging and multi-disciplinary research. 
As one of the promoters of this project, I proposed the construction of a beam transport line terminated in a dedicated bunker, which allows
performing production and research in parallel. This new facility is unique in Switzerland and is now fully operational. The first research programs are ongoing.
\item Segmented ionization chambers are suitable for the control of the position, the shape and the intensity of clinical ion beams. 
I directly contributed to the development of specific pixel and strip detectors and studied their application to {\it spot} scanning dose distribution. In particular, I led the design, construction and test
of a strip ionization chamber for the Gantry~2 at PSI.
\item Proton radiography is instrumental to determine the range of protons in tissues for an optimum treatment planning in proton therapy. I am one of the proponents of
an innovative method based on nuclear emulsion film detectors, which was successfully tested with a clinical beam.
\item The control of the beams along the transfer lines is crucial in medical accelerators. For this purpose, I proposed and contributed to develop a
beam monitor detector based on doped silica and optical fibres. In this framework, new scintillating materials have been studied. 
\end{itemize}

Under the auspices and support of the University of Bern,
the Laboratory for High Energy Physics (LHEP) 
started in 2008 a dedicated research program on medical applications of particle physics
under my guidance and coordination. During this time,
I have been involved in several research projects and collaborations.
The main focus of this activity has been the SWAN project that 
aims at the realization of a clinical and research centre of excellence for the production of radioisotopes and cancer proton therapy at the Bern University Hospital (Inselspital).
Along this line,
efforts were concentrated on the new cyclotron laboratory and, in particular, on the beam transport line dedicated to scientific research.
On this basis, research and training activities are growing and 
the Bern group is successfully and visibly assuming a leading role in the field of particle accelerators and detectors applied to medicine.

%

%
\part{Selected Publications}

This part contains the reprints of the selected publications to be considered for the Habilitation. The scientific work as experimental physicist presented here
has been carried out in the framework of the Foundation for Oncological Hadrontherapy (TERA) at CERN of which I was the Technical Director, and of the research group on medical applications of particle physics at the Albert Einstein Centre for Fundamental Physics, Laboratory for High Energy Physics (LHEP) in Bern of which I am the coordinatorr.
For clarity, this part is subdivided into three Chapters. Chapter~4 is dedicated to innovative particle accelerators for hadrontherapy based on cyclotrons and linacs, and to the new Bern cyclotron laboratory
and its research beam line.
Chapter~5 presents the design, construction and test of three innovative particle detectors for medical applications of accelerated ion beams, which were developed under my coordination. Chapter~6 contains the reprint of three review papers in which I presented the status of hadrontherapy and drew the lines of its future scientific and technological developments. 

The list of the selected publications is reported here below, together with a short comment on my contribution. 

\vspace*{0.5 cm}

{\bf Chapter~4 - Particle Accelerators for medical Diagnostics and Therapy}

\begin{itemize}

\item U. Amaldi, S. Braccini and P. Puggioni, {\it Linacs for hadrontherapy}, Reviews of Accelerator Science and Technology (RAST) Vol. 2 (2009) 111.

{\it This paper is focused on the advantages of linacs for cancer hadrontherapy and contains the original scientific work preformed under my coordination on Cyclinacs for advanced beam scanning.}

\item U. Amaldi, S. Braccini et al.,  {\it Accelerators for hadrontherapy: from Lawrence cyclotrons to linacs}, Nucl. Instrum. Meth. A 620 (2010) 563.

{\it The presented work represents an extensive summary of the achievements of the TERA Foundation obtained under my coordination as Technical Director. In particular, it contains original achievements on linacs and cyclinacs for carbon ion therapy. As current practice in particle physics, all the contributors are listed in alphabetical order. I am the corresponding author. }

\item S. Braccini et al.,  {\it The new Bern cyclotron laboratory for radioisotope production and research}, Proceedings of the second
International Particle Accelerator Conference (IPAC2011), San Sebasti\'an, Spain (2011) 3618.

{\it This paper describes the new Bern cyclotron laboratory with its external beam line for research that I proposed in 2007 and was designed and realized under my coordination. I am the corresponding author.}

\item S. Braccini,  {\it The new bern PET cyclotron, its research beam line, and the development of an innovative beam monitor detector}, American Institute of Physics (AIP) Conf. Proc. 1525 (2013) 144.

{\it This paper  followed an invited presentation at the 22nd International Conference on the Application
of Accelerators in Research and Industry (CAARI 2012) where I described the commissioning of
new Bern cyclotron laboratory with emphasis on the optimization of the beam transfer line for research activities. }

\end{itemize}
 
{\bf Chapter~5 - Particle Detectors for medical Applications of Ion Beams}

\begin{itemize}

\item S. Braccini et al.,  {\it MATRIX: an innovative pixel ionization chamber for on-line beam monitoring in hadrontherapy}, 
Astroparticle, Particle, Space Physics, Radiation Interaction, Detectors and Medical Physics Applications,
Vol. 3, World Scientific (2006) 677.

{\it The described achievements were obtained under my coordination within the collaboration involved in the construction and tests of the MATRIX detector.
I am the corresponding author. }

\item S. Braccini et al.,  {\it First results on proton radiography with nuclear emulsion detectors}, Journal of Instrumentation 2010\_JINST\_54\_P09001.

{\it Based on the worldwide recognized expertise of LHEP on nuclear emulsion technologies, the described achievements were obtained under my coordination within the research group on medical applications of particle physics. I am the corresponding author. }

 \item S. Braccini et al.,  {\it A beam monitor detector based on doped silica and optical fibres}, Journal of Instrumentation 2012\_JINST\_7\_T02001.

{\it Following my original idea, these achievements were obtained at LHEP under my coordination within the research group on medical applications of particle physics.
I am the corresponding author. }

\end{itemize}

{\bf Chapter~6 - Highlights on Hadrontherapy}

\begin{itemize}

\item S. Braccini,  {\it Progress in Hadrontherapy}, Nuclear Physics B (Proc. Suppl.) 172 (2007) 8.

{\it This paper followed an invited presentation at the 10th International Conference on Innovative Particle and Radiation Detectors where I was asked to give a review of cancer hadrontherapy, with particular focus on accelerators and detectors.}

\item S. Braccini, {\it Scientific and Technological Development of Hadrontherapy}, Astroparticle, Particle, Space Physics, Radiation Interaction, Detectors and Medical Physics Applications,
Vol. 5, World Scientific (2010) 598.

{\it This paper followed an invited talk I gave in the plenary session at the 11th ICATPP Conference on Astroparticle, Particle, Space Physics, Detectors and Medical Physics Applications. I was asked to critically review the main achievements in the field of accelerators and detectors for cancer hadrontherapy and to outline future scientific and technological developments.}

\item U. Amaldi and S. Braccini,  {\it Present challenges in hadrontherapy techniques}, Eur. Phys. J. Plus (2011) 126: 70.

{\it This invited review paper describes the main challenges of hadrontherapy from a physicist's point of view and outlines the main lines of development of the discipline. I am the corresponding author.}

\end{itemize}

%
%
%
%
%
%
%
%

\begin{thebibliography} {99}

%
%

\bibitem{Roentgen}	W. C. R\"ontgen, \"Uber eine neue Art von Strahlen, Vorl\"aufige Mitteilung. In: Aus den Sitzungsberichten der W\"urzburger Physik.-medic. Gesellschaft W\"urzburg, S 137Ð147, 1895; 
translated into English by A. Stanton: W. C. R\"ontgen, On a New Kind of Rays, Nature 53 (1896) 274-276.

\bibitem{Becquerel}	H. Becquerel, Sur les radiations \'emises par phosphorescence, Comptes rendus hebdomadaires des s\'eances de l'Acad\'emie des sciences, 122 (1896) 420Ð421.

\bibitem{Curie}	P. Curie and M. Sk\l odowska Curie, Sur une substance nouvelle radio-active, contenue dans la pechblende, Comptes rendus hebdomadaires des s\'eances de l'Acad\'emie des sciences, 127 (1898) 175Ð178 (discovery of polonium); \\
P. Curie and M. Sk\l odowska Curie, Sur une nouvelle substance fortement radio-active contenue dans la pechblende, Comptes rendus hebdomadaires des s\'eances de l'Acad\'emie des sciences, 127 (1898) 1215Ð1217 (discovery of radium).

\bibitem{LawrenceLivingston}	E.O. Lawrence and M. S. Livingston, The Production of High Speed Protons without the use of High Voltages, Phys. Rev. 38 (1931) 834; \\
E. O. Lawrence and M. S. Livingston, The Production of High Speed Light Ions without the Use of High Voltages, Phys. Rev. 40 (1932) 19 Ð 35.

\bibitem{LawrenceAebersold}	J. H. Lawrence, P. C. Aebersold and E. O. Lawrence, Comparative Effects of X-Rays and Neutrons on Normal and Tumor Tissue, Proc. Natl. Acad. Sci. U.S.A. 22(9) (1936) 543Ð557.

\bibitem{Fermi}	 E. Fermi, Radioactivity Induced by Neutron Bombardment, Nature 133 (1934) 757; \\
E. Fermi, E. Amaldi, B. Pontecorvo, F. Rasetti and E. Segr\`e, Azione di sostanze idrogenate sulla radioattivit\`a provocata da neutroni, La Ricerca Scientifica 5 (1934) 282-283; \\
E. Amaldi and E. Fermi, On the Absorption and the Diffusion of Slow Neutrons, Physical Review 50 (1936) 899-928. 

\bibitem{Hevesy} O. Chiewitz and G. Hevesy, Radioactive Indicators in the Study of Phosphorus Metabolism in Rats, Nature 136 (1935) 754-755.

\bibitem{Segre} E. Segr\`e and G. T. Seaborg, Nuclear isomerism in element 43, Phys. Rev. 54 (1938) 772.

\bibitem{Tucker} W. Tucker et al., Methods of preparation of some carrier-free radioisotopes involving sorption on alumina, Trans. Am. Nucl. Soc. 1 (1958) 160; \\
P. Richards, A survey of the production at Brookhaven National Laboratory of radioisotopes for medical research, Proceedings of the VII Rassegna Internazionale Elettronica e Nucleare, Rome (1960) 223-244.

\bibitem{Anger} H. O. Anger, Scintillation camera, Review of Scientific Instruments 29 (1958) 27-33.

\bibitem{Varian} D. Varian, The Inventor and the Pilot: Russell and Sigurd Varian, Pacific Books, 1983.

\bibitem{Maciszewski} W. Maciszewski and W. Scharf, Particle Accelerators for Radiotherapy. Present Status and Future, Physica Medica 4 (2004) 137.

\bibitem{Geiger} E. Rutherford and H. Geiger, An electrical method of counting the number of $\alpha$ particles from radioactive substances, Proceedings of the Royal Society, A 81 (1908) 141-161; \\
H. Geiger and W. M\"uller, Elektronenz\"ahlrohr zur Messung schw\"achster Aktivit\"aten, Naturwissenschaften 16 (1928) 617.

\bibitem{Braccini_ATLAS} S. Braccini, Monitored Drift Tube Chamber Production at Laboratori Nazionali di Frascati, proceedings of the 7th International Conference on Advanced Technology and Particle Physics, Como, Italy, October 2001, World Scientific, p. 368; \\
S. Braccini et al., Drift time response of the MDT chambers to a variation of the gas mixture, CERN-ATL-MUON (2004) 012; \\
C. Adorisio, S. Braccini et al., System test of the ATLAS muon spectrometer in the H8 beam at the CERN SPS, Nucl. Instr. Meth. A 593 (2008) 232; \\
ATLAS Collaboration, G. Aad, S. Braccini et al., The ATLAS Experiment at the CERN Large Hadron Collider, JINST 3 (2008) S08003.

\bibitem{HounsÞeld} G. N. HounsÞeld, Computerized transverse axial scanning (tomography): Part l. Description of system, British Journal of Radiology 46 (1973) 1016-1022; \\
A. M. Cormack, Two-Dimensional Reconstruction (CT Scanning) and Recent Topics Stemming from It, Journal of Computer Assisted Tomography 4 (1980) 658-664.

\bibitem{Damadian} R. Damadian, M. Goldsmith and L. Minkoff, NMR in cancer: XVI. Fonar image of the live human body, Physiological Chemistry and Physics 9 (1977) 97Ð100.

\bibitem{Ter-Pogossian} M. M. Ter-Pogossian, M.E. Phelps, E. J. Hoffman and N.A. Mullani, A positron-emission transaxial tomograph for nuclear imaging (PET), Radiology 114 (1975) 89Ð98; \\
M. E. Phelps, E. J. Hoffman, N. A. Mullani and M.M. Ter-Pogossian, Application of annihilation coincidence detection to transaxial reconstruction tomography, Journal of Nuclear Medicine 16 (1975) 210Ð224.

\bibitem{Townsend} T. Beyer, D.W. Townsend, T. Brun et al., A combined PET/CT scanner for clinical oncology, Journal of Nuclear Medicine 41 (2000) 1369Ð1379.

\bibitem{Wilson_1} R. R. Wilson, Radiological use of fast protons, Radiology 47 (1946) 487.

\bibitem{Wilson_2} R. R. Wilson, Range and ionization measurements of high speed protons, Physical Review 60 (1941) 749-753.

\bibitem{Bragg} W. H. Bragg, On the absorption of alpha rays, and on the classification of the alpha rays from radium, Philosophical Magazine S.6 8 (1904) 719-725.

\bibitem{AmaldiBraccini} U. Amaldi and S. Braccini, Present challenges in hadrontherapy techniques, Eur. Phys. J. Plus (2011) 126: 70; reprinted in Part~II of this work.

\bibitem{IAEA_ROP} Radiation Oncology Physics: a handbook for teachers and students, E.B. Podgorsak editor, IAEA Vienna, 2005.

\bibitem{Webb} S. Webb, Intensity-Modulated Radiation Therapy, Institute of Physics Publishing, Bristol and Philadelphia, 2001.

\bibitem{Mackie} T. R. Mackie et al., Tomotherapy: A new concept for the delivery of dynamic conformal radiotherapy, Med. Phys. 20 (1993) 1709.

\bibitem{Lagerwaard} F. Lagerwaard et al., Volumetric Modulated Arc Therapy (RapidArc) for Rapid, Noninvasive Stereotactic Radiosurgery of Multiple Brain Metastases, Int. J. Rad. Onc. Bio. Phy. 72-1 (2008) S530-S530.

\bibitem{Kooy} Proton and charged particle radiotherapy, T. F. DeLaney and H. M. Kooy editors, Lippincott Williams and Wilkins, 2008.

\bibitem{Braccini_1} S. Braccini, Scientific and Technological Development of Hadrontherapy, in Astroparticle, Particle, Space Physics, Radiation Interaction, Detectors and Medical Physics Applications,
Vol. 5, World Scientific (2010) 598-609; reprinted in Part~II of this work.

\bibitem{Braccini_2} S. Braccini, Progress in Hadrontherapy, Nuclear Physics B 172 (2007) 8; reprinted in Part~II of this work.

\bibitem{AmaldiKraft} U. Amaldi and G. Kraft, Radiotherapy with beams of carbon ions, Rep. Prog. Phys. 68 (2005) 1861Ð1882.

\bibitem{Kraft} G. Kraft, Tumor therapy with heavy charged particles, Prog. Part. Nucl. Phys. 45 (2000) S473ÐS544.

\bibitem{Jackel} O. J\"ackel, Medical physics aspects of particle therapy, Radiation Protection Dosimetry 137 (2009) 156Ð166.

\bibitem{PTCOG} Particle Therapy CoOperative Group (PTCOG), ptcog.web.psi.ch.

\bibitem{LibroBianco} The path to the Italian national centre for ion therapy, U. Amaldi and G. Magrin editors, Mercurio, 2005.

\bibitem{Pedroni} E. Pedroni et al., The PSI Gantry 2: A second generation proton scanning gantry, Z. Med. Phys. 14 (2004) 25-34.

\bibitem{Nestle} U. Nestle et al., Biological imaging in radiation therapy: role of positron emission tomography, Phys. Med. Biol. 54 (2009) R1ÐR25.

\bibitem{Braccini_SWAN} S. Braccini, D. M. Aebersold, A. Ereditato, P. Scampoli and K. von Bremen, SWAN: a combined centre for radioisotope production, proton therapy and research in Bern, Proceedings of the Workshop on Physics for Health in Europe, CERN, February 2010, p. 29.

\bibitem{Braccini_SWAN_Isot} S. Braccini, A cyclotron laboratory for radioisotope production and research located in Bern: technical specifications, document of the SWAN Project, 2007; \\
S. Braccini, A cyclotron laboratory for radioisotope production and research located in Bern: comparison, evaluation and selection of the technological equipment and its supplier, document of the SWAN Project, 2007; \\
S. Braccini, A cyclotron laboratory for radioisotope production and research located in Bern: operational concept, document of the SWAN Project, 2007; \\
S. Braccini, A cyclotron laboratory for radioisotope production and research located in Bern: technological survey, document of the SWAN Project, 2007.

\bibitem{Braccini_SWAN_Had} S. Braccini, Particle therapy: technical specifications, document of the SWAN Project, 2007; \\
S. Braccini, Particle therapy: comparison, evaluation and selection of the technological equipment and its supplier, document of the SWAN Project, 2007; \\
S. Braccini, Particle therapy: technological survey, document of the SWAN Project, 2007.

\bibitem{IAEA_465} Cyclotron produced radionuclides: principles and practice, IAEA technical report 465, Vienna, 2008.

\bibitem{IAEA_468} Cyclotron produced radionuclides: physical characteristics and production methods, IAEA technical report 468, Vienna, 2009.

\bibitem{Mettler_Guiberteau} F. A. Mettler and M. J. Guiberteau, Essentials of Nuclear Medicine Imaging, Elsevier, 2005.

\bibitem{Bohr} N. Bohr, Neutron Capture and Nuclear Constitution, Nature 137 (1936) 344-348.

\bibitem{Strangis} R. Strangis et al., Reliable fluorine-18 production at high beam power, proceedings of Cyclotrons and their applications 2007 (2007) 251-253.

\bibitem{Stokelya} M.H. Stokelya et al., 150 $\mu$A $^{18}$F$^-$ Target and Beam Port Upgrade for the IBA 18/9 Cyclotron, AIP Conference Proceedings 1509 (2012) 71.

\bibitem{Cyclotron_Directory} Directory of Cyclotrons used for Radionuclide Production in Member States, IAEA Report, IAEA-DCRP/2006.

\bibitem{Ramamoorthy} N. Ramamoorthy, Production of radioisotopes for medical applications, presented at the Workshop Physics For Health in Europe, CERN, Geneva, February 2010.

\bibitem{Stamm} H. Stamm and P. N. Gibson, Research with light ion cyclotrons in Europe, proceedings of the 5th International Conference on Isotopes, Brussels, 2005.

\bibitem{AmaldiMagrin} U. Amaldi and G. Magrin, Accelerators in medicine, in Accelerators, colliders and their application, Vol. III, S. Myers and H. Schopper eds., Springer (2011).

\bibitem{Schmor} P. Schmor, Review of Cyclotrons for the Production of Radioactive Isotopes for Medical and Industrial Applications, Reviews of Accelerator Science and Technology (RAST) Vol. 4 (2011) 103.
 
\bibitem{BracciniScampoli} S. Braccini and P. Scampoli, Potential research and training activities in physics, biophysics and related applications with SWAN cyclotrons for radioisotope production and proton therapy, document of the SWAN Project, 2009.

%
%

%
%

\bibitem{Gottschalk}	B. Gottschalk, Passive beam spreading in proton radiation therapy, 2004, available at http://huhepl. harvard. edu/~gottschalk;\\
A. M. Koelher et al., Flattening of proton dose distributions of large field radiotherapy, Med. Phys. 4 (1977) 297-301.

\bibitem{Kanematsu} N. Kanematsu et al., Treatment planning for layer-stacking irradiation system for three-dimensional conformal heavy-ion radiotherapy, Med. Phys. 29 (2002) 2823Ð2829.

\bibitem{Ishizaki} A. Ishizaki et al., Development of an irradiation method with lateral modulation of SOBP width using a cone-type filter for carbon ion beams, Med. Phys. 36 (2009) 2222-2227.

\bibitem{Haberer} T. Haberer et al., Magnetic scanning system for heavy ion therapy, Nucl. Instr. Meth. A 330 (1993) 296.

\bibitem{Pedroni_2} E. Pedroni et al., The 200-MeV proton therapy project at the Paul Scherrer Institute: conceptual design and practical realization, Med. Phys. 22 (1995) 37-53.

\bibitem{Bert_Durante} C. Bert and M. Durante, Motion in radiotherapy: particle therapy, Phys. Med. Biol. 56 (2011) R113-R144. 

\bibitem{iRCMA} D. Trbojevic et al., Lattice design of a rapid cycling medical synchrotron for carbon/proton therapy, Proceedings of IPAC2011, San Sebasti\'an, Spain (2011) 2541.

\bibitem{Braccini_RAST2} U. Amaldi, S. Braccini and P. Puggioni, Linacs for hadrontherapy, Reviews of Accelerator Science and Technology (RAST) Vol. 2 (2009) 111; reprinted in Part~II of this work.

\bibitem{Braccini_CYCLINACS} U. Amaldi, S. Braccini et al., Accelerators for hadrontherapy: from Lawrence cyclotrons to linacs, Nucl. Instr. Meth. A 620 (2010) 563; reprinted in Part~II of this work.

\bibitem{Amaldi_LIBO} U. Amaldi et al., LIBO - A linac-booster for protontherapy: construction and test of a prototype, Nucl. Instr. Meth. A 521 (2004) 512-529.

\bibitem{Aberdeen} Aberdeen Bestiary, about 1200: ``IDRA: draco multorum capitum qualis fuit in Lerna''.

\bibitem{Zennaro} R. Zennaro, IDRA: design study of a protontherapy facility, Beam Dynamics Newsletter 36 (2005) 62.

\bibitem{Tesi_Puggioni} P. Puggioni, Radiofrequency Design and Measurements of a Linear Hadron Accelerator for Cancer Therapy, Master Thesis, University of Milano-Bicocca, 2008.

\bibitem{Amaldi_Braccini_arXiv} U. Amaldi, S. Braccini et al., Cyclinacs: Fast-Cycling Accelerators for Hadrontherapy, arXiv:0902.3533 [physics.med-ph].

\bibitem{TERA_HF_NIM} A. Degiovanni et al., TERA high gradient test program of RF cavities for medical linear accelerators, 
Nucl. Instr. Meth. A 657 (2011) 55.

\bibitem{TULIP_Patent} U. Amaldi, S. Braccini, G. Magrin, P. Pearce and R. Zennaro, Ion acceleration system for medical and/or other applications, PCT/IT2006/000879, December 2006
and patent WO 2008/081480 A1.

\bibitem{Schippers} M. Scippers et al., A Next Step in Proton Therapy: Boosting to 350 MeV for Therapy and Radiography Applications, presented at the 
European Cyclotron Progress Meeting (ECPM), PSI, Villigen, 2012.

%
%

\bibitem{Braccini_Bern_Cyclotron} S. Braccini et al., The new Bern cyclotron laboratory for radioisotope production and research, Proceedings of IPAC2011, San Sebasti\'an, Spain (2011) 3618; reprinted in Part~II of this work.

\bibitem{Zhernosekov_Braccini} K. Zhernosekov, S. Braccini et al., A new R\&D radiopharmaceutical laboratory at the Insel Hospital,
Annual Report of the Laboratory for Radio and Environmental Chemistry of the University of Bern and PSI, 2010, p. 57.

\bibitem{Tuerler_Braccini} A. T\"urler, S. Braccini et al., The new GMP (Good Manufacturing Practice) compliant radiopharmaceutical laboratory at the Insel hospital in Bern,
Annual Report of the Laboratory for Radio and Environmental Chemistry of the University of Bern and PSI, 2011, p. 62.

\bibitem{Braccini_CAARI} S. Braccini, The New Bern PET Cyclotron, its Research Beam Line, and the Development of an Innovative Beam Monitor Detector, 
American Institute of Physics (AIP) Conf. Proc. 1525 (2013) 144; reprinted in Part~II of this work.

\bibitem{BeamLineSimulator} K. Dehnel and M. Dehnel, BeamLineSimulator v. 1.4, published by AccelSoft Inc., Del Mar (CA), USA, 2002.

\bibitem{Braccini_CYCLEUR} S. Braccini et al.,  The new Bern PET Cyclotron Laboratory and its Research Beam Line: Commissioning and first Results,
presented at the European Cyclotron Network Meeting (CYCLEUR), Joint Research Centre of the European Commission, Ispra, Italy, November 2012.

\bibitem{EnvAir} G. Lucconi, S. Braccini et al., An innovative system for air contamination detection with high $\beta^+$ sensitivity, 
presented at the congress of the European Association of Nuclear Medicine (EANM), Milan, Italy, October 2012.

\bibitem{PMQ_RFQ} M. Nirkko, S. Braccini et al., An adjustable focusing system for a 2~MeV H$^-$ ion beam line based on permanent magnet quadrupoles, 2013 JINST 8 P02001.

\bibitem{ECPM_PSI} S. Braccini, The new cyclotron laboratory in Bern, presented at the European Cyclotron Progress Meeting (ECPM) 2012, PSI, Villigen, Switzerland, May 2012.

\bibitem{Ion_Beams_12} S. Braccini et al., The Cyclotron Laboratory and the RFQ Accelerator in Bern, proceedings of Ion Beams 2012, International conference on
Multidisciplinary Applications of Nuclear Physics with Ion Beams, Legnaro National Laboratories of INFN, Italy, June 2012, American Institute of Physics, in press.

%
%

%
%

\bibitem{Braccini_MATRIX} S. Braccini et al., MATRIX: an innovative pixel ionization chamber for on-line beam monitoring in hadrontherapy, 
in Astroparticle, Particle, Space Physics, Radiation Interaction, Detectors and Medical Physics Applications,
Vol. 3, World Scientific (2006) 677-681; reprinted in Part~II of this work.

\bibitem{Braccini_SAMBA} S. Braccini et al., An Innovative Strip Ionization Chamber for on-line Monitoring of Hadron Beams, proceedings of the International Conference of the Particle
Therapy Co-Operative Group (PTCOG), Zurich, 2006, p. 44.

\bibitem{Ferretti} R. Ferretti, Test di un Rivelatore a Camera a Ionizzazione a Strip per il Controllo dei Fasci di Protoni in Adroterapia Oncologica, Master Thesis (in Italian), University of Piemonte
Orientale Amedeo Avogadro,
Supervisors: G. Dellacasa and S. Braccini, 2008.

\bibitem{Braccini_Pitta_readout} S. Braccini and G. Pitt\`a, Readout of the SAMBA detector, document of the TERA Foundation, October 2006.

\bibitem{Schneider_Agosteo} U. Schneider et al., Secondary neutron dose during proton therapy using spot scanning, Int. J. Radiation Oncology Biol. Phys. 53 (2002) 244Ð251. 

%
%

\bibitem{Koehler} A.M. Koehler, Proton Radiography, Science 160 (1968) 303.

\bibitem{Cormack} A.M. Cormack and A.M. Koehler, Quantitative proton tomography: preliminary experiments, Phys. Med. Biol. 21 (197) 560-569.

\bibitem{Braccini_Proton_Radiography} S. Braccini et al., First results on proton radiography with nuclear emulsion detectors, Journal of Instrumentation 2010\_JINST\_54\_P09001; reprinted in Part~II of this work.

\bibitem{Ereditato_Emulsions} G. de Lellis, A. Ereditato, K. Niwa, Nuclear Emulsions, C. W. Fabjan and H. Schopper editors, 
Springer Materials, Springer-Verlag Berlin Heidelberg, 2011.

\bibitem{Beam_Halo} S. Braccini et al., Experimental characterization of the dose distribution in the beam halo region of proton
pencil beams with emulsion film detectors, in preparation.

\bibitem{Emulsions_ICTR-PHE_2012} A. Ariga, S. Braccini et al., Research and development activities on nuclear emulsion detectors for
medical applications, presented at the ICTR-PHE 2012 conference, Geneva, February
2012, Radiotherapy and Oncology 102 (2012) S121.

\bibitem{Emulsions_Biodola} A. Ariga, S. Braccini  et al., Advances in nuclear emulsion detectors, presented at Frontier Detectors
for Frontier Physics, La Biodola, Italy, May 2012, submitted for publication to Elsevier.

\bibitem{Emulsions_AEGIS} C. Amsler, S. Braccini et al., A new application of emulsions to measure the gravitational force on
antihydrogen, 2013\_JINST\_8\_P02015.

\bibitem{Neutron_Emulsions} A. Ariga, S. Braccini et al., Experimental characterization of the dose distribution in the beam halo region of proton
pencil beams with emulsion film detectors, in preparation.

\bibitem{INET} A. Ereditato, S. Braccini et al., Innovative Nuclear Emulsion Technologies (INET) for fundamental science and
applications,  Final Scientific Report submitted to the Swiss National Science Foundation (SNSF), December 2012.

%
%

\bibitem{SLIM} L. Badano, S. Braccini et al., Laboratory and In-Beam Tests of a Novel Real-Time Beam Monitor for Hadrontherapy,
IEEE Transactions on Nuclear Science 52 (2005) 830.

\bibitem{Braccini_UniBEam} S. Braccini et al., A beam monitor detector based on doped silica and optical fibres, Journal of Instrumentation 2012\_JINST\_7\_T02001; reprinted in Part~II of this work.

\bibitem{UniBEam_ICTR-PHE_2012} S. Braccini et al., An Innovative Beam Monitor Detector for the New Bern Cyclotron Laboratory, 
presented at the ICTR-PHE 2012 conference, Geneva, February 2012, Radiotherapy and Oncology 102 (2012) S75.

\end{thebibliography}
\end{document}